\documentclass[twocolumn,nofootinbib,superscriptaddress,preprintnumbers,floatfix]{revtex4-1}
\usepackage{amsmath}
\usepackage{amsfonts}
\usepackage{amssymb}
\usepackage{graphicx}
\usepackage{epsfig}
\usepackage{subfigure}
\usepackage{color}
\usepackage{float}

\def\Li{\textrm{Li}}
\def\ln{\textrm{ln}}

\def\nn{\nonumber}
\def\MS{\overline{\rm MS}}

\newcommand{\df}{\mathrm{d}}

\begin{document}


\preprint{\vbox{\hbox{UWTHPH 2014-07}
}}

\title{\Large Variable Flavor Number Scheme for Final State Jets in Thrust}

\author{Piotr Pietrulewicz}
\affiliation{Fakult\"at f\"ur Physik, Universit\"at Wien,
Boltzmanngasse 5, 1090 Vienna, Austria}

\author{Simon Gritschacher}
\affiliation{Mathematical Institute, University of Oxford, Woodstock Road, Oxford,
OX2 6GG, United Kingdom}

\author{Andre H. Hoang}
\affiliation{Fakult\"at f\"ur Physik, Universit\"at Wien,
Boltzmanngasse 5, 1090 Vienna, Austria}
\affiliation{Erwin Schr\"odinger International Institute for Mathematical Physics,
University of Vienna, Boltzmanngasse 9, A-1090 Vienna, Austria}

\author{Ilaria Jemos}
\affiliation{Fakult\"at f\"ur Physik, Universit\"at Wien,
Boltzmanngasse 5, 1090 Vienna, Austria}

\author{Vicent Mateu}
\affiliation{Fakult\"at f\"ur Physik, Universit\"at Wien,
Boltzmanngasse 5, 1090 Vienna, Austria}

\begin{abstract}
We present results for mass effects coming from secondary radiation of heavy quark pairs related to gluon splitting
in the thrust distribution for $e^+\,e^-$ collisions. The results are given in the dijet limit where the hard
interaction scale and the scales related to collinear and soft radiation are widely separated. We account for
the corresponding fixed-order corrections at $\mathcal{O}(\alpha_s^2)$ and the summation of all logarithmic terms
related to the hard, collinear and soft scales as well as the quark mass at N$^3$LL order. We also remove the
${\cal O}(\Lambda_{\rm QCD})$ renormalon in the partonic soft function leading to an infrared evolution equation
with a matching condition related to the massive quark threshold. The quark mass can be arbitrary, ranging from
the infinitely heavy case, where decoupling takes place, down to the massless limit where the results smoothly
merge into the well-known predictions for massless quarks. Our results are formulated in the framework of
factorization theorems for $e^+\,e^-$ dijet production and provide universal threshold corrections for the
renormalization group evolution of the hard current, the jet and soft functions at the scale where the massive
quarks are integrated out. The results represent a first explicit realization of a variable flavor number scheme
for final state jets along the lines of the well-known flavor number dependent evolution of the strong coupling
$\alpha_s$ and the parton distribution functions.
\end{abstract}

\maketitle

\section{Introduction}
The systematic theoretical treatment of mass effects in collider observables represents an important area in
collider phenomenology where substantial progress is required to take full advantage of present and upcoming data.
This concerns in particular the mass of the top quark~\cite{Juste:2013dsa,Jung:2013vpa,Agashe:2013hma} affecting
physics at the Large Hadron Collider (LHC) and at a potential future linear collider, but also the masses of lighter
heavy quarks such as charm~\cite{Alekhin:2012vu,Abramowicz:1900rp,Kovarik:2012te} and bottom 
quarks~\cite{Campbell:2013qaa} relevant e.g.\ in deep-inelastic scattering and event shape analyses at LEP. In this 
context examinations for the top quark may also be considered as study cases for the treatment of new massive colored 
particles that might be discovered in the near future. For inclusive cross sections at hadron colliders a systematic 
approach to treat massive quarks from the large mass limit, where decoupling takes place, continuously down to the
small mass limit, where the description for massless quarks is approached, has been provided by Aivazis, Collins,
Olness and Tung (ACOT)~\cite{Aivazis:1993pi,Aivazis:1993kh}. Their work laid the basis of a variable flavor number
scheme (VFNS) for inclusive processes in hadron collisions which, depending on the size of the quark mass in relation
to the hard scattering and hadronization scales, allows one to factorize infrared-safe quark mass dependent hard coefficient 
corrections from flavor number dependent low-energy parton distribution functions. Since the concepts behind the work of 
ACOT are founded in the separation of close-to-mass-shell and off-shell modes, their approach is along the lines of 
effective field theory methods such as Soft-Collinear Effective Theory (SCET)~\cite{Bauer:2000ew,Bauer:2000yr} and can
be readily incorporated into it. 

\begin{figure}
 \centering
 \includegraphics[width=7cm]{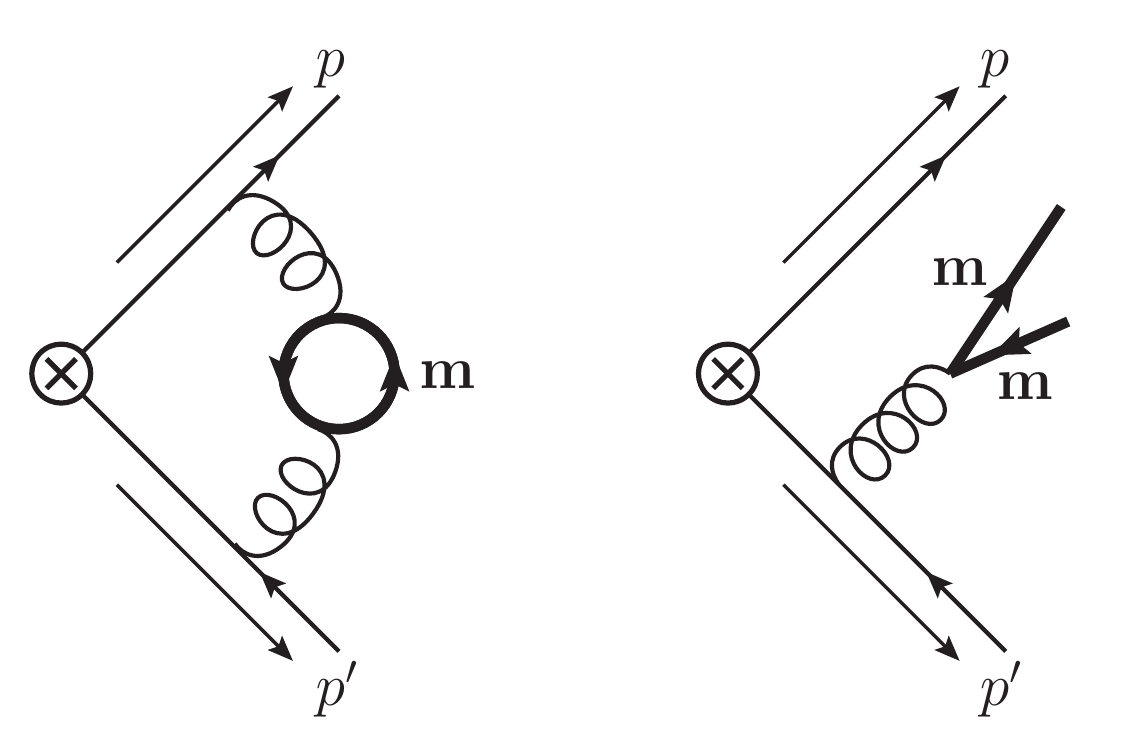}
 \caption{Diagrams at $\mathcal{O}(\alpha_s^2)$ for virtual and real secondary
   radiation of massive quark pairs in primary massless quark
   production.\label{fig:QCDdiag2}} 
\end{figure}

In this paper we present results for a VFNS for final state jets which are initiated by massless quarks and where
massive quarks are produced through the radiation of gluons that split into a massive quark-antiquark pair, see 
Fig.~\ref{fig:QCDdiag2}. We call this type of heavy quark production mechanism {\it secondary} in contrast to the case 
where the massive quarks are produced directly in the hard current interaction, which we call {\it primary}. As for the 
VFNS scheme for inclusive processes in hadron collisions the approach is valid from the large mass limit, where the heavy 
quark decouples, continuously down to the small mass limit, where the predictions approach the known results for massless 
quarks. As a concrete application used to discuss the results we consider the secondary massive quark effects in the
$e^+\, e^-$ thrust distribution, where we define the thrust variable $\tau$ via
\begin{align}
\tau \,=\, 1-T \,=\, 1 -\, \frac{ \sum_i|\vec{n} \cdot \vec{p}_i|}{\sum_j E_j} \,=\,
1-\sum_i \frac{ |\vec{n} \cdot \vec{p}_i|}{Q}\,.
\end{align}
Here $\vec{n}$ is the thrust axis, and the sum is performed over all final state particles with momenta
$\vec{p}_i$ and energies $E_i$.\,\footnote{We define the thrust variable $\tau$ normalized with respect
to the c.m.\ energy $Q$, which is the sum of all energies and also agrees with the variable 
2-jettiness~\cite{Stewart:2010tn}.} 
In the dijet limit where $\tau$ is small, the final state is governed by two narrow back-to-back jets and
the scales of the hard interaction ($\sim Q$), of collinear radiation ($\sim Q\lambda$) and of soft radiation
($\sim Q\lambda^2$) are widely separated (with $\lambda\sim {\rm max}\{\tau^{1/2},(\Lambda_{\rm QCD}/Q)^{1/2}\}$).
In this context the dominant perturbative contributions in the thrust distribution for massless quarks are related
to distributions in $\tau$ and can be factorized into a hard coefficient function, a universal jet function and
a soft function, all of which are defined in a gauge-invariant way. The latter has also a nonperturbative
component which can be parametrized through a convolution with a soft model function that can be determined
through fits to experimental data in a way free of $\mathcal{O}(\Lambda_{\rm QCD})$
renormalons~\cite{Hoang:2007vb}. The logarithmic terms within the dominant contributions
can be summed using the anomalous dimensions of the hard coefficient and the jet and 
soft functions.

Accounting for quark masses in this context adds another non-trivial twist to the factorization setup since quark masses
represent additional scales that can in principle have any hierarchy with respect to the hard, collinear and soft scales,
which themselves depend on the value of $\tau$.\,\footnote{Here we do not account for the effects of hadron masses
(see Ref.~\cite{Mateu:2012nk}).} The possible relations among these scales can therefore vary substantially
even within a single thrust distribution for a fixed c.m.\ energy $Q$. Within the factorization formalism the non-vanishing
value of the quark mass can lead to a flavor number dependent renormalization group (RG) evolution, to threshold
corrections in the evolution when crossing the quark mass scale and to additional mass-dependent fixed-order
corrections in the hard coefficient and the jet and soft functions. The conceptual setup to define these quark mass 
dependent corrections is partially guided by identifying terms that are singular in the quark mass (in the small mass 
limit). The framework of SCET -- properly extended to account for massive quarks -- provides a natural framework
to carry out this task systematically.  

\begin{figure}
 \centering
 \includegraphics[width=7cm]{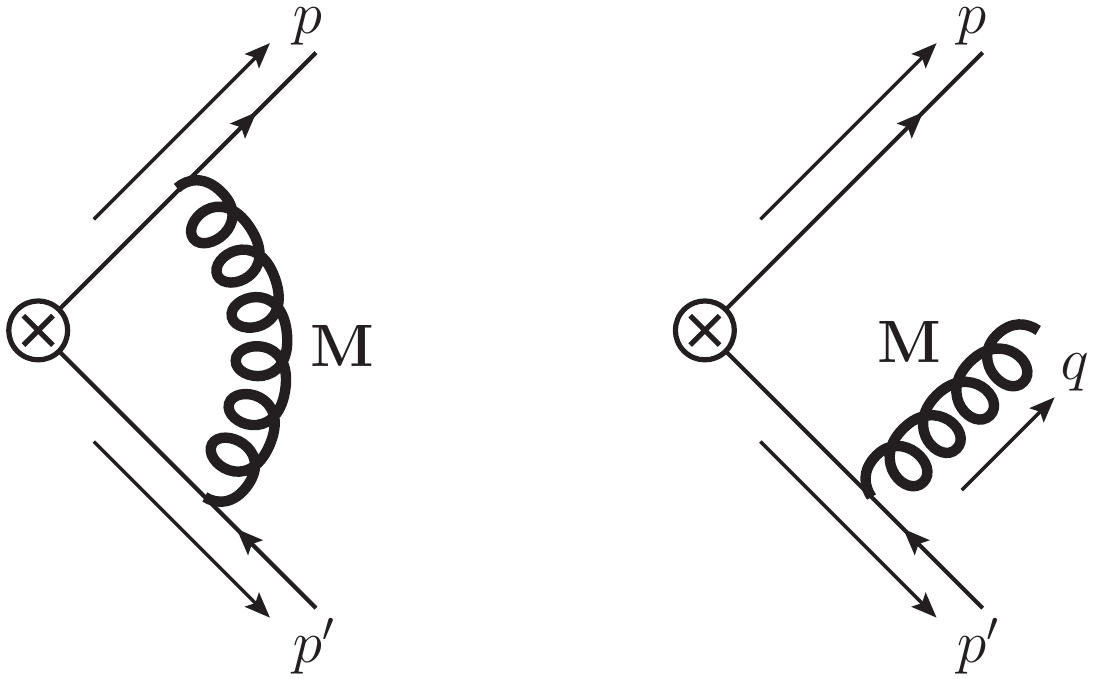}
 \caption{Diagrams at $\mathcal{O}(\alpha_s)$ for virtual and real secondary
   radiation of gluons with mass $M$ in primary massless quark
   production. \label{fig:QCDdiag1}} 
\end{figure}
\
\begin{figure}
 \centering
 \includegraphics[width=\linewidth]{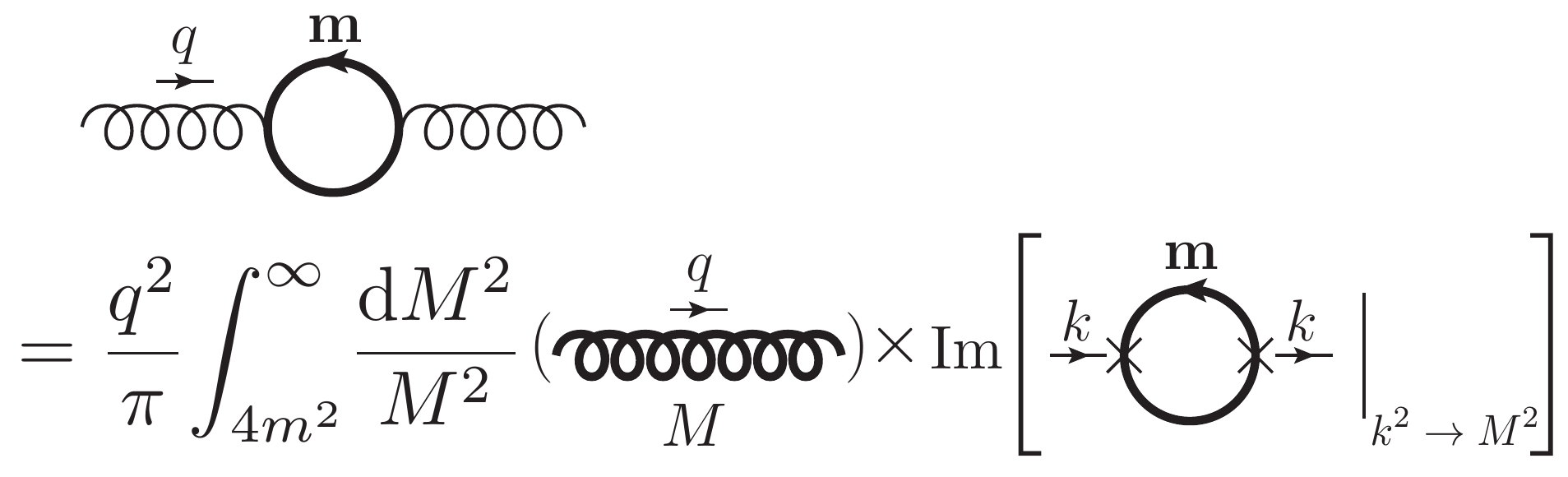}
 \caption{Figure illustrating the dispersion method for the vacuum polarization
   correction to the gluon propagator in the subtracted version with
   $\Pi(q^2=0)=0$ suitable for situations where the massive quark is not
   contributing to the renormalization group evolution. The explicit analytic
   form of the dispersion relations is 
   discussed in Sec.~\ref{sec:dispersion}.
   \label{fig:dispersion}} 
\end{figure}

The approach we propose is based on the seminal paper in Ref.~\cite{Gritschacher:2013pha} where it has been shown that 
the problem of secondary heavy quark production in the thrust distribution is closely related to the production of 
massive gauge bosons, see Fig.~\ref{fig:QCDdiag1}. The connection between these apparently different problems is related 
to the fact that the off-shell intermediate gluon that splits into the massive quark pair has an invariant mass that is 
bounded from below by twice the quark mass. So concerning the setup of the field theoretic modes needed to construct 
effective field theories for the various possible scenarios (related to the possible hierarchies with respect to the 
hard, collinear, soft and mass scales) both problems are quite similar since each of the field theoretic formulations 
has to account for (collinear and/or soft) gluonic modes with a finite typical invariant mass.  In 
Ref.~\cite{Gritschacher:2013pha} we have discussed in detail the field theoretic scenarios to 
treat all possible hierarchies involving the ``gluon mass" and the hard, collinear and soft scales, and we have provided 
the resulting form of the factorization theorems accounting for the required mass corrections, the changes of the RG
evolution above and below the mass scale and the associated threshold corrections at the mass scale (``mass mode
method").
At this level, replacing the gluonic modes by massive quark modes at $\mathcal{O}(\alpha_s^2)$ can be computationally 
more challenging, but does not lead to any additional conceptual complication. It was pointed out in 
Ref.~\cite{Gritschacher:2013pha} that for situations where the produced massive quark and antiquark enter a quantity 
coherently (i.e.\ only the sum of their momenta is relevant in the observable) one can obtain the
corresponding $\mathcal{O}(\alpha_s^2)$
massive quark corrections from the $\mathcal{O}(\alpha_s)$ ``massive gluon" result using a dispersion integration as 
illustrated in Fig.~\ref{fig:dispersion}.  It was in particular demonstrated in Ref.~\cite{Gritschacher:2013pha} that 
also the conceptual issues involving the so-called rapidity divergences, which are related to specific divergences of 
collinear and soft modes with the same typical invariant mass, and the soft-bin subtractions, which avoid double
counting between collinear and soft mass modes, can be dealt with at the level of the massive gluon results.
As was shown in Ref.~\cite{Gritschacher:2013tza}, this computational trick can be also very useful for quantities where 
the quark and antiquark enter independently as the dispersion integral might already give the bulk of the numerical
effects.

The VFNS we propose in this paper is presented and discussed on the basis of the secondary massive quark effects in the
$e^+\,e^-$ thrust distribution at N$^3$LL order in the conventional SCET counting.\,\footnote{At N$^3$LL,
one uses 4-loop cusp and 3-loop noncusp anomalous dimensions and 2-loop matrix elements and matching
conditions.} Concerning a recent thrust
distribution analysis based on the SCET factorization theorem for massless quarks~\cite{Abbate:2010xh} the dominant
secondary massive quark effects come from bottom quarks and represent minor corrections, since that analysis was carried
out with data where the bottom quark mass effects are small due to restrictions in the values of the c.m.\ energy $Q$ and
the fit range in $\tau$.\,\footnote{ In Ref.~\cite{Abbate:2010xh} some effects arising from the finite bottom mass related
to primary production were accounted for, but none related to secondary bottom quark production.} In general, however,
the effects from secondary massive quarks are sizeable for lower $Q$ values or $\tau$ values in the peak and extreme dijet
regions where the cross section depends strongly on the thrust value. Nevertheless, we consider the presentation of
the VFNS given in this work primarily as a non-trivial showcase of the method which might serve as a guideline to
apply the method to other problems involving final state jets.

The content of this work is as follows: In Sec.~\ref{sec:massless} we review the notations and the massless thrust
factorization theorem with an emphasis on the corrections related to the number of massless quarks $n_{\!f}$. In
Sec.~\ref{sec:massmode_setup} we briefly summarize the four relevant field theory scenarios needed to describe the
possible hierarchies of the quark mass w.r.\ to the hard, jet and soft scales. We describe the respective factorization
theorems, which are analogous to the ones given in Ref.~\cite{Gritschacher:2013pha} for the case of the ``massive gluon''.
We show explicitly the results of the ${\cal O}(\alpha_s^2 C_{\!F} T_F)$ massive quark corrections to the hard, jet and soft
functions including the results for the subtractions of ${\cal O}(\Lambda_{\rm QCD})$ renormalon contributions in the soft
function using the gap scheme~\cite{Hoang:2007vb,Hoang:2008fs} and the threshold corrections that arise when the RG
evolution of the hard current coefficient, the jet and soft functions, as well as the gap parameter cross the massive
quark pair flavor threshold. In Sec.~\ref{sec:computations} we describe the computation of the
${\cal O}(\alpha_s^2 C_{\!F} T_F)$ massive quark corrections to the hard and jet functions, which were the remaining unknown 
ingredients in the factorization theorems at this order. In Sec.~\ref{sec:RGconsistency} we show that the massive
threshold corrections are directly related to the matrix elements in the factorization theorems in different
renormalization schemes. This will also illustrate that at the conceptual level a separation into four different effective
field theories is in principle not necessary and that the factorization theorems merge continuously into each other.
Furthermore, we explain in this section how the freedom to set up the RG evolution leads to consistency conditions among
the various threshold corrections underlining their universality. Finally, the results of this paper allow us to predict
the singular $\mathcal{O}(\alpha_s^2 C_{\!F} T_F)$ fixed-order corrections arising from secondary massive quarks, which to
our knowledge have not been given in an explicit form in the literature before. In Sec.~\ref{sec:analysis} we carry out a
numerical analysis at N$^3$LL order for secondary massive bottom and top quarks at different c.m.\ energies. In particular
we investigate the size of the mass corrections compared to the massless limit which turn out to be small in the tail
region, but essential at the peak. Finally, Sec.~\ref{sec:conclusions} contains our conclusions.

\section{The Massless Factorization Theorem for Thrust}\label{sec:massless}
In this section we briefly review the known massless factorization theorem for the most singular contributions of
the thrust distribution in the dijet limit, which are the dominant terms for small values of $\tau$. The main purpose
of this section is to set up the notations and to collect the perturbative results at $\mathcal{O}(\alpha_s^2 C_{\!F} T_F)$
for later comparison and reference concerning the massive quark contributions discussed in later sections. Due to
consistency the massive quark results must yield the massless expressions for vanishing quark mass. The factorization 
theorem for $n_{\!f}$ massless quark flavors reads~\cite{Korchemsky:1999kt,Schwartz:2007ib,Fleming:2007qr}
\begin{align}
& \frac{1}{\sigma_0}\frac{\df\sigma}{\df\tau}\,=\, Q \,\big|C^{(n_{\!f})}(Q,\mu_H)\big|^2\,
\big|U^{(n_{\!f})}_{C}(Q,\mu_H,\mu)\big|^2 \nn \\
&  \times \int\! \df s\!\int \! \df s^\prime\, J^{(n_{\!f})}(s^\prime,\mu_J)\,U^{(n_{\!f})}_J(s-s^\prime,\mu,\mu_J)  \nn \\ 
& \times  \int \!\df\ell \, S^{(n_{\!f})}\Big(Q\,\tau-\frac{s}{Q}-\ell,\mu_S\Big)  U^{(n_{\!f})}_S(\ell,\mu,\mu_S) \, ,
\label{eq:diffsigma0}
\end{align}
where $\sigma_0$ denotes the total partonic $e^{+}e^{-}$ cross-section at tree-level, $C^{(n_{\!f})}(Q,\mu)$ is the hard
current matching condition, $J^{(n_{\!f})}(s,\mu)$ the thrust jet function and $S^{(n_{\!f})}(\ell,\mu)$ the thrust soft function.
The terms $U^{(n_{\!f})}_{C}$, $U^{(n_{\!f})}_J$ and $U^{(n_{\!f})}_S$ are the RG evolution factors for the hard current matching, the
jet and the soft functions, respectively. The superscript $(n_{\!f})$ indicates that the $\MS$ scheme with $n_{\!f}$ dynamic quark
flavors is used for all renormalized quantities, as common when heavy quarks are not involved.

The functions $C^{(n_{\!f})}$, $J^{(n_{\!f})}$ and $S^{(n_{\!f})}$ depend implicitly on $n_{\!f}$ at $\mathcal{O}(\alpha_s)$
through the strong coupling constant. The explicit dependence on $n_{\!f}$ starts at $\mathcal{O}(\alpha_s^2)$.
The expansion of the hard current matching coefficient up to this order has the form
\begin{align}\label{eq:groupstructure}
 & C^{(n_{\!f})}(Q,\mu)= 1 + C^{(n_{\!f},1)}(Q,\mu)+ \Big[C^{(n_{\!f},2)}_{C_{\!F}}(Q,\mu) \nn \\
 & +  C^{(n_{\!f},2)}_{C_{\!A}}(Q,\mu) + C^{(n_{\!f},2)}_{n_{\!f}}(Q,\mu)\Big] +\mathcal{O}(\alpha_s^3)  \, ,
\end{align}
where $C^{(n_{\!f},1)}$, $C^{(n_{\!f},2)}_{C_{\!F}}$, $C^{(n_{\!f},2)}_{C_{\!A}}$, $C^{(n_{\!f},2)}_{n_{\!f}}$ denote the contributions
at $\mathcal{O}(\alpha_s)$, $\mathcal{O}(\alpha_s^2 C_{\!F}^2)$, $\mathcal{O}(\alpha_s^2 C_{\!F} C_{\!A})$,
$\mathcal{O}(\alpha_s^2 C_{\!F} T_F n_{\!f})$, respectively. We use the analogous notation for $J^{(n_{\!f})}$
and $S^{(n_{\!f})}$ as well as for all other perturbative expressions throughout this work. The additional
dependence on a finite quark mass  will be indicated in the arguments.

The massless current matching coefficient $C^{(n_{\!f})}(Q,\mu)$ is determined by matching SCET to QCD.
The renormalized contribution at $\mathcal{O}(\alpha_s^2 C_{\!F} T_F)$ reads~\cite{Matsuura}
($\alpha_s^{(n_{\!f})}\equiv \alpha_s^{(n_{\!f})}(\mu)$)
\begin{align}
& C^{(n_{\!f},2)}_{n_{\!f}}(Q,\mu)=\frac{\big(\alpha_s^{(n_{\!f})}\big)^2 C_{\!F} T_F n_{\!f}}{16\pi^2}
\left\{-\,\frac{4}{9}\,L_{-Q}^3+\frac{38}{9}\,L_{-Q}^2 \right.\nn \\
&\left.-\,\bigg(\frac{418}{27}+\frac{4\pi^2}{9}\bigg)L_{-Q}+\frac{4085}{162}+
\frac{23\pi^2}{27}+\frac{4}{9}\,\zeta_3\right\} \, ,
\label{eq:C0}
\end{align}
where $L_{-Q}=\ln{(-Q^2/\mu^2)}$ (with $Q^2 \equiv Q^2 +i\,0$). The associated contributions to the current 
renormalization factor read
\begin{align}\label{eq:Zc}
&Z^{(n_{\!f},2)}_{C,n_{\!f}} (Q,\mu)= \frac{\big(\alpha_s^{(n_{\!f})}\big)^2 C_{\!F} T_Fn_{\!f}}{16\pi^2}
\left\{\!-\,\frac{2}{\epsilon^3}\right.  \\
&+\left.\frac{1}{\epsilon^2}\left[\frac{4}{3}\,L_{-Q}-\frac{8}{9}\right]+\frac{1}{\epsilon}
\left[-\,\frac{20}{9}\,L_{-Q}+\frac{65}{27}+\frac{\pi^2}{3}\right]\right\}\, . \nn
\end{align}

The jet function is given by a vacuum correlator of two jet fields in SCET and describes the collinear
dynamics of the two back-to-back jets. The renormalized expression for $J^{(n_{\!f})}(s,\mu)$ at
$\mathcal{O}(\alpha_s^2 C_{\!F} T_F)$ reads~\cite{Becher:2006qw}
\begin{align}
\label{eq:J0}
& \mu^2 J^{(n_{\!f},2)}_{n_{\!f}}(s,\mu)= \frac{\big(\alpha_s^{(n_{\!f})}\big)^2 C_{\!F} T_F n_{\!f}}{16\pi^2}
\left\{\left[-\,\frac{4057}{81} \right.\right.\nn \\
& +\left.\frac{136\pi^2}{27}+\frac{32}{9}\,\zeta_3\right]\!\delta(\bar{s}) + 
\left(\frac{988}{27}-\frac{16\pi^2}{9}\right)\!\left[\frac{\theta(\bar{s})}{\bar{s}}\right]_+ \nn \\
&\left.-\frac{232}{9}\!\left[\frac{\theta(\bar{s})\ln{\bar{s}}}{\bar{s}}\right]_+ 
+\frac{16}{3}\!\left[\frac{\theta(\bar{s})\ln^2{\bar{s}}}{\bar{s}}\right]_+ \right\} \, ,
\end{align}
with $\bar{s}\equiv s/\mu^2$. The corresponding contributions to the renormalization factor read
\begin{align}\label{eq:ZJ}
& \mu^2 Z^{(n_{\!f},2)}_{J,n_{\!f}}(s,\mu)=  \frac{\big(\alpha_s^{(n_{\!f})}\big)^2 C_{\!F} T_F n_{\!f}}{16\pi^2}
\left\{\left[\frac{8}{\epsilon^3}-\frac{4}{9\epsilon^2} \right.\right. \\
&\left. -\left. \frac{1}{\epsilon}\left(\frac{242}{27} +\frac{4\pi^2}{9}\right)\right]\!\delta(\bar{s})-
\left(\frac{16}{3\epsilon^2}-\frac{80}{9\epsilon}\right)\!\left[\frac{\theta(\bar{s})}{\bar{s}}\right]_+\right\} \, .
\nonumber
\end{align}

The thrust soft function $S^{(n_{\!f})}\left(\ell,\mu\right)$ describes ultrasoft radiation between the two jets.
It can be written as a convolution of the partonic soft function describing perturbative corrections at the
soft scale and the nonperturbative hadronic soft function~\cite{Hoang:2007vb}. The renormalized expression
for the partonic soft function at $\mathcal{O}(\alpha_s^2 C_{\!F} T_F)$ is~\cite{Kelley:2011ng,Monni:2011gb} 
\begin{align}
& \mu \, \hat{S}^{(n_{\!f},2)}_{n_{\!f}}(\ell,\mu) = \frac{\big(\alpha_s^{(n_{\!f})}\big)^2 C_{\!F} T_F n_{\!f}}{16\pi^2}
\left\{\left[\frac{80}{81}+\frac{74\pi^2}{27} \right.\right. \nn \\
&-\left. \frac{232}{9}\,\zeta_3 \right]\!\delta(\bar{\ell})+\left(-\,\frac{448}{27}+\frac{16\pi^2}{9}\right)
\!\left[\frac{\theta(\bar{\ell})}{\bar{\ell}}\right]_+ \nn \\ 
&\left.+\,\frac{320}{9}\!\left[\frac{\theta(\bar{\ell})\ln \,\bar{\ell}}{\bar{\ell}}\right]_+
-\frac{64}{3}\!\left[\frac{\theta(\bar{\ell})\ln^2{\bar{\ell}}}{\bar{\ell}}\right]_+\right\}
\,.
\label{eq:S0}
\end{align}
with $\bar{\ell}\equiv \ell/\mu$. The corresponding contributions to the renormalization factor read
\begin{align}\label{eq:ZS}
 & \mu\, Z^{(n_{\!f},2)}_{S,n_{\!f}}(\ell,\mu)= \frac{\big(\alpha_s^{(n_{\!f})}\big)^2 C_{\!F} T_F n_{\!f}}{16\pi^2}
 \left\{\left[-\,\frac{4}{\epsilon^3}+\frac{20}{9\epsilon^2} \right.\right. \\
 &\left.+\left.\frac{1}{\epsilon}\left(\frac{112}{27}-\frac{2\pi^2}{9}\right)\right]\! \delta(\bar{\ell}) +
 \left(\frac{16}{3\epsilon^2}-\frac{80}{9\epsilon}\right)\!
 \left[\frac{\theta(\bar{\ell})}{\bar{\ell}}\right]_+\right\} \, .\nn
\end{align}
The overlap between the partonic and the nonperturbative contributions in dimensional regularization
leads to an infrared sensitivity of the perturbative corrections implying factorially enhanced
coefficients (``renormalon''). One can eliminate the renormalon problem for the leading
${\cal O}(\Lambda_{\rm QCD})$ power correction that arises in the operator production expansion (OPE) of the soft
function for $\ell\gg \Lambda_{\rm QCD}$ by introducing a gap parameter $\Delta\sim\Lambda_{\rm QCD}$ in the
hadronic soft model function related to a minimal hadronic energy deposit together with properly defined
perturbative subtractions in the partonic soft function. This cancels the linear sensitivity to small momenta
in the partonic soft function order-by-order in perturbation theory~\cite{Hoang:2007vb,Hoang:2008fs}.
The complete function including the renormalon subtractions has the form
\begin{align}\label{eq:Spartmod}
S^{(n_{\!f})}(\ell,\mu) = & \int \!\df\ell'\, \hat{S}^{(n_{\!f})}\Big(\ell-\ell'-2\,\delta^{(n_{\!f})}(R,\mu),\mu\Big) \nn \\
&\times F\Big(\ell'-2\,\bar{\Delta}^{(n_{\!f})}(R,\mu)\Big) \,,
\end{align}
where $\delta^{(n_{\!f})}(R,\mu)$ is the subtraction series, $\bar{\Delta}^{(n_{\!f})}(R,\mu)$ is the gap parameter which
is free of the $\mathcal{O}(\Lambda_{\rm QCD})$ renormalon and $F$ is the soft model function. A convenient
definition for $\delta^{(n_{\!f})}(R,\mu)$ with consistent RG properties has been given in Ref.~\cite{Hoang:2008fs}
and has the form
\begin{align}\label{eq:delta_thrust}
 \delta^{(n_{\!f})}(R,\mu)=  \frac{R}{2} \,e^{\gamma_E} \frac{\df}{\df\,\ln(ix)} \!\left. \ln \, 
 \tilde{S}^{(n_{\!f})}(x,\mu)\right|_{x=(i R e^{\gamma_E})^{-1}} \, ,
\end{align}
where $\tilde{S}^{(n_{\!f})}$ is the partonic soft function in configuration space,
$\tilde{S}^{(n_{\!f})}(x,\mu)=\int \df \ell \,\hat{S}^{(n_{\!f})}(\ell,\mu) \, e^{-i \ell x}$.
The $\mathcal{O}(\alpha_s^2 C_{\!F} T_F)$ correction reads
\begin{align}\label{eq:delta0}
 &\delta^{(n_{\!f},2)}_{n_{\!f}}(R,\mu)= \frac{\big(\alpha_s^{(n_{\!f})}\big)^2 C_{\!F} T_F n_{\!f}}{16\pi^2} \,R\, e^{\gamma_E}\!
 \left[\frac{8}{3}\, \ln^2\bigg(\frac{\mu^2}{R^2}\bigg) \right. \nn \\  
 & +\left. \frac{80}{9}\,\ln\bigg(\frac{\mu^2}{R^2}\bigg) + \frac{224}{27} +\frac{8\pi^2}{9}\right] \, .
\end{align}
The renormalon-free gap parameter $\bar\Delta^{(n_{\!f})}(R,\mu)$ is related to the ambiguous, but
scale-independent ``bare'' gap parameter $\Delta$ by the relation\footnote{The ``bare'' gap parameter $\Delta$ is 
conceptually analogous to the heavy quark pole mass parameter, so all renormalon-free gap schemes can be
related to each other unambiguously through their relation to the bare $\Delta$. Frequently $\Delta$ is also
called the ``$\MS$ gap parameter''.}
\begin{align}
\Delta = \bar\Delta^{(n_{\!f})}(R,\mu)+\delta^{(n_{\!f})}(R,\mu) \, 
\end{align}
Thus $\bar\Delta^{(n_{\!f})}(R,\mu)$ is scale- and subtraction scheme-dependent.
The natural choice for the scale $R$ of the renormalon-free gap parameter is $R\gtrsim \Lambda_{\rm QCD}$.
On the other hand, the renormalon subtraction $\delta^{(n_{\!f})}(R,\mu)$ should be evaluated for $\mu=\mu_S$,
which is much larger than $\Lambda_{\rm QCD}$ in the tail region, in order to 
achieve a proper cancellation with the IR sensitive terms in the soft function. This requires a resummation if the 
logarithm $\ln(\mu_S/\Lambda_{\rm QCD})$ is large, which can be performed by solving the 
evolution equations~\cite{Hoang:2007vb,Hoang:2008yj,Hoang:2009yr,Hoang:2008fs}
\begin{align}\label{eq:gamma_R}
R\,\frac{\df}{\df R}\, \bar{\Delta}^{(n_{\!f})}(R,R)& = -\,R\, \frac{\df}{\df R}\, \delta^{(n_{\!f})}(R,R) \nn \\
&  \equiv -\,R \,\gamma^{(n_{\!f})}_R [\alpha_s^{(n_{\!f})}(R)]   \, , \\
\mu\,\frac{\df}{\df \mu}\, \bar{\Delta}^{(n_{\!f})}(R,\mu)& \equiv -\, R \,\gamma^{(n_{\!f})}_{\Delta,\mu} \nn \\
&= 2\, R\, e^{\gamma_E} \Gamma_{\rm cusp}^{(n_{\!f})}[\alpha_s^{(n_{\!f})}(\mu)] \, . \label{eq:gamma_Deltamu}
\end{align}
Note that the R-anomalous dimension $\gamma^{(n_{\!f})}_R$, which is responsible for relating $\bar{\Delta}^{(n_{\!f})}$
at different values of $R$ to each other in a way free of the $\mathcal{O}(\Lambda_{\rm QCD})$ renormalon and
free of large IR logarithms, happens to vanish at $\mathcal{O}(\alpha_s)$. Thus the leading anomalous dimension
at $\mathcal{O}(\alpha_s^2)$ depends both linearly and via the strong coupling constant on the number of active
flavors $n_{\!f}$. The $\mathcal{O}(\alpha_s^2 C_{\!F} T_F)$ contribution reads
\begin{align}
 \gamma^{(n_{\!f},2)}_{R,n_{\!f}} = \frac{\big(\alpha_s^{(n_{\!f})}\big)^2 C_{\!F} T_F n_{\!f}}{16\pi^2}\, e^{\gamma_E} 
\! \left(\frac{224}{27}+\frac{8\pi^2}{9}\right) . \label{eq:gamma0_R}
\end{align}

Note that the first moment of the soft model function $2\,\Omega_1$ also becomes a scheme- and scale-dependent
quantity once we employ a renormalon-free gap scheme. The moment parameter $\Omega^{(n_{\!f})}_1 (R,\mu)$ is
then related to the gap parameter $\bar{\Delta}^{(n_{\!f})}(R,\mu)$ via
\begin{align}\label{eq:omega1}
\Omega^{(n_{\!f})}_1 (R,\mu)& \equiv \frac{1}{2}\int_0^\infty\! {\rm d} \ell \, \ell\, 
F\big(\ell-2\,\bar{\Delta}^{(n_{\!f})}(R,\mu)\big) \nn \\
& =  \bar{\Delta}^{(n_{\!f})}(R,\mu) + \frac{1}{2}\int_0^\infty\! {\rm d} \ell \, \ell\, F(\ell) \, .
\end{align}

Large logarithms between the characteristic scales of each sector, $\mu_H$, $\mu_J$ and $\mu_S$, and the final,
common renormalization scale of the factorization theorem $\mu$ are summed by the evolution factors $U^{(n_{\!f})}_{C}$,
$U^{(n_{\!f})}_J$ and $U^{(n_{\!f})}_S$. They satisfy the RG equations
\begin{align}
\label{eq:currentRGE_massless}
 &\mu\,\frac{\df}{\df\mu}\,U^{(n_{\!f})}_{C}(Q,\mu_H,\mu) \nn \\
 &=\gamma^{(n_{\!f})}_C(Q,\mu) \, U^{(n_{\!f})}_{C}(Q,\mu_H,\mu) \, , \\
\label{eq:jetRGE_massless}
 &\mu\,\frac{\df}{\df\mu}\,U^{(n_{\!f})}_{J}(s,\mu,\mu_J) \nn \\
 &=\int\! \df s' \,\gamma^{(n_{\!f})}_J(s-s',\mu)\,U^{(n_{\!f})}_{J}(s',\mu,\mu_J) \, , \\
\label{eq:softRGE_massless}
 & \mu\,\frac{\df}{\df\mu}\,U^{(n_{\!f})}_{S}(\ell,\mu,\mu_S) \nn \\
 &=\int\! \df\ell'\, \gamma^{(n_{\!f})}_S(\ell-\ell',\mu)\,U^{(n_{\!f})}_{S}(\ell',\mu,\mu_S) \, .
\end{align}
The evolution factors are already at LL sensitive to the number of active flavors $n_{\!f}$ due to the running of 
$\alpha_s^{(n_{\!f})}$. Thus, modifying the number of active quark flavors in the evolution affects the thrust
distribution already at LL through its dependence on $\alpha_s$, which happens when a mass threshold is crossed.
The explicit dependence of the anomalous dimensions on $n_{\!f}$ starts at $\mathcal{O}(\alpha_s^2)$, and the
corresponding $\mathcal{O}(\alpha_s^2 C_{\!F} T_F)$ terms read ($Q^2 \equiv Q^2+i\,0$)
\begin{align}
 &\gamma^{(n_{\!f},2)}_{C,n_{\!f}}(Q,\mu)= \frac{\big(\alpha_s^{(n_{\!f})}\big)^2 C_{\!F} T_F n_{\!f}}{16\pi^2}
 \left[\Gamma^{(2)}_{n_{\!f}} \, \ln\bigg(\!\!-\frac{Q^2}{\mu^2}\bigg) \right. \nn \\
 & + \left. \frac{260}{27} + \frac{4\pi^2}{3} \right] ,\\
 &\mu^2 \gamma^{(n_{\!f},2)}_{J,n_{\!f}}(s,\mu)= \frac{\big(\alpha_s^{(n_{\!f})}\big)^2 C_{\!F} T_F n_{\!f}}{16\pi^2}
 \left\{\!-\,4 \,\Gamma^{(2)}_{n_{\!f}} \left[\frac{\theta(\bar{s})}{\bar{s}}\right]_+ \right.\nn \\
 & - \left. \left(\frac{968}{27} + \frac{16\pi^2}{9}\right)\! \delta(\bar{s}) \right\} ,\\
 &\mu \gamma^{(n_{\!f},2)}_{S,n_{\!f}}(\ell,\mu)= \frac{\big(\alpha_s^{(n_{\!f})}\big)^2 C_{\!F} T_F n_{\!f}}{16\pi^2}  \left\{4 
 \,\Gamma^{(2)}_{n_{\!f}} \left[\frac{\theta(\bar{\ell})}{\bar{\ell}}\right]_+ \right. \nn \\
 &+ \left. \left(\frac{448}{27} - \frac{8\pi^2}{9}\right)\!\delta(\bar{\ell}) \right\},
\end{align}
where $\Gamma^{(2)}_{n_{\!f}}=-\,80/9$ denotes the $\mathcal{O}(\alpha_s^2 C_{\!F} T_F n_{\!f})$ coefficient of
the cusp anomalous dimension $\Gamma^{(n_{\!f})}_{{\rm cusp}}$.

In Eq.~(\ref{eq:diffsigma0}) the choice of $\mu$ is arbitrary, and the dependence on $\mu$ cancels exactly working
to any given order in perturbation theory. In the following we will present our results adopting the choice
$\mu=\mu_S$, such that the evolution factor $U_S^{(1)}(\ell,\mu_S,\mu_S)=\delta(\ell)$ and can be dropped from 
Eq.~(\ref{eq:diffsigma0}). The fact that any other choice for $\mu$ can be implemented leads to a consistency relation 
between the renormalization group factors~\cite{Fleming:2007qr}, which reads
\begin{align}\label{eq:consistency_ML}
 Q \,\big|U_C^{(n_{\!f})}(Q,\mu_0,\mu)\big|^2\, U_J^{(n_{\!f})}(Q\ell,\mu,\mu_0) = U_S^{(n_{\!f})}(\ell,\mu_0,\mu) \, .
\end{align}
It can also be written as a relation for the $\mu$-anomalous dimensions, 
\begin{align}\label{eq:consistency_ML2}
2\,{\rm Re}\big[\gamma^{(n_{\!f})}_C(Q,\mu)\big]
\delta(\bar{\ell})+Q\,\mu\,\gamma_J^{(n_{\!f})}(Q\ell,\mu) = -\,\mu\,\gamma_S^{(n_{\!f})}(\ell,\mu) .
\end{align}
In the massive quark case this consistency relation remains intact since the UV divergences are mass independent.
However, since the quark mass represents an additional relevant scale the factorization theorem exhibits a richer 
structure due to the increased number of scales, and additional consistency relations emerge.

\section{Mass Mode Setup and Summary of Results}\label{sec:massmode_setup}
In this section we briefly review the mass mode setup of Ref.~\cite{Gritschacher:2013pha}, which is based on four 
different effective field theory scenarios associated to the hierarchies between the hard, jet and soft scales and the 
quark mass. We also discuss the form of the resulting factorization theorems and present the final results for all 
mass-dependent perturbative corrections. The explicit calculations are described in detail in Sec.~\ref{sec:computations}. 
An alternative conceptual (and likely more practical)
view based only on a single effective theory below the hard scale, but
with renormalization schemes for the different components of the factorization theorem that vary according to the relation 
between the hard, jet and soft scales and the quark mass is described in Sec.~\ref{sec:RGconsistency}. 

For the discussion of the mass mode method we consider a generic setup with {\it one} massive quark flavor with mass $m$ 
in addition to $n_l$ massless flavors, and our notation is set up accordingly. It is convenient to define the ratio
\begin{align}
 \lambda_m =\frac{m}{Q}\,,
\end{align}
in addition to the regular power counting parameter $\lambda\sim {\rm max}\{\tau^{1/2},(\Lambda_{\rm QCD}/Q)^{1/2}\}$
that is already present in the purely massless setup. From the field theoretic point of view we consider $n$-, 
$\bar{n}$-collinear and soft mass modes. If kinematically allowed, these can have the momentum scaling and virtualities of 
their massless counterparts, but in addition one has to account for the fluctuations around their mass-shell which have 
the scaling \mbox{$p_n^\mu \sim Q(\lambda_m^2,1,\lambda_m)$}, $ p^\mu_{\bar{n}} \sim Q(1,\lambda_m^2,\lambda_m)$ for the $n$-, 
$\bar{n}$-collinear mass modes, respectively, and \mbox{$p_s^\mu \sim Q(\lambda_m, \lambda_m, \lambda_m)$} for the soft mass modes. 
Since the typical invariant masses of the mass modes are bounded from below by $p_n^2 \sim p_{\bar{n}}^2 \sim p_s^2 \sim 
Q^2 \lambda_m^2 \sim m^2$, dynamic real radiation effects can only occur if the typical collinear or soft scales are 
bigger than $m^2$.  This means that depending on the relative sizes of $\lambda_m$ and $\lambda$, collinear and soft 
mass mode fluctuations might not both contribute at the same time to the matrix elements, i.e.\ the jet and the soft 
functions, respectively. Since the hierarchy between the hard scale $Q$, the jet scale $Q \lambda$ and the soft scale
$Q \lambda^2$ and their relation to the quark mass $m$ can vary substantially, there are also different scenario-dependent
threshold corrections when the RG evolution crosses the mass scale and the massive quark flavor is integrated 
out.\,\footnote{Throughout this work we adopt the convention that the effects of the massive quark flavor in the 
factorization theorems are integrated out globally at the scale $\mu_m~\sim m$.} 

Concerning Feynman rules, the collinear massive quark interactions are determined from a massive quark collinear 
Lagrangian~\cite{Leibovich:2003jd} which is a straightforward generalization of the massless collinear Lagrangian. In 
practice, since the collinear sector is essentially just a boosted version of usual QCD, the effects of the secondary 
massive quarks in the collinear sector can be calculated using regular QCD Feynman rules. We emphasize, however, that the 
consistency for calculations in the collinear sector with massive modes involves additional (non-vanishing) soft mass mode 
bin subtractions~\cite{Chiu:2009yx} in the collinear loop integrations to avoid double counting with the soft sector and 
to maintain collinear gauge invariance. As we have shown in Ref.~\cite{Gritschacher:2013pha}, these soft mass mode bin 
subtractions are essential to obtain meaningful and gauge-invariant results. Concerning the interactions within the soft 
sector, the Feynman rules are anyway given by the usual QCD interactions 
and Feynman rules. This is sufficient for the treatment of the secondary soft massive quarks in this work.

\begin{figure*}
  \includegraphics[width=0.9\linewidth]{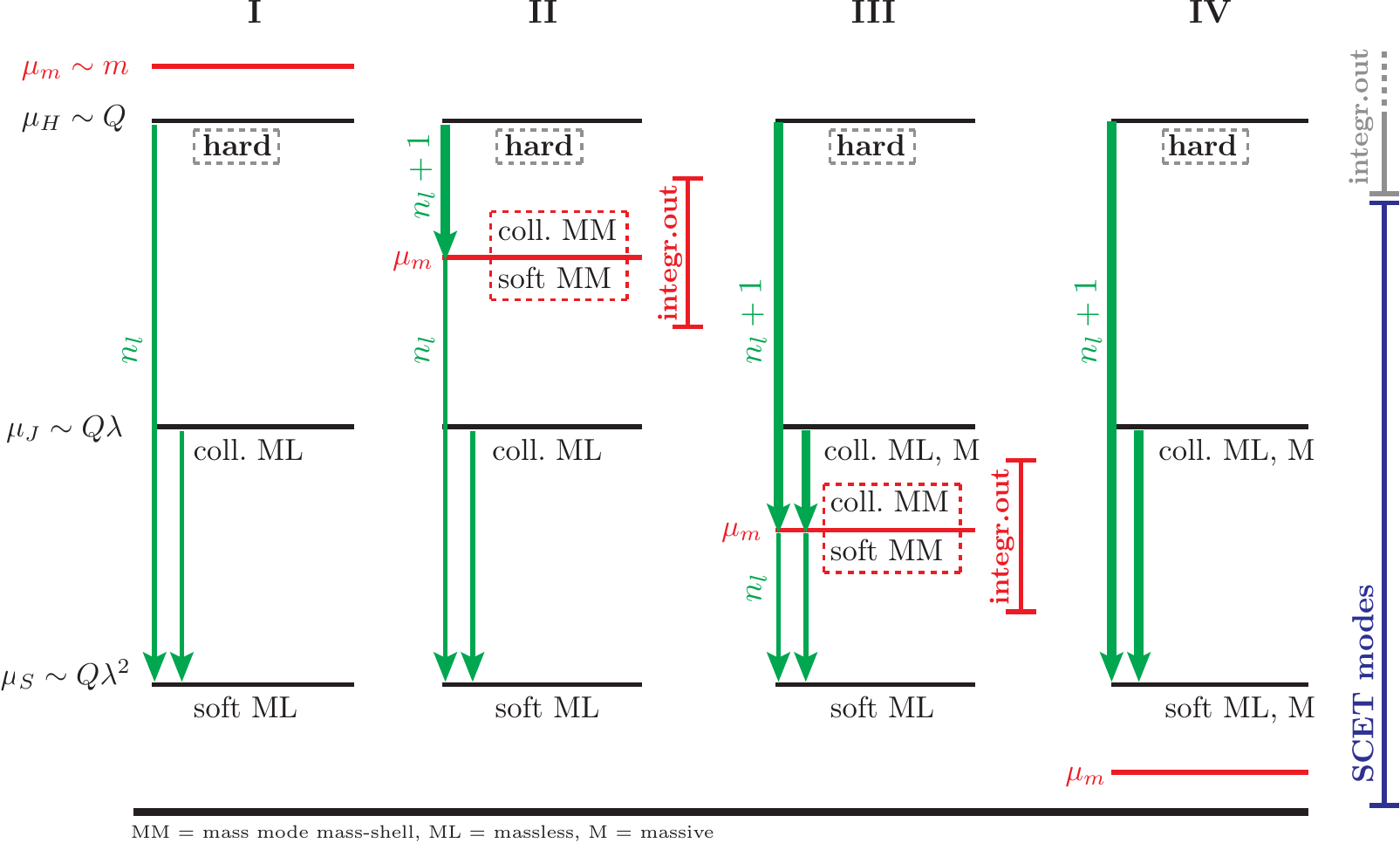}
  \caption{\label{fig:scenarios}
The different scenarios depending on the hierarchy between the mass scale $\mu_m$ and the hard, jet and ultrasoft
scales. MM indicates mass-shell scaling, ML the massless one. With M we denote modes that have a mass $m$ but scale as 
their massless counterparts. The renormalization group evolution is also shown in the top-down evolution from the hard 
scale $\mu_H$ down to $\mu=\mu_S$. When the mass scale is crossed the mass-shell fluctuations are integrated out
(dashed box). This leads to a matching condition and to a change in the evolution factor.} 
\end{figure*}

Note that some of the notation, the formulation of the factorization theorems and the organization of the RG evolution we 
use for the presentation of the results in this section are related to the choice that the final renormalization scale 
$\mu$ is set to be equal to the soft scale, $\mu=\mu_S$ such that the soft evolution factor $U_S(\ell,\mu_S,\mu_S)=\delta(\ell)$ 
and can be dropped. Thus only the current and jet function RG evolution factors $U_C$ and $U_J$, respectively, appear. The 
RG pattern of this ``top-down" approach together with a graphical display of how the collinear and soft massive modes give 
contributions is illustrated in Fig.~\ref{fig:scenarios}. Clearly, any other choice for $\mu$ is possible and can 
significantly affect the form as well as the interpretation of the various components of the factorization theorems in the 
different scenarios. Since the number of possibilities in connection with the different scenarios and the choices for 
$\mu$ in the factorization theorems proliferates strongly, we 
postpone a more general discussion to Sec.~\ref{sec:RGconsistency}, where we focus more on the RG properties of the hard 
coefficient and the jet and soft functions rather than on the full factorization theorem. This allows us to streamline the 
discussion significantly and to generalize our results to other processes.

\subsection{Scenario I: $m>Q>Q\lambda>Q\lambda^2$}
When the mass $m$ is larger than the hard scale $Q$ the massive quark is not described in SCET, but integrated out when 
SCET is matched to QCD. The factorization theorem is the one for $n_l$ massless fermions in analogy to 
Eq.~(\ref{eq:diffsigma0}) up to the hard current matching coefficient which acquires an additional contribution due to
the heavy quark,
\begin{align}
 & \frac{1}{\sigma_0} \frac{\df\sigma}{\df\tau}= Q\,\big|C^{(n_l)}(Q,m,\mu_H)\big|^2\,
 \big|U^{(n_l)}_{C}(Q,\mu_H,\mu_S)\big|^2 \nn \\
 & \times \int \!\df s\!  \int\! \df s'\, J^{(n_l)}(s',\mu_J)\, U^{(n_l)}_J(s-s',\mu_S,\mu_J) \nn \\
 & \times S^{(n_l)}\Big(Q\,\tau-\frac{s}{Q},\mu_S\Big)\,,
\label{eq:diffsigmaI}
\end{align}
where
\begin{align}
C^{(n_l)}(Q,m,\mu)= C^{(n_l)}(Q,\mu)+F^{(n_l,2)}_{\rm QCD} (Q,m) \, .
\label{eq:hardcoeffI}
\end{align}
The term $F^{(n_l,2)}_{\rm QCD}$ represents the massive quark bubble contribution to the QCD current form factor,
see the diagram in Fig.~\ref{fig:QCDdiag2}(a). Scenario I is designed to show manifest decoupling in the infinite
mass limit, i.e.\
\begin{align}
 C^{(n_l)}(Q,m\rightarrow \infty,\mu) \rightarrow C^{(n_l)}(Q,\mu) \, .
\end{align}
This is achieved in $F^{(n_l,2)}_{\rm QCD}$ by two ingredients. First, the on-shell condition for the external
quarks is related to a subtraction of the form factor at $Q^2=0$ concerning the virtual secondary massive quark
effects. Second, the massive quark bubble contribution to the strong coupling constant is renormalized in the
on-shell scheme rather than in $\MS$.\,\footnote{The massless quark bubble contributions are still renormalized
in the $\MS$ scheme as usual.} So the massive quark is not an active dynamic flavor and does not contribute to
the RG evolution of the strong coupling. One can derive the expression by first calculating the corresponding
one-loop diagram with a massive gauge boson in Fig.~\ref{fig:QCDdiag1}(a) and then using the subtracted form
of the dispersion relation according to Fig.~\ref{fig:dispersion}. This yields\,\footnote{Throughout the paper
we suppress the dependence on the renormalization scale $\mu$ in the arguments of terms which are just implicitly
depending on the renormalization scale through $\alpha_s$ and the mass $m$.}
($\alpha_s^{(n_l)}=\alpha_s^{(n_l)}(\mu)$)
\begin{align}\label{eq:F_QCD}
F^{(n_l,2)}_{\rm QCD} (Q,m)&\equiv F^{(2,\rm OS)}_{\rm QCD}(Q,m) \nn \\
& =\frac{\big(\alpha_s^{(n_l)}\big)^2 C_{\!F} T_F}{16\pi^2} f^{(2)}_{\rm QCD}(m/Q) \, ,
\end{align}
where the function $f^{(2)}_{\rm QCD}(x)$ is given by~\cite{Kniehl1990,Hoang:1995fr}
\begin{align}\label{eq:f_QCD}
& \!\!f^{(2)}_{\rm QCD}(x)= \Big(\frac{46}{9}\,r^3+\frac{10}{3}\,r\Big)\!
\bigg[\Li_2\bigg(\frac{r-1}{r+1}\bigg) - \Li_2\bigg( \frac{r+1}{r-1}\bigg)\bigg] \nn\\
&\!\!+ \Big(\!-r^4 + 2\, r^2+\frac{5}{3}\Big)\!
\bigg[\Li_3\bigg(\frac{r-1}{r+1}\bigg) + \Li_3\bigg(\frac{r+1}{r-1}\bigg)-2\,\zeta_3\bigg]\nonumber\\
&\!\!+\Big(\frac{110}{9}\,r^2+\frac{200}{27}\Big)\ln\Big(\frac{1-r^2}{4}\Big)
+ \frac{238}{9}\,r^2+\frac{1213}{81}\, ,
\end{align}
and we have defined
\begin{align}
 x^2=\frac{m^2}{Q^2+i\,0} \,, \quad r= \sqrt{1+4\,x^2}\,.
\end{align}
In the limit $m\to\infty$ the massive quark decouples indeed,
i.e.\ $f^{(2)}_{\rm QCD}(x) \rightarrow 0$ for $x \to \infty$. For light fermions, i.e.\ $x\rightarrow 0$, we find
\begin{align} \label{eq:f_QCD0}
& \left.f^{(2)}_{\rm QCD}(x)\right|_{x \rightarrow 0} = \frac{4}{9}\,\ln^3{(-x^2)}+\frac{38}{9}\,\ln^2{(-x^2)} \\
& +\bigg(\frac{530}{27}+\frac{4\pi^2}{9}\bigg)\ln{(-x^2)}+\frac{3355}{81}+\frac{38\pi^2}{27}-\frac{16}{3}\,\zeta_3 \, ,
\nn
\end{align}
with subleading corrections going as $\mathcal{O}(x^2)$. Eq.~(\ref{eq:f_QCD0}) does not bear any similarity to the
massless result of Eq.~(\ref{eq:C0}) and further exhibits large unresummed mass logarithms. Thus the QCD result of
Eq.~(\ref{eq:F_QCD}) is not suitable for taking the massless limit. This is because Eq.~(\ref{eq:F_QCD}) still contains
mass mode on-shell contributions which must be subtracted prior to taking the limit $m \ll Q$. This procedure is
described in scenario II.

\subsection{Scenario II: $Q>m>Q\lambda>Q\lambda^2$}
The mass $m$ is below the hard scale, but still above the jet and the soft scales. It is our aim (i) to resum the mass
logarithms in Eq.~(\ref{eq:f_QCD0}) and (ii) to determine the hard current matching coefficient such that it contains
no mass-singularities and in particular approaches the massless limit for $m\rightarrow 0$. The collinear and soft mass
modes are included into the SCET setup, so that they render the hard coefficient IR safe by subtracting the mass-shell
contributions in the matching procedure. They contribute as dynamic degrees of freedom to the RG evolution above $m$.
In the RG evolution of the current from the hard to the jet scale the mass-shell fluctuations are finally integrated
out at the scale $m$. The mass mode effects are purely virtual because the jet scale $Q\lambda$, the typical invariant
mass for real collinear particle radiation is below $m$. Therefore, the jet and soft functions as well as their RG 
evolution factors towards scales smaller than $m$ coincide with the ones for $n_l$ massless quarks. The factorization 
theorem reads
\begin{align}
 & \frac{1}{\sigma_0}\frac{\df\sigma}{\df\tau}= Q\, \big|C^{(n_l+1)}(Q,m,\mu_H)\big|^2\,
 \big|U^{(n_l+1)}_{C}(Q,\mu_H,\mu_m)\big|^2 \nn \\
 & \times \big|{\mathcal{M}_{C}(Q,m,\mu_m)}\big|^2 \,\big|U^{(n_l)}_{C}(Q,\mu_m,\mu_S)\big|^2
 \nonumber\\
&\times \int\! \df s \! \int \!\df s'\,
 J^{(n_l)}(s',\mu_J)\, U^{(n_l)}_J(s-s',\mu_S,\mu_J)  \nn \\
 & \times S^{(n_l)}\Big(Q\,\tau-\frac{s}{Q},\mu_S\Big)\,.
\label{eq:diffsigmaII}
\end{align}
Compared to $C^{(n_l)}(Q,m,\mu_H)$ in Eq.~(\ref{eq:hardcoeffI}) the hard current coefficient $C^{(n_l+1)}(Q,m,\mu_H)$ 
acquires a subtractive contribution arising from the non-vanishing SCET diagrams involving virtual collinear and soft
mass modes and contributions related to the use of the $\MS$ renormalization prescription for the strong coupling
rather than the OS one concerning the massive quark bubble. The latter correction means that the massive quark now 
contributes to the RG evolution, and that we employ $\alpha_s^{(n_l+1)}$. The result for $C^{(n_l+1)}(Q,m,\mu)$ reads
\begin{align}
C^{(n_l+1)}(Q,m,\mu)= C^{(n_l+1)}(Q,\mu)+ \delta F^{(n_l+1,2)} (Q,m)\,.
\label{eq:hardcoeffII}
\end{align}
The term $\delta F^{(n_l+1,2)}$ represents the corrections due to the non-vanishing mass of the heavy quark and can
be written as ($\alpha_s^{(n_l+1)}=\alpha_s^{(n_l+1)}(\mu)$)
\begin{align}\label{eq:F_II}
 &\delta F^{(n_l+1,2)} (Q,m) =  \frac{\big(\alpha_s^{(n_l+1)}\big)^2 C_{\!F} T_F}{16\pi^2} \nn \\
 &\times \left[f^{(2)}_{\rm QCD}(m/Q) - \left.f^{(2)}_{\rm QCD}(m/Q)\right|_{m\rightarrow 0}\right] \, ,
\end{align}
which can be read off Eqs.~(\ref{eq:f_QCD}) and~(\ref{eq:f_QCD0}). All calculational steps are explained in detail
in Sec.~\ref{sec:hardmatching}. One can check explicitly that all of the IR divergent mass-shell contributions are
removed and that the massless limit for Eq.~(\ref{eq:hardcoeffII}) is recovered for $m \rightarrow 0$, i.e.\
$\delta F^{(n_l+1,2)} (Q,m) \stackrel{m\rightarrow 0}{\longrightarrow} 0$. We note that in the results of the
collinear, the soft and the soft mass mode bin contributions there are rapidity divergences that cancel in the sum
of all terms. We stress, however, that for $\mu \sim Q$ no large (rapidity) logarithm remains in the hard current
matching.

Note that the UV divergences of the bare SCET form factor are insensitive to the non-vanishing quark mass, such
that we get in total the UV divergences from Eq.~(\ref{eq:Zc}) for $n_{\!f} = n_l + 1$. This argument does not rely on
a specific order in $\alpha_s$, so the evolution factor $U^{(n_l+1)}_{C}$ obeys the RG equation
\begin{align}
 &\mu\,\frac{\df}{\df\mu}\,U^{(n_l+1)}_{C}(Q,\mu_H,\mu) \nn \\
 &=\gamma^{(n_l+1)}_C(Q,\mu) \, U^{(n_l+1)}_{C}(Q,\mu_H,\mu) \, 
\end{align}
to all orders in perturbative QCD for scales $\mu>\mu_m$.

At the scale $\mu_m$ the mass-shell fluctuations of the collinear and soft mass modes are integrated out.
This leads to the current mass mode matching coefficient $\mathcal{M}_C(Q,m,\mu_m)$, 
which is the analogue of the well-known matching correction between the strong coupling schemes with $n_l+1$ and $n_l$ 
running dynamic flavors, $\alpha_s^{(n_l+1)}$ and $\alpha_s^{(n_l)}$ respectively. The result reads ($\alpha_s\, 
\ln(m^2/Q^2) \sim \mathcal{O}(1)$) 
\begin{widetext}
\begin{align}
& \mathcal{M}_C(Q,m,\mu_H,\mu_m) =  1+\left[\frac{\big(\alpha_s^{(n_l+1)}\big)^2
C_{\!F} T_F}{(4\pi)^2}\ln\left(\frac{\mu_H^2}{\mu_m^2}\right)
\left\{-\frac{4}{3}\,L_m^2-\frac{40}{9}\,L_m-\frac{112}{27}\right\}\right]_{\mathcal{O}(\alpha_s)} \nn \\
& + \left[\frac{\big(\alpha_s^{(n_l+1)}\big)^2 C_{\!F} T_F}{(4\pi)^2}
\left\{\frac{4}{9}\,L_m^3+\frac{38}{9}\,L_m^2+ \left(\frac{242}{27}+\frac{2\pi^2}{3}\right)\!L_m - 
\ln\bigg(\!\!-\frac{Q^2}{\mu_H^2}\bigg)\!\left\{\frac{4}{3}\,L_m^2+\frac{40}{9}\,L_m+\frac{112}{27}\right\} 
+\frac{875}{54}+\frac{5\pi^2}{9}-\frac{52}{9}\,\zeta_3\right\} \right. \nn \\
& + \frac{\big(\alpha_s^{(n_l+1)}\big)^3 C_{\!F} T_F}{(4 \pi)^3} \,\ln\bigg(\frac{\mu_H^2}{\mu_m^2}\bigg)\!
\left\{L_m^3\!\left[\frac{88}{27}\,C_{\!A}-\frac{64}{27}\,T_F-\frac{32}{27}\,T_F n_l\right] 
+L_m^2\!\left[\left(-\frac{92}{9}+\frac{8}{9} \pi ^2\right)\! C_{\!A} + 12\,C_{\!F} -\frac{160}{27}\,T_F\right] \right. \nn \\
 &+ \left.L_m\left[\left(-\frac{620}{81}+\frac{80\pi^2}{27}-\frac{112}{3}\, \zeta_3 \right)\!C_{\!A} 
 +\left(-\frac{4}{3}+32\,\zeta_3\right)\!C_{\!F} -\frac{1088}{81}\,T_F n_l -\frac{992}{81}\,T_F \right] -
 \frac{\mathcal{M}_3^{C,+}}{C_{\!F} T_F} \right\} \nn \\
 & + \left. \frac{\big(\alpha_s^{(n_l+1)}\big)^4 C_{\!F}^2 T_F^2}{(4\pi)^4}\, 
 \ln^2\bigg(\frac{\mu_H^2}{\mu_m^2}\bigg)\!\left\{\frac{8}{9}\,L_m^4+
 \frac{160}{27}\,L_m^3+\frac{416}{27}L_m^2+\frac{4480}{243}\,L_m+
 \frac{6272}{729}\right\} \right]_{\mathcal{O}(\alpha_s^2)} \, ,
\label{eq:matchingII}
\end{align}
\end{widetext}
where $Q^2=Q^2+i0$, $L_{m}\equiv \ln{\left(m^2/\mu_m^2\right)}$ and $\alpha_s^{(n_l+1)}=\alpha_s^{(n_l+1)}(\mu_m)$.
The quark mass $m\equiv \overline{m}(\mu_m)$ is given in the $\MS$ scheme. We see that the result contains large 
logarithms $\ln(\mu_m^2/\mu_H^2)$, which are not summed by the RG $\mu$-evolution of the current. These logarithms
are related to rapidity singularities that arise in the overlap region between the collinear and soft mass modes, all 
having invariant masses of order $m^2$. Analogous logarithmic terms were also found for the $\mathcal{O}(\alpha_s)$ 
massive gluon results discussed in Ref.~\cite{Gritschacher:2013pha}. There are several approaches to resum these rapidity 
logarithms~\cite{Becher:2011dz,Chiu:2012ir,Chiu:2011qc}, but the outcome is just a simple exponentiation which yields the 
term at $\mathcal{O}(\alpha_s^4\, \ln^2(m^2/Q^2))$ in Eq.~(\ref{eq:matchingII}). We refer to Ref.~\cite{Hoang:2015vua} for 
a calculation of the anomalous dimension in rapidity space. We note that through the rapidity 
RG evolution $\mathcal{M}_C$ depends at each order on two rapidity scales. For simplicity we  correlate them with the two 
invariant mass scales $\mu_H$ and $\mu_m$. We stress, however, that the dependence of $\mathcal{M}_C$ on the hard matching 
scale $\mu_H$ is actually spurious and cancels in an expansion at fixed-order in $\alpha_s$. The existence of the large 
logarithms has the important consequence that the $\mathcal{O}(\alpha_s^2 \, \ln(m^2/Q^2))$ corrections in 
Eq.~(\ref{eq:matchingII}) enter at the same order as the $\mathcal{O}(\alpha_s)$ fixed-order corrections based on the 
counting $\alpha_s\, \ln(m^2/Q^2) \sim \mathcal{O}(1)$ and thus contribute already at N$^2$LL order where one-loop 
fixed-order corrections to the hard coefficient and the jet and the soft functions are accounted for. At N$^3$LL
order we therefore need the terms at $\mathcal{O}(\alpha_s^3\, \ln(m^2/Q^2))$ and
$\mathcal{O}(\alpha_s^4\, \ln^2(m^2/Q^2))$.\,\footnote{In the primed counting one might still need to
distinguish between terms enhanced by rapidity logarithms (and related to terms summed by the rapidity RGE)
and the remaining terms in the series for the mass mode threshold factors.}
We have indicated this counting by using the subscripts $``\mathcal{O}(\alpha_s)"$ and 
$``\mathcal{O}(\alpha_s^2)"$ in the result of Eq.~(\ref{eq:matchingII}). From the $\mathcal{O}(\alpha_s^3\, \ln(m^2/Q^2))$ 
terms the contributions explicitly depending on $\mu_m$ can be inferred using the $\mu_m$-independence 
of the factorization theorem and the explicit form of the current evolution factors $U_C^{(n_l+1)}$ and $U_C^{(n_l)}$. In 
Eq.~(\ref{eq:matchingII}) the full form of the $\mathcal{O}(\alpha_s^3\, \ln(m^2/Q^2))$ term is displayed with the constant 
$\mathcal{M}_3^{C,+} $ which cannot be determined from RG arguments. This 
constant corresponds to a rapidity logarithm that is physically unrelated to logarithms of $\mu_m$. Details of these
computations can be found in Sec.~\ref{sec:currentmassmodematching}. 

\subsection{Scenario III: $Q>Q\lambda>m>Q\lambda^2$}
The mass is between the jet and the soft scales. The current evolution is the same as the one in scenario II and the soft 
function still includes only the effects of the $n_l$ massless quarks. Since the massless as well as the massive collinear 
modes both can now fluctuate in the collinear sector the difference to scenario II concerns the jet function, where 
additional massive real and virtual contributions arise. The setup is constructed such that it (i) sums all mass 
logarithms that arise in the evolution of the jet function and (ii) ensures that the jet function approaches the known 
massless result for $n_l+1$ flavors in the limit $m\rightarrow 0$. In analogy to the current, the RG evolution of the jet 
function is performed with $n_l+1$ flavors above the mass threshold. Collinear mass-shell fluctuations are integrated out 
at the mass scale yielding a collinear mass mode matching coefficient $\mathcal{M}_J$, and the evolution continues with
$n_l$ light quarks down to the soft scale. Overall, the factorization theorem in this scenario has the form
\begin{align}
 & \frac{1}{\sigma_0} \frac{\df\sigma}{\df\tau}= Q\,\big|C^{(n_l+1)}(Q,m,\mu_H)\big|^2\,
 \big|U^{(n_l+1)}_{C}(Q,\mu_H,\mu_m)\big|^2 \nn \\
 & \times \big|{\mathcal{M}_{C}(Q,m,\mu_m)}\big|^2\, \big|U^{(n_l)}_{C}(Q,\mu_m,\mu_S)\big|^2
 \nonumber\\
&\times\int\! \df s  \!\int\! \df s' \!\int \!\df s'' \!\int\! \df s''' J^{(n_l+1)}(s''',m,\mu_J)  \nn \\
&\times U^{(n_l+1)}_J(s''-s''',\mu_m,\mu_J) \, \mathcal{M}_J(s'-s'',m,\mu_m) \nn \\
& \times U^{(n_l)}_J(s-s',\mu_S,\mu_m) \,  S^{(n_l)}\Big(Q\,\tau-\frac{s}{Q},\mu_S\Big),
\label{eq:diffsigmaIII}
\end{align}
where the matching coefficients $C^{(n_l+1)}(Q,m,\mu_H)$ and $\mathcal{M}_{C}(Q,m,\mu_m)$ are the same as in scenario~II,
see Eqs.~(\ref{eq:hardcoeffII}) and (\ref{eq:matchingII}). The jet function $J^{(n_l+1)}(s,m,\mu)$ contains contributions
related to virtual and real radiation of the massive secondary quarks and has the form
\begin{align}
J^{(n_l+1)}(s,m,\mu)= & \, J^{(n_l+1)}(s,\mu) + \delta J^{\rm{dist}}_{m}(s,m,\mu) \nn \\
& +  \delta J^{\rm{real}}_{m}(s,m)\,,
\label{eq:jetmassive}
\end{align}
where the two latter terms represent corrections due to the quark mass. The computation is described in 
Sec.~\ref{sec:jetfunction}. The expression for $\delta J^{\rm{dist}}_{m}(s,m,\mu)$ contains
only distributions and corresponds to collinear massive virtual corrections (including soft-bin subtractions) as well as
terms related to the subtraction of the massless quark result contained in $J^{(n_l+1)}(s,\mu)$ (see Eq.~(\ref{eq:J0})). 
Its renormalized expression reads ($\bar s = s/\mu^2, \, \alpha_s^{(n_l+1)}=\alpha_s^{(n_l+1)}(\mu)$)
\begin{align}
& \mu^2 \delta J^{\rm{dist}}_{m}(s,m,\mu)= \frac{\big(\alpha_s^{(n_l+1)}\big)^2 C_{\!F} T_F}{16\pi^2}
\left\{\left[\frac{16}{9}\,L_m^3 \right. \right. \nn \\
&+\frac{116}{9}\,L_m^2+\left(\frac{1436}{27}-\frac{16\pi^2}{9}\right)\!L_{m}+\frac{8650}{81}-\frac{116\pi^2}{27} \nn\\
&-\left.\frac{64}{3}\,\zeta_3\right]\!\delta(\bar{s})+\left(-\frac{16}{3}L_{m}^2-
\frac{232}{9}\,L_{m}-\frac{1436}{27}\right.\nn \\ 
&+\left.\frac{16\pi^2}{9}\right)\!\left[\frac{\theta(\bar{s})}{\bar{s}}\right]_+ 
+\left(\frac{32}{3}\,L_{m}+\frac{232}{9}\right)\!\left[\frac{\theta(\bar{s})\ln\,{\bar{s}}}{\bar{s}}\right]_+ \nn \\
&- \left.\frac{16}{3}\!\left[\frac{\theta(\bar{s})\ln^2\,{\bar{s}}}{\bar{s}}\right]_+\right\} \, .
\label{eq:J_virt}
\end{align}
The term $\delta J^{\rm{real}}_{m}(s,m)$ in Eq.~(\ref{eq:jetmassive}) contributes only when the jet invariant
mass is above the threshold $4m^2$ and thus corresponds to real production of the massive quarks.
It is given by
\begin{align}\label{eq:J_real}
& \mu^2 \delta J^{\rm{real}}_{m}(s,m)= \frac{\big(\alpha_s^{(n_l+1)}\big)^2 C_{\!F} T_F}{16\pi^2}
\frac{1}{\bar{s}}\, \theta(s-4m^2)  \nn \\
& \times \bigg\{\!-\frac{64}{3}\,\Li_2\bigg(\frac{b-1}{b+1}\bigg) + 
\frac{32}{3}\,\ln{\left(\frac{1-b^2}{4}\right)}\ln\bigg(\frac{1-b}{1+b}\bigg) \nn\\
& -\frac{16}{3}\,\ln^2\bigg(\frac{1-b}{1+b}\bigg)  +\Big(b^4-2\,b^2+\frac{241}{9}\Big)\ln\bigg(\frac{1-b}{1+b}\bigg) \nn\\
&- \frac{10}{27}\,b^3+\frac{482}{9}\,b- \frac{16\pi^2}{9} \bigg\}\,,
\end{align}

with

\begin{align}
\qquad b = \sqrt{1-\frac{4m^2}{s}} \, .
\label{eq:bdef}
\end{align}
Due to its physical character it is UV-finite and does not contain any explicit logarithmic $\mu$-dependence.
Furthermore, $\delta J^{\rm{real}}_{m}$ and its first two derivatives in $s$ vanish at the threshold, so that no
discontinuity arises due to real radiation. Note that the range in $\tau$ where scenario~III is employed may be
chosen such that it fully includes the domain for collinear massive real radiation, namely $\tau \geq 4m^2/Q^2$,
so that the threshold is properly accounted for through the analytic form of $\delta J^{\rm{real}}_{m}(s,m,\mu_m)$.
For $m\rightarrow 0$ the jet function $J^{(n_l+1)}(s,m,\mu)$ yields correctly the fully massless jet function
at $\mathcal{O}(\alpha_s^2)$, i.e.\
\begin{align}
 J^{(n_l+1)}(s,m,\mu) \stackrel{m\rightarrow 0}{\longrightarrow} J^{(n_l+1)}(s,\mu) \, .
\end{align}
We note that in the calculation of $\delta J_m^{\rm{dist}}$ rapidity divergences arise which cancel  in the sum of the 
collinear diagrams and the corresponding soft-bin subtractions. We stress that for $\mu^2 \sim s$ all associated 
logarithms cancel completely in the sum of $\delta J_m^{\rm{real}}$ and $\delta J^{\rm{dist}}_{m}$, so that no (large) 
rapidity logarithm remains in the jet function.

The UV divergences of the bare jet function $J^{(n_l+1)}_{\rm bare}(s,m,\mu)$ are mass independent and agree with
the known massless ones for $n_l+1$ dynamic flavors. The $\mathcal{O}(\alpha_s^2 C_{\!F} T_F)$ contributions to the jet
function counterterm are the ones from Eq.~(\ref{eq:ZJ}) for $n_{\!f} = n_l +1$. This statement holds to any order in
$\alpha_s$, so that the jet function evolution factor $U^{(n_l+1)}_{J}$ obeys
\begin{align}
 &\mu\,\frac{\df}{\df\mu}\,U^{(n_l+1)}_{J}(s,\mu,\mu_J) \nn \\
 &=\int \!\df s' \gamma^{(n_l+1)}_J(s-s',\mu)\, U^{(n_l+1)}_{J}(s',\mu,\mu_J) \, .
\end{align}

At the scale $\mu_m$ the mass-shell fluctuations of the collinear mass modes are integrated out. These contributions are encoded in the jet mass mode matching coefficient
$\mathcal{M}_J(s,m,\mu_m)$ and contain all virtual effects of the massive flavor such that for the scales $\mu<\mu_m$ 
massive collinear effects decouple. The result up to $\mathcal{O}(\alpha_s^2)$ reads 
\begin{widetext}
\begin{align}
& \mu_J^2\, \mathcal{M}^{(2)}_{J}(s,m,\mu_J,\mu_m)= \delta(\tilde{s})+
\left[\frac{\big(\alpha_s^{(n_l+1)}\big)^2 C_{\!F} T_F}{(4\pi)^2}\, \delta(\tilde{s})
\,\ln\bigg(\frac{\mu_J^2}{\mu_m^2}\bigg)\!\!
\left(\frac{16}{3}\,L_m^2+\frac{160}{9}\,L_m+\frac{448}{27}\right)\right]_{\mathcal{O}(\alpha_s)} \nn \\
& + \Bigg[ \frac{\big(\alpha_s^{(n_l+1)}\big)^2 C_{\!F} T_F}{(4\pi)^2}
\left\{\left[-\,\frac{16}{9}\,L_m^3 - \frac{116}{9}\,L_m^2+
\left(-\frac{932}{27}-\frac{8\pi^2}{9}\right)\!L_m -\frac{1531}{27}-\frac{20\pi^2}{27}+
\frac{160}{9}\,\zeta_3 \right] \!\delta(\tilde{s})\right. \nn \\
& + \left. \left[\frac{16}{3}\,L_m^2+\frac{160}{9}\,L_m+\frac{448}{27}\right]\!
\left[\frac{\theta(\tilde{s})}{\tilde{s}}\right]_+\right\}  + \frac{\big(\alpha_s^{(n_l+1)}\big)^3 C_{\!F} T_F}{(4 \pi)^3} 
\,\delta(\tilde{s}) \,\ln\bigg(\frac{\mu_J^2}{\mu_m^2}\bigg)\!
\left\{L_m^3\!\left[-\,\frac{352}{27}\,C_{\!A}+\frac{256}{27}\,T_F+\frac{128}{27}\,T_F n_l\right] \right. \nn \\
& +L_m^2\!\left[\left(\frac{368}{9}-\frac{32\pi^2}{9}\right)\! C_{\!A} - 48\, C_{\!F} +\frac{640}{27}\,T_F \right]+ 
L_m\!\left[\left(\frac{448}{3}\,\zeta_3+\frac{2480}{81}-\frac{320\pi^2}{27} \right)\! C_{\!A}+
\left(\frac{16}{3}-128\,\zeta_3\right)\!C_{\!F} \right.\nn \\
 & \left.\left. +\frac{4352}{81}\,T_F\, n_l +\frac{3968}{81}\,T_F\right] + \frac{\mathcal{M}^{J,+}_{3}}{C_{\!F} T_F} \right\}
 +\frac{\big(\alpha_s^{(n_l+1)}\big)^4 C_{\!F}^2 T_F^2}{(4 \pi)^4}\,
 \delta(\tilde{s})\, \ln^2\bigg(\frac{\mu_J^2}{\mu_m^2}\bigg)\!\!\left(\frac{128}{9}\,L_m^4+\frac{2560}{27}\,L_m^3+
 \frac{6656}{27}\,L_m^2 \right. \nn \\
 &+ \left.\frac{71680}{243}\,L_m+\frac{100352}{729}\right)\Bigg]_{\mathcal{O}(\alpha_s^2)} \, ,
\label{eq:matchingIII}
\end{align}
\end{widetext}
where $\tilde{s} \equiv s/\mu^2_J$, $L_{m}\equiv \ln{\left(m^2/\mu_m^2\right)}$ and 
$\alpha_s^{(n_l+1)}=\alpha_s^{(n_l+1)}(\mu_m)$. The quark mass $m=\overline{m}(\mu_m)$ is given in the $\MS$ scheme.
It is interesting to note that the result is a non-trivial distributive function of the jet invariant mass $s$ and
thus differs substantially from the local mass mode matching coefficient of the current (see Eq.~(\ref{eq:matchingII}))
or the strong coupling which do not depend explicitly on any kinematic scale. As for the case of the current mass mode matching 
coefficient, $\mathcal{M}_J$ contains large logarithms involving the ratio of the jet scale $s\sim \mu_J$ and the mass 
scale $m\sim \mu_m$ which are not summed by the RG $\mu$-evolution of the jet function. They are related to rapidity-type 
singularities that arise in the massive virtual corrections in the overlap region between the collinear mass mode 
contributions and  their soft-bin subtractions. These logarithms exponentiate as in the case for the current mass mode 
matching coefficient. We note that, through the rapidity RG evolution, $\mathcal{M}_J$ depends at each order on two rapidity 
scales, which we correlate to the jet and the mass scales $\mu_J$ and $\mu_m$, respectively. We stress, however, that the 
dependence of $\mathcal{M}_J$ on the jet scale $\mu_J$ cancels in a fixed-order expansion. Using the counting
$\alpha_s\, \ln(m^2/s) \sim \mathcal{O}(1)$ concerning rapidity logarithms the $\mathcal{O}(\alpha_s^2\, \ln(m^2/s))$ 
corrections in Eq.~(\ref{eq:matchingIII}) are counted as $\mathcal{O}(\alpha_s)$, while at $\mathcal{O}(\alpha_s^2)$ one has 
to include the terms of $\mathcal{O}(\alpha_s^4\, \ln^2(m^2/s))$ and $\mathcal{O}(\alpha_s^3\, \ln(m^2/s))$. From the latter 
terms the contributions explicitly depending on $\mu_m$ can be inferred using the $\mu_m$ independence of the 
factorization theorem and the explicit form of the jet function evolution factors $U_J^{(n_l+1)}$ and $U_J^{(n_l)}$.  In 
Eq.~(\ref{eq:matchingIII}) we have displayed all terms that are counted as $\mathcal{O}(\alpha_s)$ and 
$\mathcal{O}(\alpha_s^2)$ as well as the constant $\mathcal{M}^{J,+}_3$ which is not constrained by RG arguments. The 
computations are described in detail in Sec.~\ref{sec:jetmassmodematching}.

\subsection{Scenario IV: $Q>Q\lambda>Q\lambda^2>m$}\label{sec:scenarioIV}
The mass is below the ultrasoft scale. There is no separation between the collinear and soft mass modes and the 
corresponding collinear and soft massless modes since the RG evolution following the top-down approach of 
Fig.~\ref{fig:scenarios} never crosses the massive quark threshold and all evolution is carried out for $n_l+1$ active 
dynamic flavors. So compared to scenario III there are no mass mode matching coefficients, and the soft function accounts 
for the secondary massive contributions. The factorization theorem reads
\begin{align}
 &\frac{1}{\sigma_0}\frac{\df\sigma}{\df\tau}= Q\,\big|{C^{(n_l+1)}(Q,m,\mu_H)}\big|^2\,
 \,\big|U^{(n_l+1)}_{C}(Q,\mu_H,\mu_S)\big|^2 \nn \\
 & \times \int\! \df s\!  \int\! \df s' \, U^{(n_l+1)}_J(s-s',\mu_S,\mu_J)  \, J^{(n_l+1)}(s',m,\mu_J) \nn \\
 & \times S^{(n_l+1)}\Big(Q\,\tau-\frac{s}{Q},m,\mu_S\Big)\,,
\label{eq:diffsigmaIV}
\end{align}
where the hard current matching coefficient $C^{(n_l+1)}(Q,m,\mu_H)$ is the same as in scenarios II and III, see 
Eq.~(\ref{eq:hardcoeffII}), and the jet function $J^{(n_l+1)}(s,m,\mu_J)$ is the same as in scenario III, see 
Eq.~(\ref{eq:jetmassive}). The soft function $S^{(n_l+1)}(\ell,m,\mu_S)$ contains virtual as well as real radiation 
contributions related to the massive quark. The partonic contribution can be written as
\begin{align}
 &\hat{S}^{(n_l+1)}(\ell,m,\mu)= \hat{S}^{(n_l+1)}(\ell,\mu) + \delta S^{\rm dist}_m(\ell,m,\mu) \nn \\
 &+ \delta S^{{\rm real},\theta}_{m}(\ell,m) + \delta S^{{\rm real},\Delta}_m(\ell,m) \, ,
\label{eq:softmassive}
\end{align}
where $\hat{S}^{(n_l+1)}(\ell,\mu)$ is the partonic soft function for $n_l+1$ massless quark flavors.
The other terms represent the $\mathcal{O}(\alpha_s^2 C_{\!F} T_F)$ corrections due to the non-zero quark
mass and were computed in Ref.~\cite{Gritschacher:2013tza}. For the convenience of the reader we briefly review
these results in the following.

The expression for $\delta S^{\rm dist}_m$ contains only distributions and corresponds to virtual massive
quark radiation as well as to the terms related to the subtractions of the massless quark result
(see Eq.~(\ref{eq:S0})) to avoid double counting with the full massless result in the first term of
Eq.~(\ref{eq:softmassive}). The renormalized expression reads
($\bar \ell = \ell/\mu, \, \alpha_s^{(n_l+1)}=\alpha_s^{(n_l+1)}(\mu)$) 
\begin{align}
&\!\mu\, \delta S^{\rm dist}_m (\ell,m,\mu)=  \frac{\big(\alpha_s^{(n_l+1)}\big)^2 C_{\!F} T_F}{16\pi^2}
\left\{\!\left[-\frac{8}{9}\,L_m^3-\frac{40}{9}\,L_m^2 \right.\right.\nn \\
&\!+\!\left.\left(-\frac{448}{27}+\frac{8\pi^2}{9}\right)\!L_m -\frac{2048}{81}-
\frac{64\pi^2}{27}+32\,\zeta_3 \right]\!\delta(\bar{\ell}) \nn \\
 &\!+\! \left(\frac{16}{3}\,L_m^2+\frac{160}{9}\,L_m+\frac{896}{27}-\frac{16\pi^2}{9}\right)\!
 \left[\frac{\theta(\bar{\ell})}{\bar{\ell}}\right]_+ \nn \\
 &\!-\! \left.\left(\frac{64}{3}\,L_m+\frac{320}{9}\right)\!\!\left[\frac{\theta(\bar{\ell})\ln\,
 \bar{\ell}}{\bar{\ell}}\right]_+  \!\!+
 \frac{64}{3}\!\left[\frac{\theta(\bar{\ell})\ln^2\, \bar{\ell}}{\bar{\ell}}\right]_+ \!\right\}\!.  \label{eq:S_virt}
\end{align} 
The term $\delta S^{{\rm real},\theta}_{m}$ describes real massive quark radiation for the prescription that
the coherent sum of the massive quark and antiquark momentum (i.e.\ the virtual gluon momentum) enters the thrust
definition.\,\footnote{It was demonstrated in Ref.~\cite{Gritschacher:2013tza} that this prescription can be easily calculated
analytically and agrees with the regular thrust prescription except when the quark and antiquark enter different hemispheres.}
It contains a threshold at $\ell=2m$ and reads
\begin{align}
&\mu\, \delta S^{{\rm real},\theta}_{m} (\ell,m)=  \frac{\big(\alpha_s^{(n_l+1)}\big)^2 C_{\!F} T_F}{16\pi^2} \, 
\theta(\ell-2m)\,\frac{1}{\bar{\ell}} \nn \\
& \times \bigg\{\frac{64}{3}\, \Li_2\bigg(\frac{w-1}{w+1}\bigg)
 -\frac{32}{3}\,\ln \Big(\frac{1-w^2}{4}\Big)\ln\bigg(\frac{1-w}{1+w}\bigg)\nn\\
&+\frac{16}{3}\,\ln^2\bigg(\frac{1-w}{1+w}\bigg)-\frac{160}{9}\, \ln \bigg(\frac{1-w}{1+w}\bigg)
+\frac{64}{27}\,w^3 \nn \\
 & -\frac{320}{9}\,w + \frac{16\pi^2}{9} \bigg\} \, ,
\label{eq:S_real}
\end{align} 
with
\begin{align}
\qquad w = \sqrt{1-\frac{4m^2}{\ell^2}}\, .
\label{eq:wdef}
\end{align}
$\delta S^{{\rm real},\theta}_{m}$ and its first two derivatives in $\ell$ vanish at the threshold, so that no 
discontinuity arises due to real radiation. Since the momenta of the quark and antiquark enter the thrust
prescription as different respective projections on one of the two light-cone axes, if they enter different
hemispheres, $\delta S^{{\rm real},\theta}_{m}$ does not represent the complete real radiation contribution. For the
part of the phase space where the massive quark and antiquark go into opposite hemispheres, one has to account for the
additional, numerically small {\it hemisphere mismatch contribution} $\delta S^{{\rm real},\Delta}_m$ that
has also been computed in Ref.~\cite{Gritschacher:2013tza}. This correction does not have a threshold and is
nonvanishing for all positive thrust momenta $\ell$. In the massless limit $\delta S^{{\rm real},\Delta}_m$
approaches a $\delta$-distribution. In Ref.~\cite{Gritschacher:2013tza} a parametrization for
$\delta S^{{\rm real},\Delta}_m$ was given that approximates this contribution up to better than
$2\%$ relative accuracy:
\begin{align}\label{eq:DeltaS_parametrization}
 & \delta S^{{\rm real},\Delta}_{m} (x\, m,m) \Big|_{\rm fit} =
\frac{\big(\alpha_s^{(n_l+1)}\big)^2 C_{\!F} T_F}{16\pi^2} \,  \frac{1}{m} \nn \\
 & \times\frac{x^5\left[a \, \ln^2\left(1+x^2\right)+b \, \ln\left(1+x^2\right) +
 c\,\right]}{d\, x^8 +e x^7 +f x^6 +g x^4 + h x^3 + j x^2 +1} \, ,
\end{align}
with $a=8\,d$, $b=-\,80\,d$, $c=8/15$ and $d=6/(2400+360\pi+73\pi^2)$ being fixed from imposing the correct
asymptotic behavior for $m \gg \ell$ and $m \ll \ell$. The remaining 5 parameters were obtained using a fit
with the constraint of satisfying the correct normalization corresponding to the massless analytic limit:
\begin{align}
 &  e = 0.0117 \,, && f = \,0.100 \, ,   && g = -\,0.502 \,, \nn \\
 & h = 0.747 \,,  &&  j \,= -\,0.180 \,. &&
\end{align}
Note that both real radiation contributions are UV finite. For $m\rightarrow 0$ the soft function 
$\hat{S}^{(n_l+1)}(\ell,m,\mu)$ yields correctly the fully massless partonic soft function at
$\mathcal{O}(\alpha_s^2)$, i.e.\
\begin{align}
 \hat{S}^{(n_l+1)}(\ell,m,\mu) \stackrel{m\rightarrow 0}{\longrightarrow} \hat{S}^{(n_l+1)}(s,\mu) \, .
\end{align}
We note that in the calculation of $\delta S_m^{\rm{dist}}$ rapidity divergences arise in the contributions coming
from the different hemispheres which cancel in the sum of the terms. We stress, however, that for $\mu \sim \ell$
all associated logarithmic mass-singularities cancel in the sum of $\delta S_m^{\rm{real},\theta}$ and
$\delta S^{\rm{dist}}_{m}$, so that no (large) rapidity logarithm remains in the soft function.

The UV divergences of the bare soft function $\hat{S}^{(n_l+1)}_{\rm bare}(\ell,m,\mu)$ are mass independent and agree
with the known massless ones for $n_l+1$ dynamic flavors in Eq.~(\ref{eq:ZS}) with the replacement $n_{\!f} = n_l+1$.
The evolution factor $U^{(n_l+1)}_{S}$ obeys
\begin{align}
 & \mu\,\frac{\df}{\df\mu}\,U^{(n_l+1)}_{S}(\ell,\mu,\mu_S)\nn \\
 &=\int\! \df\ell'\, \gamma^{(n_l+1)}_S(\ell-\ell',\mu)\, U^{(n_l+1)}_{S}(\ell',\mu,\mu_S) \, ,
\end{align}
which holds to any order in the strong coupling.

\subsection{Gap Subtraction, Evolution and Matching}\label{sec:gap}
In scenarios I to III the quark mass is above the soft scale, and therefore the massive quark does not affect the soft
function. Thus the gap subtraction agrees with the one from the factorization theorem for $n_{\!f}=n_l$ massless quarks
as described in Sec.~\ref{sec:massless}. In scenario IV, for $m > \Lambda_{\rm QCD}$ the finite quark mass provides
an infrared cutoff for the virtuality of the exchanged gluon in the partonic soft function such that the factorial
growth of the coefficients related to the massive flavor at large orders in perturbation theory is suppressed and,
in principle, a corresponding subtraction in the gap series $\delta(R,\mu)$ might be unnecessary. However,
implementing the gap scheme along the lines of Eqs.~(\ref{eq:Spartmod}) and (\ref{eq:delta_thrust}) including
the effects of the secondary massive quarks is useful in order to have a smooth interpolation of the gap scheme
parameters to the massless quark limit. Since the resulting subtraction series $\delta^{(n_l+1)}(R,m,\mu)$ encodes
infrared-sensitive perturbative contributions, it now becomes mass dependent. Thus the complete soft function
in scenario IV reads
\begin{align}\label{eq:SpartmodIV}
& S^{(n_l+1)}(\ell,m,\mu) \nn\\
&= \!\int\! \df\ell'\, \hat{S}^{(n_l+1)}(\ell-\ell'-2\,\delta^{(n_l+1)}(R,m,\mu),m,\mu) \nn\\
&\times F(\ell'-2\,\bar{\Delta}^{(n_l+1)}(R,m,\mu)) \,.
\end{align}
The renormalon subtractions $\delta^{(n_l+1)}(R,m,\mu)$ can be written as
\begin{align}
\delta^{(n_l+1)}(R,m,\mu)=& \, \delta^{(n_l+1)}(R,\mu) + \delta_m(R,m) \, ,
\label{eq:renormalonmassive}
\end{align}
where $\delta^{(n_l+1)}(R,\mu)$ is the series for $n_l+1$ massless quark flavors and $\delta_m(R,m)$ represents
the correction to the massless result due to the finite quark mass. In Ref.~\cite{Gritschacher:2013tza} the 
$\mathcal{O}(\alpha_s^2 C_{\!F} T_F)$ correction to $\delta_m(R,m)$ was calculated according to
Eq.~(\ref{eq:delta_thrust}), and the result can be parametrized by\,\footnote{This parametrization differs
from the one given in Ref.~\cite{Gritschacher:2013tza}. It has a better precision and interpolates the R-anomalous 
dimension more smoothly for small values of $m/R$.} ($\alpha_s^{(n_l+1)}=\alpha_s^{(n_l+1)}(\mu)$) 
\begin{align}\label{eq:delta_m}
  & \delta_m(R,y R) = \frac{\big(\alpha_s^{(n_l+1)}\big)^2C_{\!F} T_F}{16\pi^2}  R\, e^{\gamma_E} \tilde{g}(y) \, , \\
  & \tilde{g}(y) = \tilde{h}(y)-\frac{\tilde{h}(y) +a y }{1 + b y + c y^2}\, e^{-\alpha y^{\beta}} \, ,
\end{align}
where
\begin{align}
 & \alpha = 0.634 \,,&&\beta = 1.035 \,, && a = \,23.6 \, ,\nn \\
 & b \,= -\,0.481 \,, && c \,= 1.19 \,,&&
\end{align}
and
\begin{align}
 \tilde{h}(y)= -\,\frac{8}{3}\, \ln^2 y^2-\frac{80}{9}\,\ln\, y^2-\frac{448}{27}-\frac{8\pi^2}{9}\, .
\end{align}
The expression in Eq.~(\ref{eq:delta_m}) provides an approximation that is much better than 1\% and that is
constructed such that the massless limit in Eq.~(\ref{eq:renormalonmassive}) is recovered for $m \rightarrow 0$,
i.e.\ $\delta_m(R,m) \stackrel{m\rightarrow 0}{\longrightarrow} 0$. Moreover, for $m/R\rightarrow \infty$ the 
parametrization yields the correct limit, 
\begin{align}
 \!\!\delta_m (R, y R) \stackrel{y\rightarrow \infty }{\longrightarrow} \,\frac{\big(\alpha_s^{(n_l+1)}\big)^2 C_{\!F} T_F}{16\pi^2} \,R \,e^{\gamma_E} \,\tilde{h}(y) \,.
\end{align}

The $\mu$-evolution of the gap parameter $\bar{\Delta}^{(n_l+1)}(R,m,\mu)$ is mass independent and thus the same as
for the massless gap parameter as given in Eq.~(\ref{eq:gamma_Deltamu}) with the replacement  $n_{\!f} = n_l+1$. With
the quark mass dependent gap subtraction at $\mathcal{O}(\alpha_s^2 C_{\!F} T_F)$, however, the gap evolution in $R$
becomes mass dependent, and one can determine the R-evolution equation directly from Eq.~(\ref{eq:renormalonmassive}) 
using Eq.~(\ref{eq:gamma_R}). The R-anomalous dimension can then be written as ($\alpha_s^{(n_l+1)}=\alpha_s^{(n_l+1)}(R)$)
\begin{align}\label{eq:gammaRmassive}
& \gamma^{(n_l+1)}_{R}(m/R) =\gamma^{(n_l+1)}_{R} + \gamma_{R,m}(m/R)\, ,\\
& \gamma_{R,m}(y) =  \frac{\big(\alpha_s^{(n_l+1)}\big)^2 C_{\!F} T_F}{16\pi^2}\, e^{\gamma_E}
 \bigg[1 - y\,\frac{\rm d}{{\rm d} y}\,\bigg]\tilde{g}(y)\,,
\end{align}
where $\gamma^{(n_l+1)}_{R}$ denotes the R-anomalous dimension with $n_l+1$ massless quarks. Using the parametrization of Eq.~(\ref{eq:delta_m}) the result for the $\mathcal{O}(\alpha_s^2 C_{\!F} T_F)$ massive quark 
correction $\gamma_{R,m}(m/R)$ can be easily computed. It approximates the exact result within 2\% (except for
$m/R <0.1$, where the correction is anyway tiny) and yields the correct massless limit in Eq.~(\ref{eq:gammaRmassive})
for $m \rightarrow 0$, i.e. $\gamma_{R,m}(m/R) \stackrel{m\rightarrow 0}{\longrightarrow} 0$. 
The explicit solution for the $\mu$- and R-evolution for $\bar{\Delta}$ with massless quarks can be found in Eq.~(41) of 
Ref.~\cite{Abbate:2010xh}. The quark mass just modifies the R-evolution terms of that solution. It affects the function 
$D^{(k)}(\alpha_s (R_1),\alpha_s (R_0))$, defined for massless quarks in Eq.~(A31) of Ref.~\cite{Abbate:2010xh}, where 
$R_0$ ($R_1$) is the initial (final) scale of the R-evolution. Mass effects start contributing at N$^2$LL order and modify 
$D^{(2)}(\alpha_s(R_1),\alpha_s(R_0))$ in the following way\,:
\begin{align}\label{eq:mass-R-running}
&D^{(2)}(\alpha_s(R_1),\alpha_s(R_0), m) = D^{(2)}(\alpha_s(R_1),\alpha_s(R_0), m = 0) \nn \\
&+\,\frac{1}{4\beta_0^2}\int_{t_0}^{t_1} {\rm d}t\, e^{-t}(-t)^{-2-\frac{\beta_1}{2\beta_0^2}}\,
\tilde{\gamma}_{R,m}\bigg(\frac{m\,e^{G(t)}}{\Lambda^{(2)}_{\rm QCD}} \!\bigg) \, ,
\end{align}
with $t_i = -\,2\pi/(\alpha_s(R_i)\beta_0)$ and $\beta_i$ being the coefficients of the perturbative expansion of the 
$\beta$ function as defined in Eq.~(\ref{eq:beta}) and $\gamma_{R,m}(m/R)=\alpha_s^2/(16\pi^2) \,\tilde{\gamma}_{R,m}(m/R)$. Here the strong coupling $\alpha_s$ is understood to be in the 
($n_\ell+1$)-flavor scheme. The function $G(t)$ is given by
\begin{align}
G(t) = t + \frac{\beta_1}{2\beta_0^2}\,\ln(-\,t) - \frac{\beta_1^2 - \beta_0\beta_2}{4\beta_0^4} \frac{1}{t}\,.
\end{align}
Eq.~(\ref{eq:mass-R-running}) can be obtained following the changes of variables as explained in 
Ref.~\cite{Hoang:2008yj}. A generalization to higher orders is straightforward. 

To complete the discussion on the evolution of the gap parameter, we have to consider the matching relation between 
$\bar{\Delta}^{(n_l+1)}(R,m,\mu)$ for $R,\mu > \mu_m \sim m$, where the massive quark is an active dynamic flavor, and 
$\bar{\Delta}^{(n_l)}(R,\mu)$ for $R,\mu < \mu_m \sim m$, where the massive quark is integrated out. 
The matching relation is most easily derived using the fact that the ``bare" gap parameter is scheme independent, very
much like the massive quark pole mass. This gives the relation
\begin{align}
 \Delta &=\bar{\Delta}^{(n_l)}(R,\mu)+\delta^{(n_l)}(R,\mu) \nn \\
 &=\bar{\Delta}^{(n_l+1)}(R,m,\mu)+\delta^{(n_l+1)}(R,m,\mu) \, ,
\end{align}
and we thus obtain ($L_m=\ln(m^2/\mu^2)$)
\begin{align}\label{eq:gap_decoupling}
 & \bar{\Delta}^{(n_l)}(R,\mu)=\bar{\Delta}^{(n_l+1)}(R,m,\mu)+\delta^{(n_l+1,2)}_{n_{\!f}=1}(R,\mu)\nn \\
 & +\delta_{m}(R,m)- \frac{\alpha_s^{(n_l+1)} T_F}{3\pi}\, L_m\,\delta^{(n_l+1,1)}(R,\mu) \, .
\end{align}
with the one-loop gap subtraction
\begin{align}
 \!\delta^{(n_l+1,1)}(R,\mu) = \frac{\alpha_s^{(n_l+1)} C_{\!F}}{4\pi} \,R \,e^{\gamma_E}
 \left[-\,4 \,\ln\bigg(\frac{\mu^2}{R^2}\bigg) \right] .
\end{align}
The latter term arises from the matching relation of the strong coupling between the $n_l$- and $(n_l+1)$-scheme.
To avoid large logarithms, the gap matching relation should be employed for $R\sim \mu\sim m$. 

\begin{figure}
 \centering
 \includegraphics[width=\linewidth]{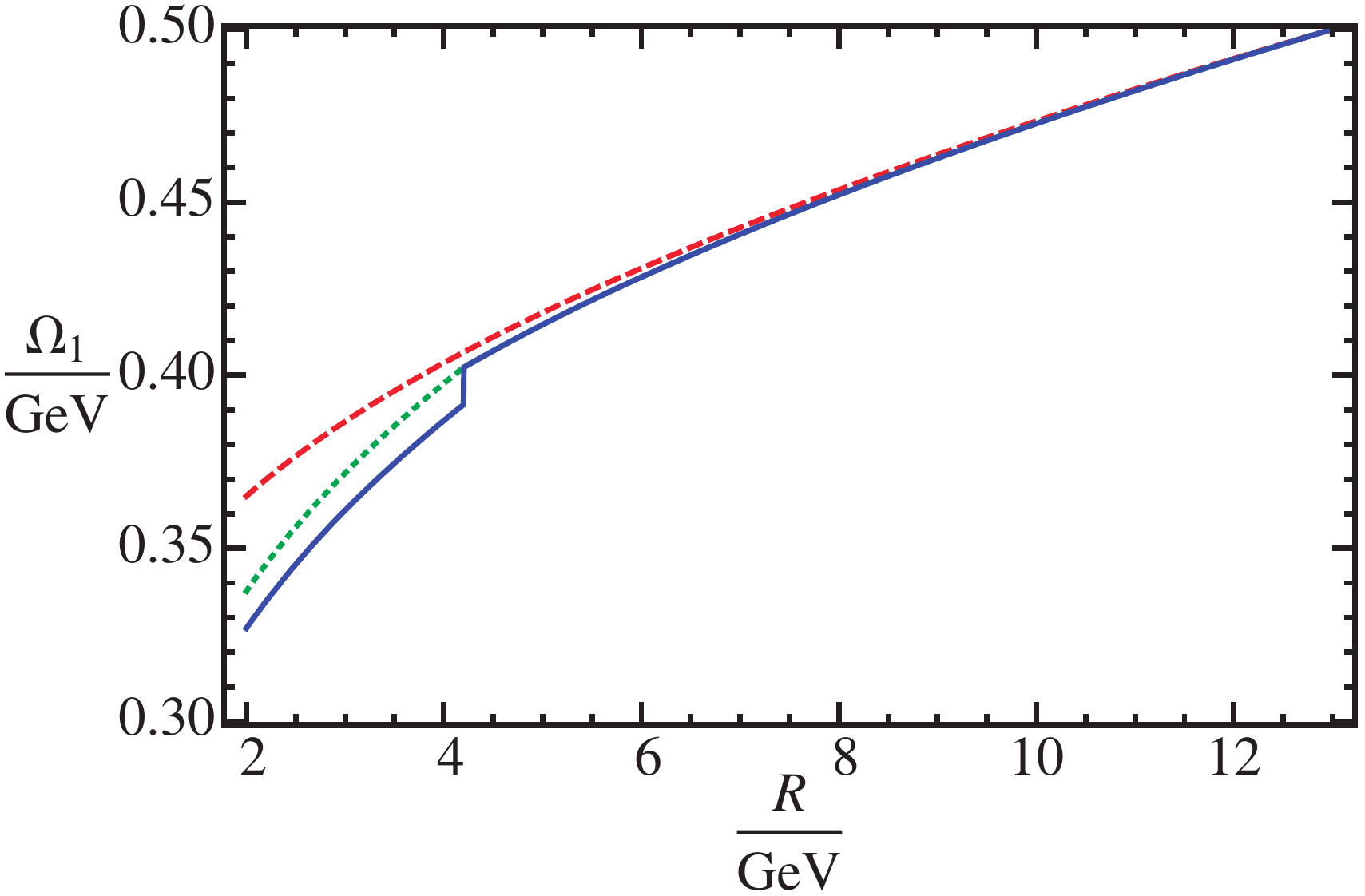}
 \caption{R-evolution of $\Omega_1(R,\mu=R)$ with a massive bottom quark at $\mathcal{O}(\alpha_s^3)$ as described
 in the text. The curves represent purely massless evolution (red, dashed), massive evolution including threshold
 matching at $\overline{m}_b(\overline{m}_b)$ (blue, solid) and massive evolution without threshold matching
 (green, dotted) \label{fig:Revo_bottom}}
\end{figure}

\begin{figure}
 \centering
 \includegraphics[width=\linewidth]{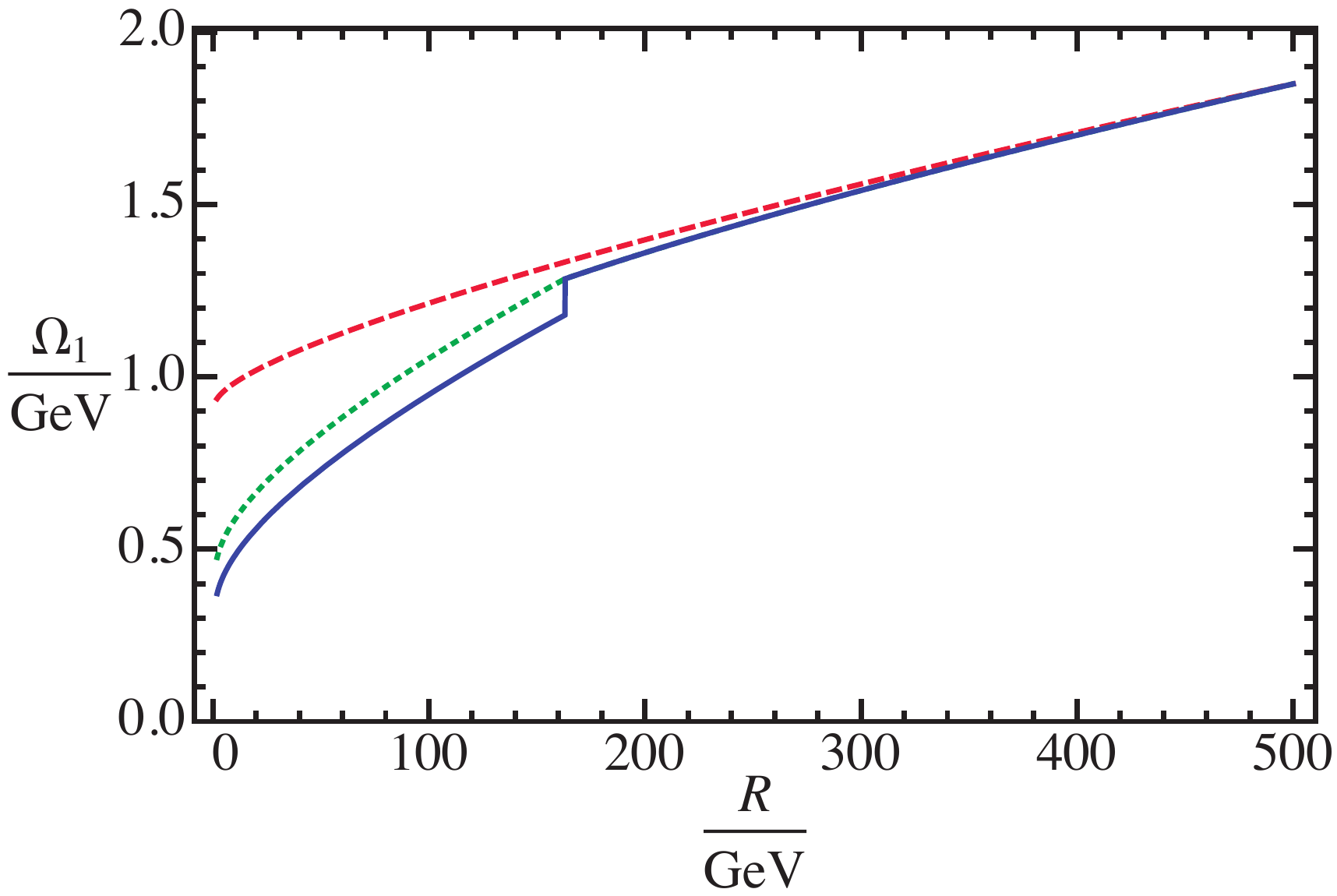}
 \caption{R-evolution of $\Omega_1(R,\mu=R)$ with a massive top quark at $\mathcal{O}(\alpha_s^3)$ as described in
 the text. The curves represent purely massless evolution (red, dashed), massive evolution including threshold matching
 at $\overline{m}_t(\overline{m}_t)$ (blue, solid) and massive evolution without threshold matching (green, 
 dotted)\label{fig:Revo_top}}  
\end{figure}

In Fig.~\ref{fig:Revo_bottom}  we show $\Omega_1(R,\mu=R)$ (see Eq.~(\ref{eq:omega1})) as a function of $R$ in the
range between 2 and 13 GeV using $\Omega_1^{(5)}(13 \, {\rm GeV}, 13 \, {\rm GeV})=0.5 \, {\rm GeV}$ and 
$\alpha_s^{(5)}(m_Z)=0.114$ as initial conditions. The choice of these initial conditions is motivated by recent fits
for $\alpha_s$ and $\Omega_1$ in Refs.~\cite{Abbate:2010xh,Abbate:2012jh}, which involved only experimental data related
to R-scale values above 10 GeV despite the fact that values for $\Omega_1$ at $R=2 \, {\rm GeV}$ were quoted in the
final result. The red, dashed curve shows the purely massless evolution using the R-anomalous dimension at 
$\mathcal{O}(\alpha_s^3)$. The blue, solid curve shows the R-dependence accounting for the finite bottom quark mass
taking $\overline{m}_b(\overline{m}_b) =4.2 \, {\rm GeV}$ as an input for the $\MS$ bottom quark mass and using the 
threshold matching relation of Eq.~(\ref{eq:gap_decoupling}) at $R=\mu= \overline{m}_b(\overline{m}_b)$ when 
switching from the $n_{\!f}=5$ to the $n_{\!f}=4$ flavor scheme for the gap parameter. The difference between the blue and the red 
curve illustrates the impact of the finite bottom mass corrections on the R-dependence. We see that the mass effects are 
relatively small for $R>\overline{m}_b(\overline{m}_b)$, which indicates that the mass corrections in the anomalous 
dimension in $R$ represent only a minor effect. On the other hand, for $R<\overline{m}_b(\overline{m}_b)$, the bottom mass 
effects, which arise from the threshold matching corrections and from using the $n_{\!f}=4$ flavor anomalous dimension, are 
quite sizeable. This indicates that the latter two effects represent the most important effect due to the finite bottom 
mass. To visualize the impact of the bottom mass on the R-evolution alone we have also displayed the dependence on R when 
the threshold matching correction is ignored (green, dotted curve).  Overall, we see that the impact due to the finite 
bottom quark mass is sizeable and non-negligible particularly for scales below the bottom quark mass. 

In Fig.~\ref{fig:Revo_top} we display $\Omega_1(R,\mu=R)$ as a function of $R$ in the range up to 500 GeV showing the
same type of curves as in Fig.~\ref{fig:Revo_bottom} in order to illustrate the impact of the finite top quark mass.
All curves have the common input value $\Omega_1^{(6)}(500 \, {\rm GeV}, 500 \, {\rm GeV}) = 1.85 \, {\rm GeV}$ using 
again $\alpha_s^{(5)}(m_Z)=0.114$
and the switch from the $n_{\!f}=6$ to the $n_{\!f}=5$ flavor scheme has been carried out 
exactly at the top quark mass $\overline{m}_t(\overline{m}_t)=163 \, {\rm GeV}$. We can make observations that are
very similar to the ones already discussed for the bottom quark threshold region. The difference is that the impact of the 
finite top quark mass effects are even more dramatic than for the bottom quark case leading to a
discrepancy of a factor of two between the appropriate mass dependence and the evolution for a massless top quark when 
$\Omega_1$ is evolved down to the bottom quark scale. This is related to the fact that the threshold matching relation
at $R \sim m_t$ involves the top quark mass (see Eq.~(\ref{eq:gap_decoupling})) and that the R-evolution involves a linear 
dependence on $R$.

Note that the mass dependence of the R-evolution equations at $\mathcal{O}(\alpha_s^3)$ is currently unknown. We have 
therefore employed at $\mathcal{O}(\alpha_s^3)$ the known massless corrections with the appropriate number of flavors. As 
the $\mathcal{O}(\alpha_s^3)$ corrections amount to at most 25 \% of the $\mathcal{O}(\alpha_s^2)$ terms, and  -- as we 
have just shown above -- the mass dependence in the R-evolution equation only represents a minor effect, this approach is 
certainly justified. We have checked that these $\mathcal{O}(\alpha_s^3)$ contributions in the R-evolution lead to a total 
numerical impact in the bottom quark mass corrections for the thrust distribution that is less than half of the one generated 
by the variation of $\mu_m$ discussed in our numerical analysis of Sec.~V.  This indicates that the missing quark mass 
corrections at $\mathcal{O}(\alpha_s^3)$ might be safely ignored at this stage.

\section{Computations for the Hard Current Coefficient and Jet Function}\label{sec:computations}
In this section we give details on the calculations of the secondary massive quark corrections at
$\mathcal{O}(\alpha_s^2 C_{\!F} T_F)$ to the hard current coefficient and the jet function, for masses below the
hard and jet scales, respectively, i.e. for cases where the massive quark represents an active dynamic flavor.
The massive quark corrections to the partonic soft function for masses below the soft scale have been already
computed in Ref.~\cite{Gritschacher:2013tza} and the corresponding results have been reviewed in 
Sec.~\ref{sec:scenarioIV}. For all of these results the scheme with $n_l+1$ running flavors is employed (also for the 
strong coupling) allowing us to recover the known results for massless quarks in the limit $m\to 0$. For the calculations we 
use the dispersion relation method which enables us to obtain the secondary massive quark corrections at 
$\mathcal{O}(\alpha_s^2 C_{\!F} T_F)$ from the \mbox{$d$-dimensional} results for a massive gauge boson at $\mathcal{O}(\alpha_s)$ 
via an integration over the imaginary part of the gluon vacuum polarization due to the massive quark-antiquark 
bubble~\cite{Gritschacher:2013pha}. The dispersion relation method facilitates in particular the treatment of the rapidity 
singularities and the soft-bin subtractions since they can be dealt with completely at the level of the 
$\mathcal{O}(\alpha_s)$ diagrams with the massive gluon propagator. This allows us to separate these issues conveniently 
from the effects of the gluon splitting, which simplifies the calculations considerably. 

\subsection{Dispersion Relations}\label{sec:dispersion}
We explain the dispersive method for a secondary massive quark-antiquark pair starting from the gluonic vacuum 
polarization $\Pi(m^2,p^2)$ due to a massive quark-antiquark bubble, 
\begin{align}
 \Pi^{AB}_{\mu \nu}(m^2,p^2) & = -\,i\,\big(p^2 g_{\mu \nu}-p_{\mu} p_{\nu}\big) \Pi(m^2,p^2) 
\,\delta^{AB} \nn \\
& \equiv \int \!\df^4 x \,e^{i p x}\, \langle 0\lvert T \, J_{\mu}^A(x)J_{\nu}^B(0) \lvert 0 \rangle \, ,
\end{align}
with the vector current $J^A_\mu(x)=ig_s\bar{q}(x)T^A\gamma_{\mu}q(x)$. The vacuum polarization function
$\Pi(m^2,p^2)$ can be rewritten as a dispersion integral over its absorptive part. The unsubtracted
(unrenormalized) dispersion
integral reads
\begin{equation}
\Pi(m^2,p^2) = -\, \frac{1}{\pi} \int\!\df M^2 \, \frac{\mathrm{Im}\! \left[\Pi(m^2,M^2)\right]}{p^2-M^2+i \epsilon}  \, ,
\label{eq:dispersion1unsubtracted}
\end{equation}
and the subtracted (on-shell and finite) dispersion relation has the form 
\begin{align}
\Pi^{\rm OS}(m^2,p^2)& = 
\Pi(m^2,p^2)-\Pi(m^2,0) \nn \\
&=-\frac{p^2}{\pi} \int {\frac{\df M^2}{M^2} \, \frac{\mathrm{Im}\! \big[\Pi(m^2,M^2)\big]}{p^2-M^2+i \epsilon}
  } \, .
\label{eq:dispersion1subtracted}
\end{align}
The absorptive part in $d$ dimensions reads
\begin{align}\label{eq:Im_Pi}
&\mathrm{Im}\!\left[\Pi(m^2,p^2)\right] = \theta(p^2-4m^2) \, g^2 T_F \tilde{\mu}^{2\epsilon}(p^2)^{(d-4)/2} \nn \\
&\!\!\times\! \frac{2^{3-2d}\pi^{(3-d)/2}}{\Gamma\Big(\frac{d+1}{2}\Big)} 
\Big(d-2+\frac{4m^2}{p^2}\Big)\!\Big(1-\frac{4m^2}{p^2}\Big)^{(d-3)/2} \! ,
\end{align}
where $\tilde{\mu}^2 = \mu^2\,e^{\gamma_E}/(4\pi)$. Eqs.~(\ref{eq:dispersion1unsubtracted}) 
and~(\ref{eq:dispersion1subtracted}) are valid for any $d$.
The subtracted vacuum polarization function $\Pi^{\rm OS}(m^2,p^2)$ has the important feature that its
insertion into the gluon line can be rewritten as a dispersion integration over a ``massive gluon" propagator,
\begin{align}
& \frac{-\,i\,g^{\mu\rho}}{p^2 +i \epsilon}\, \Pi^{\rm OS}_{\rho\sigma}(m^2,p^2)
\,\frac{-\,i\,g^{\sigma\nu}}{p^2 +i \epsilon} \nn \\
  &=\,\frac{1}{\pi} \int\frac{\df M^2}{M^2}\,\frac{-\,i\,\Big(g^{\mu\nu}-\frac{p^\mu
        p^\nu}{p^2}\Big)}{p^2-M^2+i \epsilon} \,\mathrm{Im}\! \left[\Pi(m^2,M^2)\right] \,, 
\label{eq:propagatorsubtracted}
\end{align}
where $p^\mu$ denotes the external gluon momentum, and we have dropped the overall color conserving Kronecker 
$\delta^{AB}$. Note that in Eq.~(\ref{eq:propagatorsubtracted}) the propagator becomes transverse from the insertion
of the vacuum polarization. In our calculations the contributions from the additional $p^\mu p^\nu$ term vanish due
to gauge invariance and can be ignored. The insertion of the full unsubtracted vacuum polarization function
$\Pi(m^2,p^2)$ can be recovered by subtracting a term with the massless gluon propagator times the zero-momentum
vacuum polarization function,
\begin{align}\label{eq:propagatorunsubtracted}
 & \frac{-\,i\,g^{\mu\rho}}{p^2 +i \epsilon}\, \Pi_{\rho\sigma}(m^2,p^2)
 \, \frac{-\,i\,g^{\sigma\nu}}{p^2 +i \epsilon} \nn \\
 &=\,\frac{1}{\pi} \int\!\frac{\df M^2}{M^2}\,\frac{-\,i\,\Big(g^{\mu\nu}-\frac{p^\mu p^\nu}{p^2}\Big)}
 {p^2-M^2+i \epsilon} \,\mathrm{Im}\! \left[\Pi(m^2,M^2)\right]\nn \\
 & \, -\frac{-\,i\,\Big(g^{\mu\nu}-\frac{p^\mu p^\nu}{p^2}\Big)}{p^2+i \epsilon}\,\Pi(m^2,0) \, .
\end{align}
Note that Eqs.~(\ref{eq:propagatorsubtracted}) and (\ref{eq:propagatorunsubtracted})
hold in any gauge employed on the LHS of the equalities. The zero-momentum vacuum polarization function at 
$\mathcal{O}(\alpha_s)$ in $d$ dimensions reads
\begin{align}
\Pi(m^2,0) \,=\,  \frac{\alpha_s T_F}{3\pi} \bigg(\frac{\mu^2 e^{\gamma_E}}{m^2}\bigg)^{2-\frac{d}{2}} 
\Gamma\bigg(2-\frac{d}{2}\bigg) \, .
\label{eq:vacpolzero}
\end{align}
Using the on-shell vacuum polarization insertion via Eq.~(\ref{eq:propagatorsubtracted}) automatically implements
the on-shell subtraction for the renormalization of the strong coupling with respect to the effects of the massive
quark. So using Eq.~(\ref{eq:dispersion1subtracted}) implies that we employ the strong coupling in the $n_l$-flavor
scheme, i.e.\ $\alpha_s^{(n_l)}$. The subtracted dispersion relation has the computational advantage that the
integration over the virtual gluon mass is suppressed by an additional inverse power of $M^2$. This can make the 
dispersion integration UV finite and may allow us to carry out the integral directly in $d=4$ dimensions. Using
the full vacuum polarization insertion of Eq.~(\ref{eq:propagatorunsubtracted}) implies that the strong coupling is
still unrenormalized with respect to the effects of the massive quark flavor. 

The relations in Eqs.~(\ref{eq:propagatorsubtracted}) and~(\ref{eq:propagatorunsubtracted}) show explicitly that
we can obtain the result for the massive quark-antiquark pair from a dispersion integral over the corresponding
result for a gluon with mass $M$. We note that the dispersion relation method may not only be used to determine
the effects of secondary virtual massive quarks, but also for real radiation corrections as long as it is only
the sum of the quark and antiquark momenta (i.e.\ the momentum of the gluon that splits into the massive quark
pair) that enters the phase space constraint in the computation. Even if this is not the case the dispersion
integration may be useful to determine the dominant corrections or to deal with singular or divergent parts
of the result, see e.g.\ Ref.~\cite{Gritschacher:2013tza} for such an application in the calculation of the 
$\mathcal{O}(\alpha_s^2 C_{\!F} T_F)$ massive quark contributions to the soft function. 

\subsection{Hard Current Matching Coefficient for $m<Q$}\label{sec:hardmatching}

\begin{figure}
 \centering
 \includegraphics[width= \linewidth]{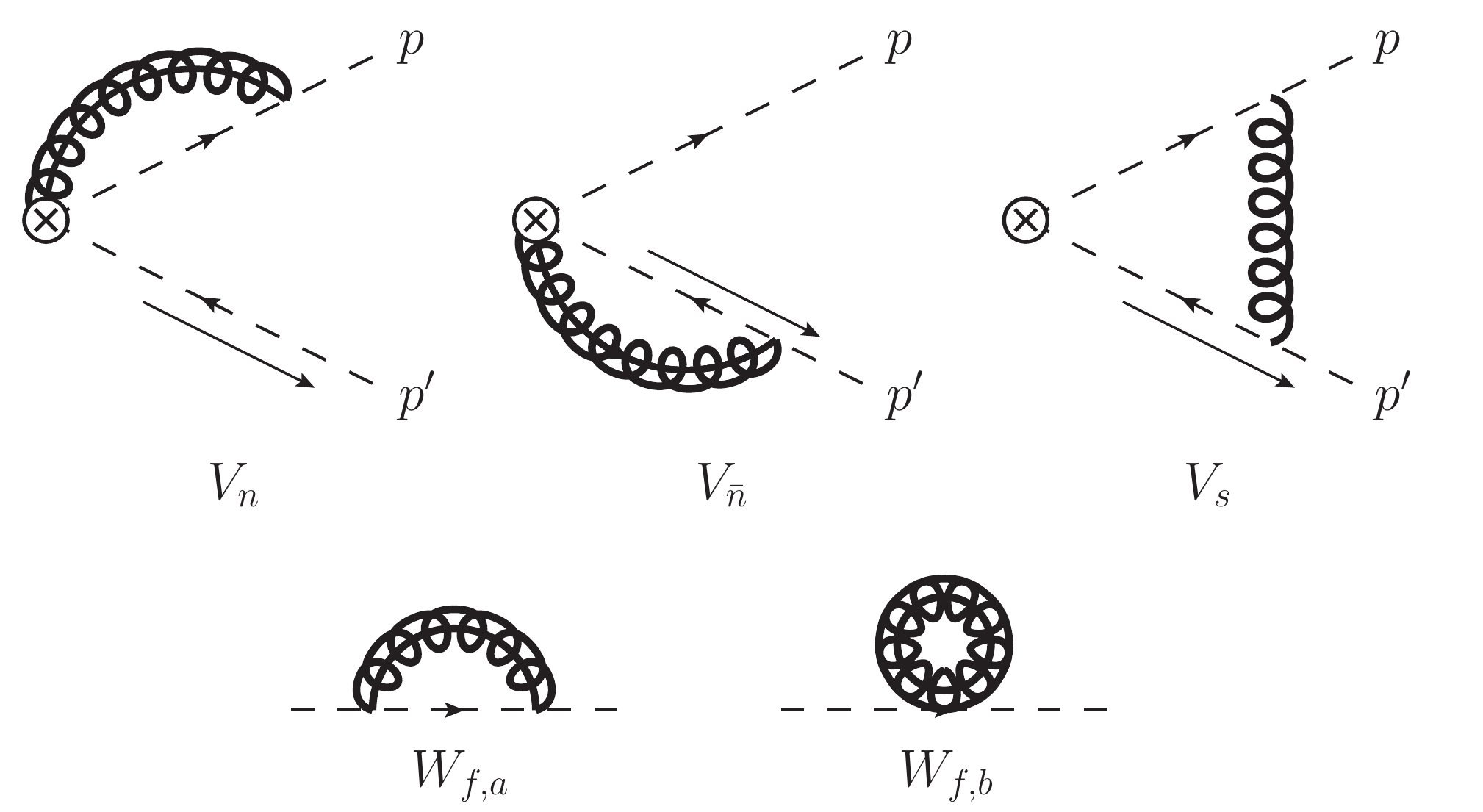}
 \caption{Non-vanishing EFT diagrams for the computation of the hard matching coefficient,
   soft mass mode bin subtractions are implied for the collinear
   diagrams. \label{fig:hardfunction_diagrams}}  
\end{figure}

Following Eq.~(\ref{eq:propagatorsubtracted}) we can obtain the ${\cal O}(\alpha_s^2 C_{\!F} T_F)$ secondary
massive quark form factor corrections relevant for the hard current matching calculation with the on-shell
subtraction for the strong coupling by the relation
\begin{align}
  F^{(2,\rm OS)}_{\rm QCD (SCET)}\!(Q,m,\mu) = & \, \frac{1}{\pi}\! \int \!\frac{\df M^2}{M^2}
  F^{(1)}_{M,\rm QCD (SCET)}\!(Q,M,\mu) \nn \\
 & \times \mathrm{Im}\! \left[\Pi(m^2,M^2)\right] , \label{eq:F_SCET_OS}
\end{align}
where $F^{(1)}_{M,\rm QCD}$ ($F^{(1)}_{M,\rm SCET}$) denotes the one-loop massive gluon form factor in QCD (SCET). 
$F^{(2,\rm OS)}_{\rm QCD}$ is both IR- and UV-finite, has been computed in Refs.~\cite{Kniehl1990,Hoang:1995fr} and is 
equivalent to $F^{(n_l,2)}_{\rm QCD}$ given in Eq.~(\ref{eq:F_QCD}). The massive gluon form factor diagrams in SCET are 
displayed in Fig.~\ref{fig:hardfunction_diagrams}, have been computed in 
Refs.~\cite{Gritschacher:thesis,Gritschacher:2013pha} 
and read in $d$ dimensions\footnote{Here we have corrected a typo in Eq.~(71) of Ref.~\cite{Gritschacher:2013pha} concerning 
the factor $(-1)^{2-d/2}$ appearing in Eq.~(\ref{eq:FMd}). For $d \rightarrow 4$ both expressions give the same terms up to 
terms of $\mathcal{O}(\epsilon)$.} 
\begin{align}\label{eq:FMd}
& F^{(1)}_{M,\rm SCET}(Q,M,\mu)= \frac{\alpha_s C_{\!F}}{2\pi}  \Gamma\bigg(2-\frac{d}{2}\bigg)
\Big(\frac{\mu^2 e^{\gamma_E}}{M^2}\Big)^{2-\frac{d}{2}} \nn \\
& 
\times\!\bigg[H_{\frac{d}{2}-1}-(-1)^{2-d/2}\,\Gamma\bigg(\frac{d}{2}\bigg)\Gamma\bigg(1-\frac{d}{2}\bigg)\nn \\
& - \frac{4-6\,d+d^2}{d\,(d-2)}+\ln{\bigg(\frac{M^2}{Q^2}\bigg)}\bigg] \, ,
\end{align}
where $H_n$ denotes the $n$-th Harmonic Number. To avoid double counting and achieve gauge invariance it is crucial
to subtract the soft-bin contributions, which arise from the soft scaling regions of the collinear diagrams.
The collinear diagrams $V_n$, $V_{\bar{n}}$, their soft-bin subtractions and the soft diagram $V_s$ in 
Fig.~\ref{fig:hardfunction_diagrams} are for themselves not fully regularized in dimensional regularization due
to rapidity divergences. These cancel in the sum of all diagrams and leave behind the rapidity logarithm
$\ln(M^2/Q^2)$. Due to the finite gluon mass the soft-bin contributions
are essential and non-vanishing for a general regularization of the rapidity singularities.
Interestingly, this logarithm cancels in the difference
of $F^{(1)}_{M,\rm QCD}$ and $F^{(1)}_{M,\rm SCET}$ so that there is no corresponding rapidity logarithm in the
$\mathcal{O}(\alpha_s^2 C_{\!F} T_F)$ secondary massive quark corrections to the hard current matching coefficient at
the scale $\mu_H \sim Q$. Thus the rapidity singularities that arise in the SCET form factor computation do not
leave any trace in the hard current matching coefficient.

Carrying out the convolution in Eq.~(\ref{eq:F_SCET_OS}) in \mbox{$d=4-2\epsilon$} dimensions and expanding in
$\epsilon$ we obtain \mbox{($x^2\equiv m^2/(Q^2+i\,0)$}, $L_{-Q}\equiv \ln[-(Q^2+i\,0)/\mu^2]$,
\mbox{$\alpha_s=\alpha_s^{(n_l)}(\mu)$})
\begin{align}\label{eq:F2_SCET}
& F^{(2,\rm OS)}_{\rm SCET}(Q,m,\mu)=\frac{\alpha_s^2 C_{\!F} T_F}{(4\pi)^2}
\left\{ \frac{2}{\epsilon^3}+\frac{1}{\epsilon^2}\left[-\frac{8}{3}\, \ln{(-x^2)} \right.\right.\nn \\
&-\left.4\, L_{-Q}+\frac{8}{9}\right] +\frac{1}{\epsilon}\left[\frac{4}{3}\, \ln^2{(-x^2)}+\frac{16}{3}\,\ln(-x^2)\,
L_{-Q} \right.\nn \\
&+\!\left.4 \,L_{-Q}^2-4 \,\ln{(-x^2)}-\frac{16}{9}\,L_{-Q}-\left(\frac{65}{27}+\frac{\pi^2}{9}\right)\!\right]\!-
\frac{8}{3}\,L_{-Q}^3 \nn \\
&-\frac{16}{3}\,\ln{(-x^2)} L_{-Q}^2-\frac{8}{3}\,\ln^2{(-x^2)} L_{-Q}+\frac{56}{9}\ln^2{(-x^2)} \nn \\
&+8 \, \ln{(-x^2)} L_{-Q}+\frac{16}{9}L_{-Q}^2 +\left(\frac{242}{27}+\frac{4\pi^2}{9}\right)\!\ln{(-x^2)} \nn \\
&+\left.\left(\frac{130}{27}+\frac{2\pi^2}{9}\right)\!L_{-Q}+\frac{875}{54}+\frac{8\pi^2}{9}-
\frac{20}{3}\,\zeta_3\right\}\, .
\end{align}
Since $F^{(2,\rm OS)}_{\rm QCD}$ and $F^{(2,\rm OS)}_{\rm SCET}$ have been computed with the subtracted dispersion 
relation they correspond to expressions in the $n_l$-flavor scheme for the strong coupling. To switch to the 
($n_l+1$)-flavor scheme one has to add the $\MS$-subtracted vacuum polarization function at zero-momentum times
the corresponding one-loop form factor,
\begin{align}
 &F^{(2)}_{\rm QCD (SCET)}(Q,m,\mu,\Delta) = F^{(2, \rm OS)}_{\rm QCD (SCET)}(Q,m,\mu) \nn \\
 & - \left(\Pi(m^2,0)-\frac{\alpha_s T_F}{3\pi} \,\frac{1}{\epsilon}\right)\! F^{(1)}_{\rm QCD (SCET)}(Q,\mu,\Delta) \, ,
\end{align}
where $F^{(1)}_{\rm QCD}$ ($F^{(1)}_{\rm SCET}$) is the massless gluon one-loop QCD (SCET) form factor calculated with
an IR regulator $\Delta$. To obtain the matching coefficient we should in principle first renormalize both quantities
and then calculate their difference where the dependence on $\Delta$ cancels. Since the QCD current is UV finite, it
is convenient to revert this procedure, i.e.\ to first determine the difference of the unrenormalized quantities and
renormalize the UV divergences in the SCET contribution at the very end. In this way the cancellation of the IR divergences can be made explicit from the beginning. The difference of the massless gluon one-loop
QCD and SCET form factors has the form\,\footnote{Using dimensional regularization for both UV and IR divergences the
SCET form factor for massless gluons vanishes identically.}
\begin{align} 
 & F^{(1)}_{\rm QCD}(Q,\mu) -F^{(1)}_{\rm SCET}(Q,\mu) = \frac{\alpha_s C_{\!F}}{4 \pi}\!
 \left(\!-\,\frac{\mu^2e^{\gamma_E}}{Q^2}\right)^{\!\!2-\frac{d}{2}} \nn \\
 &\times \frac{d^2- 7d+16}{d-4} \,
 \frac{\Gamma\!\left(2-\frac{d}{2}\right)\Gamma\!\left(\frac{d}{2}-1\right)^2}{\Gamma(d-2)}\,.
\end{align}
The additional term corresponding to the change from the $n_l$- to the ($n_l+1$)-flavor scheme thus reads
\begin{align}\label{eq:FII_OSMSbar}
 & \delta F^{{\rm OS} \rightarrow \MS}(Q,m,\mu) \nn \\
 &= -\left(\Pi(m^2,0)-\frac{\alpha_s T_F}{3\pi} \,\frac{1}{\epsilon}\right)\!
 \left(F^{(1)}_{\rm QCD} - F^{(1)}_{\rm SCET}\right) \nn\\
 & =\frac{\alpha_s^2 C_{\!F} T_F}{(4\pi)^2}\bigg\{ \frac{1}{\epsilon^2}
 \left[-\frac{8}{3}\,\ln (-x^2)-\frac{8}{3}\,L_{-Q}\right] \nn \\
 &+\frac{1}{\epsilon}\bigg[\frac{4}{3}\,\ln^2(-x^2)+\frac{16}{3}\,\ln (-x^2)
 L_{-Q}+4\,L_{-Q}^2 \nn \\
 & - 4 \,\ln(-x^2) - 4 \,L_{-Q} +\frac{2\pi^2}{9}\bigg]\!-\frac{4}{9}\, \ln^3(-x^2)  \nn \\
&- \frac{8}{3}\, \ln^2(-x^2)L_{-Q}-\frac{16}{3}\, \ln(-x^2) L_{-Q}^2 -
 \frac{28}{9}\,L_{-Q}^3 \nn \\
&+2 \,\ln^2(-x^2) +8 \,\ln (-x^2)L_{-Q}+6\, L_{-Q}^2  \nn \\
&-\frac{32}{3}\, \ln(-x^2)-\bigg(\frac{32}{3}+\frac{2\pi^2}{9}\bigg)L_{-Q}+\frac{\pi^2}{3}-\frac{8}{9}\,\zeta_3\bigg\}\,.
\end{align}
Combining all contributions and including the $\MS$ current counterterm contribution $Z^{(n_l+1,2)}_{C,n_{\!f}=1}$
given in Eq.~(\ref{eq:Zc}), the result for the $\mathcal{O}(\alpha_s^2 C_{\!F} T_F)$ secondary massive quark
contributions to the hard current coefficient in the \mbox{($n_l+1$)-flavor} scheme reads
($\alpha_s=\alpha_s^{(n_l+1)}(\mu)$)
\begin{align}
 & \delta C^{(n_l+1)}(Q,m,\mu)  =    F^{(2,\rm OS)}_{\rm QCD}(Q,m) - F^{(2,\rm OS)}_{\rm SCET}(Q,m,\mu) \nn \\
 &+ \delta F^{{\rm OS} \rightarrow \MS}(Q,m,\mu)  - Z^{(n_l+1,2)}_{C,n_{\!f}=1}(Q,\mu) \, ,
\label{eq:FII_MSbar}
\end{align}
Inserting Eqs.~(\ref{eq:Zc}),~(\ref{eq:F2_SCET}), and (\ref{eq:FII_OSMSbar}) and subtracting from Eq.~(\ref{eq:FII_MSbar})
the massless limit of Eq.~(\ref{eq:C0}) for one single flavor we obtain the mass corrections to the form factor given
in Eq.~(\ref{eq:hardcoeffII}). We see from the result of Eq.~(\ref{eq:F_II}) that the SCET matching procedure in
the ($n_l+1$)-flavor scheme does in principle nothing other than exactly subtracting the asymptotic massless limit from
the full QCD on-shell form factor correction.

\subsection{Thrust Jet Function}\label{sec:jetfunction}

\begin{figure}
 \centering
 \includegraphics[width=\linewidth]{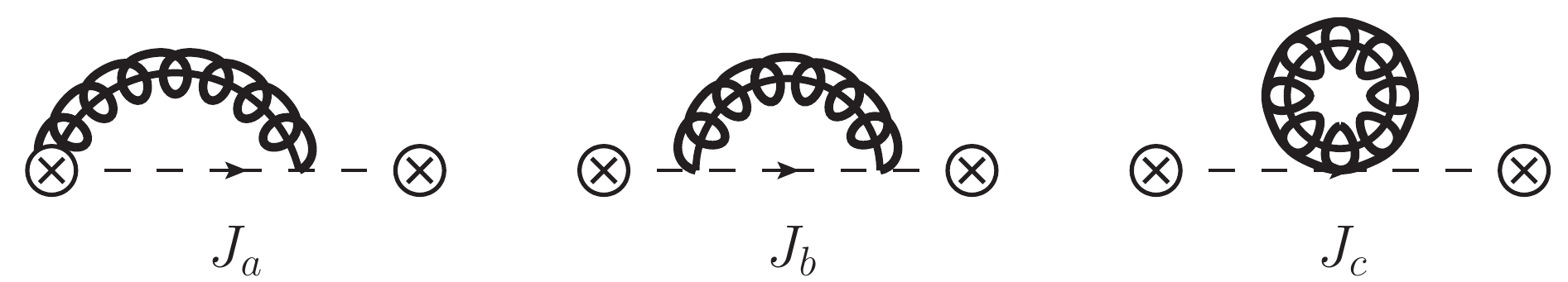}
 \caption{Non-vanishing EFT diagrams for the computation of the jet function. The required
   soft mass mode bin subtractions are implicit. Concerning $J_a$, the
   right-symmetric diagram also has to be taken into
   account.\label{fig:jetfunction_diagrams}}  
\end{figure}

The calculation of the ${\cal O}(\alpha_s^2 C_{\!F} T_F)$ secondary massive quark corrections to the jet function in the 
($n_l+1$)-flavor scheme goes along the lines of the hard current coefficient. The $\mathcal{O}(\alpha_s)$ corrections
to the jet function due to a massive gauge boson with QCD vector coupling have the 
form~\cite{Gritschacher:thesis,Gritschacher:2013pha}\,\footnote{We consider directly the corrections to the total
thrust jet function, which are exactly twice the contributions for the function of a single jet.}
\begin{align}
\!\!\delta J_M^{(1)}(s,M,\mu) = \delta J_{M, \rm virt}^{(1)}(s,M,\mu) + \delta J_{M, \rm real}^{(1)}(s,M).
\end{align}
The distributive part $\delta J_{M, \rm virt}^{(1)}$ corresponds to virtual radiation of the massive gauge boson
and the full expression in $d$ dimensions reads
\begin{align}
& \mu^2 \delta J_{M, \rm virt}^{(1)}(s,M,\mu)= \frac{2\alpha_s C_{\!F}}{\pi} \Gamma\bigg(2-\frac{d}{2}\bigg)\!
\bigg(\frac{\mu^2 e^{\gamma_E}}{M^2}\bigg)^{2-\frac{d}{2}} \nn \\
& \times \left\{\left[H_{\frac{d}{2}-1}-H_{1-\frac{d}{2}}+
\ln{\left(\frac{M^2}{\mu^2}\right)}+\frac{2-d}{2d}\,\right]\!\delta(\bar{s}) \right. \nn \\
& \left.- \left[\frac{\theta(\bar{s})}{\bar{s}}\right]_+ \right\} \, .
\label{eq:jetfunction1_virt}
\end{align}
The UV- and IR-finite real radiation contribution $\delta J_{M, \rm real}^{(1)}$ can for our purposes be evaluated
for $d=4$ since it does not require any regularization for the convolution in the subtracted dispersion relation.
It reads
\begin{align}
&\mu^2 \delta J_{M, \rm real}^{(1)}(s,M)=\frac{\alpha_s C_{\!F}}{2\pi} \mu^2\theta\left(s-M^2\right) \nn \\
& \times \left\{\frac{(M^2-s)(3s+M^2)}{s^3}+\frac{4}{s}\,\ln{\left(\frac{s}{M^2}\right)}\right\}.
\label{eq:jetfunction1_real}
\end{align}
The calculation of $\delta J_{M}^{(1)}$ involves the collinear diagrams in Fig.~\ref{fig:jetfunction_diagrams},
where the corresponding soft-bin subtractions are implied and contribute only to $\delta J_{M, \rm virt}^{(1)}$.~The 
soft-bin subtractions are crucial for gauge invariance as well as for a cancellation of all rapidity 
singularities~\cite{Gritschacher:2013pha}. As a remnant of this cancellation we get a rapidity logarithm $\ln(M^2/\mu^2) 
\sim \ln(M^2/s)$ in $\delta J_{M, \rm virt}^{(1)}$ in Eq.~(\ref{eq:jetfunction1_virt}). For $M^2 \ll s$ a corresponding 
logarithm arises in the real radiation term $\delta J_{M, \rm real}^{(1)}$ in Eq.~(\ref{eq:jetfunction1_real}) which 
cancels the rapidity logarithm from $\delta J_{M, \rm virt}^{(1)}$. We emphasize, however, that in the calculation of 
$\delta J_{M, \rm real}^{(1)}$ rapidity divergences do not arise anywhere. These properties are also inherited to the 
$\mathcal{O}(\alpha_s^2 C_{\!F} T_F)$ massive quark corrections discussed in the following.

The ${\cal O}(\alpha_s^2 C_{\!F} T_F)$ unrenormalized massive quark corrections to the jet function in the $n_l$-flavor
scheme for $\alpha_s$ can be obtained with the subtracted dispersion relation
\begin{align}
 & \delta J^{(2,\rm OS)}_{m}(s,m,\mu) =  \delta J^{(\rm OS, virt)}_{m}(s,m,\mu) +
 \delta J^{\rm real}_{m}(s,m) \nn \\
 & =  \frac{1}{\pi} \int \frac{\df M^2}{M^2} \, \delta J_M^{(1)}(s,M,\mu) \,  \mathrm{Im}\! \left[\Pi(m^2,M^2)\right] \, .
\end{align}
The convolution is performed separately for the \mbox{$d$-dimensional} virtual terms in Eq.~(\ref{eq:jetfunction1_virt}) and
the four-dimensional threshold term in Eq.~(\ref{eq:jetfunction1_real}), where for the latter no divergences arise
in the $M$-integration and thus the $d=4$ version of the absorptive part of the vacuum polarization function in 
Eq.~(\ref{eq:Im_Pi}) can be used. This yields Eq.~(\ref{eq:J_real}) for the real radiation term
$\delta J^{\rm real}_{m}$ and ($L_m=\ln(m^2/\mu^2), \alpha_s=\alpha_s^{(n_l)}(\mu)$)
\begin{align}\label{eq:jet_OS}
& \mu^2 \delta J^{(\rm OS,\rm virt)}_{m}(s,m,\mu)= \frac{\alpha_s^2 C_{\!F} T_F}{(4\pi)^2}
\left\{ \left[\frac{8}{\epsilon^3} \right. \right. \nn \\
& +\frac{1}{\epsilon^2}\left(-\frac{32}{3}\,L_{m}-\frac{4}{9}\right)+\frac{1}{\epsilon}
\left(\frac{16}{3}\,L_{m}^2 - 8\, L_m-\frac{242}{27}+\frac{4\pi^2}{9}\right) \nn \\
&\left.+\frac{152}{9}\,L_{m}^2+\frac{932}{27}\,L_{m}+\frac{1531}{27}+\frac{38\pi^2}{27}-\frac{64}{3}\,\zeta_3\right]
\!\delta(\bar{s}) \nn \\
& + \left[-\frac{16}{3\epsilon^2}+
\frac{1}{\epsilon}\left(\frac{32}{3}\,L_{m}+\frac{80}{9}\right)-\frac{32}{3}\,L_{m}^2-\frac{160}{9}\,L_{m} \right. \nn \\
&-\left.\left.\frac{448}{27}-\frac{8\pi^2}{9}\right]\!\!\left[\frac{\theta(\bar{s})}{\bar{s}}\right]_+ \right\} \, .
\end{align}

We switch to the ($n_l+1$)-flavor scheme for $\alpha_s$ by adding the $\MS$-renormalized $\Pi(0)$ times
the (unrenormalized) massless one loop contribution to the jet function which reads
\begin{align}
\!\!\!\! J^{(1)}_{\rm bare}(s,\mu)=\frac{\alpha_s C_{\!F}}{2\pi} \frac{1}{s} \Big(\frac{\mu^2 e^{\gamma_E}}{s}\Big)^{2-\frac{d}{2}}
 \frac{d+4}{d-4} \, \frac{\Gamma\left(\frac{d}{2}\right)}{\Gamma\left(d-2\right)} \, .
\end{align}
Thus the corresponding contribution needed to change from the $n_l$ to the ($n_l+1$)-flavor scheme reads
\begin{align}\label{eq:jet_OSMS}
& \delta J^{{\rm OS} \rightarrow \MS}_m (s,m,\mu) \nn \\
& = -\,\bigg(\Pi(m^2,0)-\frac{\alpha_s T_F}{3\pi}  \,\frac{1}{\epsilon}\bigg)\, J^{(1)}_{\rm bare}(s,\mu) \nn \\
& =  \frac{\alpha_s^2 C_{\!F} T_F}{(4\pi)^2} \left\{ \left[\frac{32}{3 \epsilon^2}\, L_m +
\frac{1}{\epsilon}\left(- \frac{16}{3}\,L_m^2+8\, L_m -\frac{8 \pi ^2}{9}\right)\right.\right. \nn \\
&+\left.\!\frac{16}{9}L_m^3-4\, L_m^2+\bigg(\frac{56}{3}-\frac{16 \pi^2}{9}\bigg)L_m-\frac{2 \pi^2}{3}+
\frac{32}{9}\,\zeta_3 \right]\! \delta(\bar{s})\nn \\
& + \left[-\frac{32}{3\epsilon}\,L_{m}+\frac{16}{3}\,L_{m}^2-
8\,L_{m} +\frac{8\pi^2}{9}\right]\!\!\left[\frac{\theta(\bar{s})}{\bar{s}}\right]_+ \nn \\
&+\left.\frac{32}{3}\,L_m \left[\frac{\theta(\bar{s})\,\ln\, \bar s}{\bar{s}}\right]_+\right\} \, .
\end{align}
Combining all contributions and renormalizing the result with the jet counterterm contribution
$Z_{J,n_{\!f}=1}^{(n_l+1,2)}$ in Eq.~(\ref{eq:ZJ}) finally gives
\begin{align}\label{eq:JvirtMSbar}
 \delta J_m^{\rm virt}(s,m,\mu) = & \,  \delta J^{({\rm OS, virt})}_{m}(s,m,\mu)+
 \delta J^{{\rm OS} \rightarrow \MS}_m(s,m,\mu) \nn \\ 
 & - Z_{J,n_{\!f}=1}^{(n_l+1,2)}(s,\mu) \, .
\end{align}
Inserting Eqs.~(\ref{eq:ZJ}),~(\ref{eq:jet_OS}),~(\ref{eq:jet_OSMS}) and subtracting from Eq.~(\ref{eq:JvirtMSbar})
the massless limit of Eq.~(\ref{eq:J0}) for one single flavor we obtain the virtual massive quark corrections to the
jet function given in Eq.~(\ref{eq:J_virt}).

\section{Renormalization Conditions, Threshold Corrections and Consistency Relations}\label{sec:RGconsistency}

In this section we discuss the RG properties of the individual ingredients of the factorization theorem, namely the
hard current coefficient, the jet function and the soft function, rather than the factorization theorem as a whole.
Since the hard coefficient and the jet and soft functions are gauge-invariant quantities, they can also be renormalized 
independently. This fact can be used to determine the threshold correction factors $\mathcal{M}_C$ for the hard 
coefficient (see Eq.~(\ref{eq:matchingII})) and $\mathcal{M}_J$ for the jet function (see Eq.~(\ref{eq:matchingIII})) as 
well as the threshold correction factor $\mathcal{M}_S$ for the soft function (see Eq.~(\ref{eq:matchingIV})). The latter 
becomes relevant if one sets the final renormalization scale $\mu$ above the soft scale and the RG evolution of the soft 
function crosses the massive quark threshold. Instead of using different effective theories that follow the strict 
guideline of having the massive quark modes either as fluctuating fields contributing to the RG 
evolution in the same way as the massless quarks or excluded completely (i.e.\ integrated out), we use only a single theory 
which contains the massive quark modes, but employs different renormalization conditions for the quantum corrections that 
arise from the massive quark modes. These renormalization conditions are either the $\MS$ prescription or an on-shell (or 
low-energy momentum subtraction) prescription. The former leads to the usual $\MS$ feature that massless quarks and the 
massive flavor all contribute to the RG evolution in the same way, so one uses the ($n_l+1$) running flavor scheme. The 
latter also subtracts finite and scale-dependent contributions such that the massive flavor does not lead to any 
contribution in the RG evolution, so there are only $n_l$ running flavors. This concerns the strong coupling $\alpha_s$ 
(see Sec.~\ref{sec:dispersion}) as well as the hard coefficient and the jet and soft functions. 

Obviously the $\MS$ prescription is suitable to cover the situation where the quark mass becomes small (where ``suitable'' 
means that no large mass logarithms arise in the massless limit) and, as already demonstrated in 
Sec.~\ref{sec:computations}, leads to results which give the known results for massless quarks in the limit $m\rightarrow 
0$. The on-shell prescription is suitable to cover the decoupling limit, such that the effects of the massive quark vanish 
in the infinite mass limit. The decoupling condition renders the finite subtraction unique for all calculations within 
SCET. This method of using different renormalization conditions for the RG evolution schemes with ($n_l+1$) and $n_l$ 
running flavors also has the advantage that the kinematic thresholds of the jet and soft functions due to the quark mass
are fully contained in them regardless of which type of renormalization scheme is used. This is unlike the case 
of using the effective theory method, where the massive quark is completely excluded from the $n_l$-flavor theory,
and one is forced to take care of the fact that the real radiation thresholds are always located in the
($n_l+1$)-flavor theory.

The differences of the renormalized quantities w.r.\ to both of these renormalization prescriptions constitute threshold 
matching conditions that uniquely define the mass mode matching threshold correction factors $\mathcal{M}_C$, 
$\mathcal{M}_J$ and $\mathcal{M}_S$. Since the hard current coefficient, the jet and the soft functions are independent 
and in principle not tied to the particular factorization theorem for thrust, the important outcome is that the threshold 
correction factors can be determined from these quantities and do not rely on a separate perturbative calculation of the 
thrust distribution in full QCD. From the form of the factorization theorem for thrust we can therefore {\it predict} the 
singular $\mathcal{O}(\alpha_s^2 C_{\!F} T_F)$ massive quark corrections to the thrust distribution in full fixed-order QCD. 
To our knowledge they have not been calculated in an explicit form before in the literature.

The fact that the hard coefficients, the jet and soft functions and the massive quark threshold corrections factors that 
appear in the factorization theorems in the four scenarios (in schemes with either $n_l$ or $n_l+1$ running flavors) are 
conceptually connected through different choices of renormalization schemes and not related in any way to expansions in 
either small or large quantities makes it evident that the predictions of the different factorization theorems at their 
respective borders of validity have an overlap region and are continuous.\,\footnote{We mean continuity up to higher order 
perturbative corrections which are not enhanced by large logarithms.}

\begin{figure*}
  \begin{center}
  \subfigure{\epsfig{file=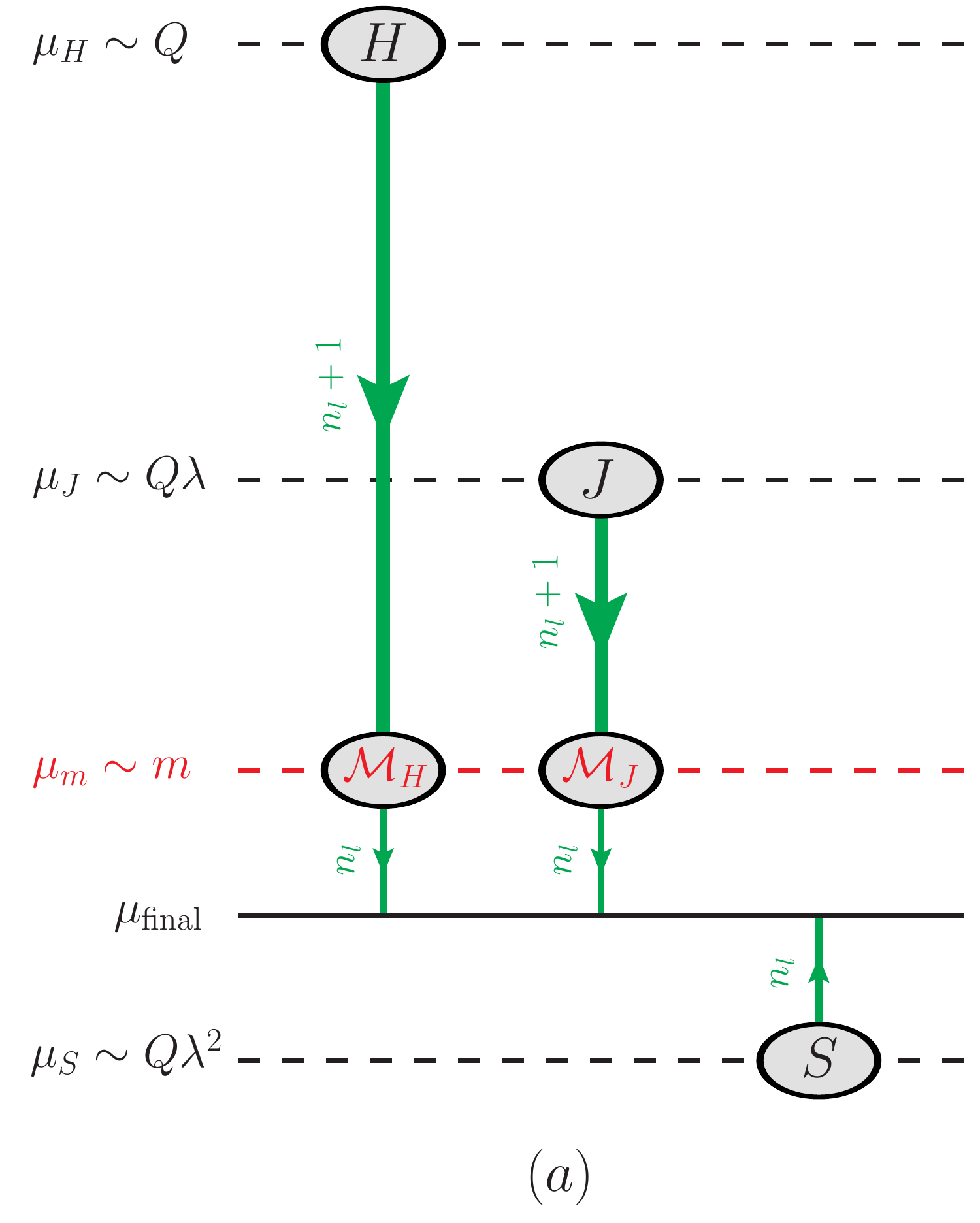,height=0.40\linewidth,clip=}}
  \subfigure{\epsfig{file=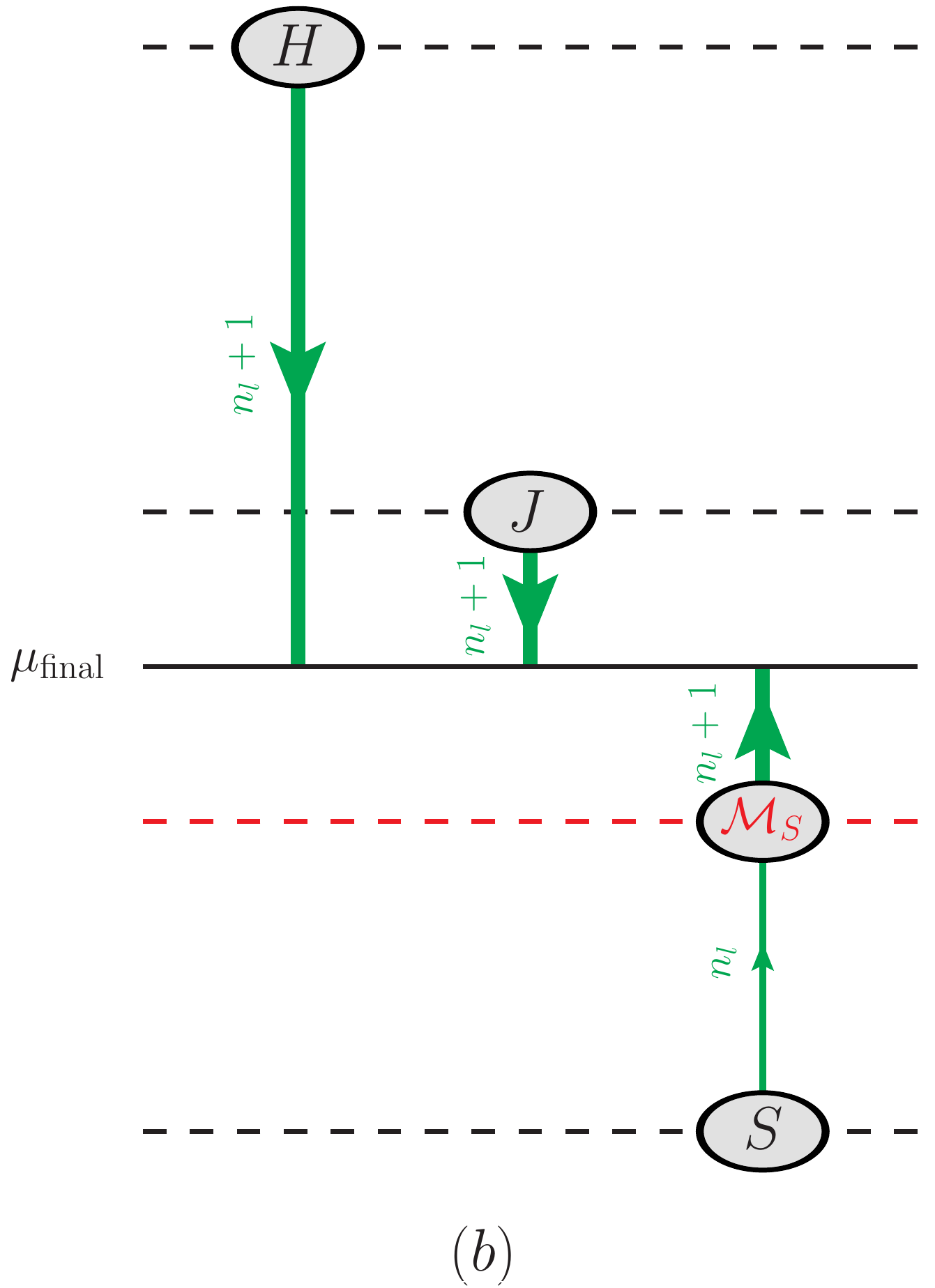,height=0.40\linewidth,clip=}}
  \subfigure{\epsfig{file=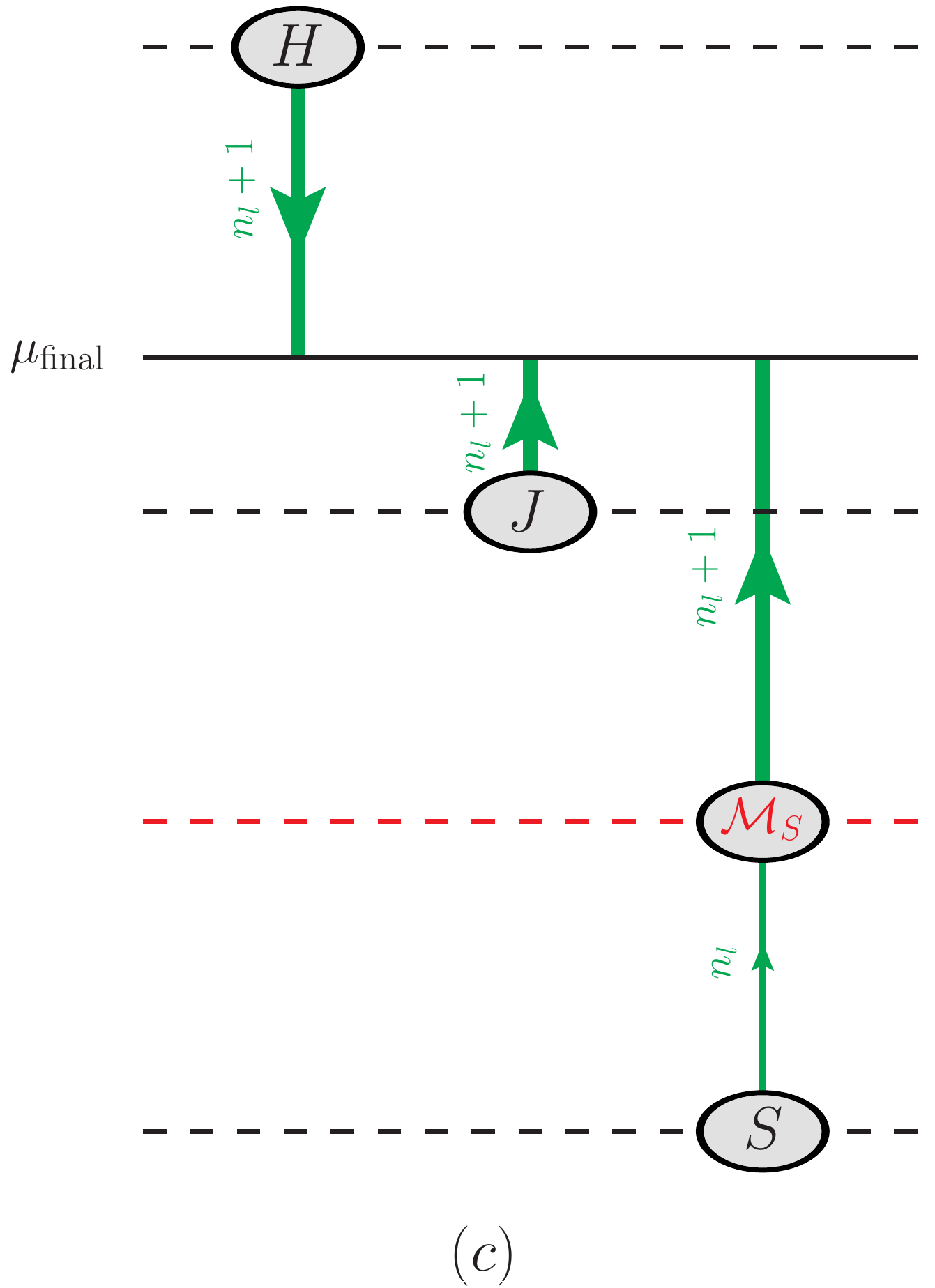,height=0.40\linewidth,clip=}}
   \caption{Illustration of the different RG setups for scenario III ($\mu_J>\mu_m>\mu_S$) leading to the consistency
relations mentioned in the text.
Shown are the cases where the final renormalization scale $\mu_{\rm final}$ satisfies (a) 
$\mu_m>\mu_{\rm final}>\mu_S$, (b) $\mu_J>\mu_{\rm final}>\mu_m$ and (c) $\mu_H>\mu_{\rm final}>\mu_J$.}
  \label{fig:RGconsistency}
  \end{center}
\end{figure*}

We believe that it is worth discussing this issue briefly in the following. In Sec.~\ref{sec:massmode_setup} we have 
discussed four scenarios one needs to distinguish for treating the possible hierarchies between the quark mass $m$ on the 
one side and the hard scale $Q$, the jet scale $\sim Q\lambda$ and the soft scale $Q\lambda^2$ on the other. 
Concerning the power counting we may assume the canonical strong hierarchy between these scales (such as
$Q \gg m \gg Q\lambda \gg Q\lambda^2$) when discussing the effective theory setup, the RG evolution, the results
for hard current coefficient, jet and soft functions and the mass mode threshold corrections that result when the
massive modes are integrated out. However, since the components of the factorization theorems are related simply by 
different choices of renormalization conditions, each factorization theorem also applies in cases where hierarchies 
involving the mass are only marginal or non-existent (such as $Q \gg m\gtrsim Q\lambda \gg Q\lambda^2$ or
$Q \gtrsim m \gg Q\lambda \gg Q\lambda^2$). In these cases the RG evolution between close-by scales is equivalent to a 
perturbative treatment, so that the factors of concern might be simply expanded out. Since this might be as well
applied to the two neighboring scenarios within some range, we have continuity between the two 
descriptions, and the transition point where one switches between them can be picked freely within this range. This 
feature is very similar to (but distinguished from) the property that the scale $\mu_m$ where the quark mass $m$ is 
integrated out can be picked freely within some range in a practical application. The freedom in these choices causes 
changes in the numerical predictions due to the truncation of the perturbative expansion and might contribute to 
estimating the remaining perturbative uncertainty. 

The fact of having four different scales relevant for setting up the RG evolution ($\mu_H$, $\mu_J$, $\mu_S$ and $\mu_m$) 
leads to another interesting feature related to the proliferation of possibilities to pick the final renormalization scale 
$\mu = \mu_{\rm final}$ to which the hard current coefficient, the jet function and the soft function are being evolved in 
the different factorization theorems. See Fig.~\ref{fig:RGconsistency} for an illustration of three equivalent choices for 
scenario III, where we display the situations in which $\mu_{\rm final}$ lies (a) between mass and soft scales,
(b) between the
jet and mass scales, and (c) between the hard and jet scales. The different possibilities and their equivalence concerning 
predictions is one of the deeper conceptual properties of factorization theorems. On the one hand, they imply the 
well-known consistency conditions between the RG evolution factors $U_C^{(n_{\!f})}$, $U_J^{(n_{\!f})}$ and $U_S^{(n_{\!f})}$ for 
$n_{\!f}=n_l$ and $n_{\!f}=n_l+1$, see Eq.~(\ref{eq:consistency_ML}). However, 
in the context of the RG evolution crossing a massive quark threshold they also imply a consistency relation between the 
mass mode threshold matching correction factors $\mathcal{M}_C$, $\mathcal{M}_J$ and $\mathcal{M}_S$. This can be used to 
gain interesting general insights into properties of mass singularities, and at the practical level, may be used as a 
non-trivial tool for consistency checks. 

Apart from providing consistency checks of theoretical calculations, these relations also have computational power, as 
they can be used to calculate properties of gauge-invariant and independent field theoretic objects once it has become 
clear that they represent building blocks of a factorization theorem. In the case of thrust these building blocks are the 
hard (vector or axial) current coefficient, the jet function and the soft function. Hereby, one of the most interesting 
aspects is that the various building blocks can appear in different factorization theorems and one may gain insights into 
the mass-singularities of apparently unrelated quantities. 

In the following subsections A-C we discuss the evolution and the mass mode threshold corrections for the hard current 
coefficient, the jet function and the soft function using the renormalization method described above. The calculations are 
fast and concise and are based on the $\mathcal{O}(\alpha_s^2 C_{\!F} T_F)$ massive quark results given in 
Sec.~\ref{sec:computations}. Using RG-invariance w.r.\ to the scale where one switches between $n_l$- and ($n_l+1$)-flavor 
schemes we also examine terms of $\mathcal{O}(\alpha_s^3)$ and $\mathcal{O}(\alpha_s^4)$ which are enhanced by rapidity 
logarithms and may be counted as $\mathcal{O}(\alpha_s^2)$ in the  logarithmic counting
$\alpha_s \ln \sim \mathcal{O}(1)$. In subsection~D we discuss the consistency conditions among the mass mode
matching corrections $\mathcal{M}_C$, $\mathcal{M}_J$ and $\mathcal{M}_S$, and we also show that they are also
relevant for the perturbative equivalence of the factorization theorems of neighboring scenarios in
their overlap region. In subsection E we present the explicit result for the singular
$\mathcal{O}(\alpha_s^2 C_{\!F} T_F)$ massive quark correction of the thrust distribution in full QCD in the
fixed-order expansion.

\subsection{Current mass mode matching coefficient}\label{sec:currentmassmodematching}
The mass mode threshold factor $\mathcal{M}_C(Q,m,\mu_m)$ arises when the RG evolution of the hard current coefficient
crosses the massive quark threshold. In the following we describe how it is related to the renormalization conditions
for the hard current. The bare and the renormalized current coefficients $C^{(0)}(Q,m,\mu)$ and $C^{(n_{\!f})}(Q,m,\mu)$
are related to each other via
\begin{align}
 C^{(0)}(Q,m,\mu) = Z_C^{(n_{\!f})}(Q,m,\mu) \, C^{(n_{\!f})}(Q,m,\mu) \, ,
\end{align}
where $Z_C^{(n_{\!f})}(Q,m,\mu)$ is the renormalization factor in a $n_{\!f}$-flavor scheme. In the following we will omit all
$\mathcal{O}(\alpha_s^2 C_{\!F}^2)$ and $\mathcal{O}(\alpha_s^2 C_{\!F} C_{\!A})$ terms, as they are irrelevant for our considerations.
For scales $\mu> m$ we use the $(n_l+1)$-flavor scheme, so we employ the $\MS$-subtractions for the UV divergent 
contributions to the strong coupling and the current. The counterterm is mass independent and reads with the
notation of section~\ref{sec:massless} 
\begin{align}\label{eq:ZC_MS}
 Z_C^{(n_l+1)}(Q,\mu)= 1+ Z_C^{(n_l+1,1)}(Q,\mu) + Z_{C,n_l+1}^{(n_l+1,2)}(Q,\mu) \, .
\end{align}
The $\mathcal{O}(\alpha_s^2 C_{\!F} T_F)$ contribution $Z_{C,n_l+1}^{(n_l+1,2)}(Q,\mu)$ is given in Eq.~(\ref{eq:Zc})
with $n_{\!f} = n_l+1$, whereas the $\mathcal{O}(\alpha_s)$ term reads ($L_{-Q}=\ln(-(Q^2+i\,0)/\mu^2)$)
\begin{align}\label{eq:ZC1}
 Z_C^{(n_l+1,1)}(Q,\mu)\,=\, \frac{\alpha_s^{(n_l+1)}(\mu) C_{\!F}}{4\pi}
 \bigg(\!-\frac{2}{\epsilon^2}-\frac{3}{\epsilon}+\frac{2}{\epsilon}L_{-Q}\bigg)\, .
\end{align}
Note that the contribution from the massive quark agrees with the one related to a massless flavor.
The renormalized current coefficient reads
\begin{align}
 C^{(n_l+1)}(Q,m,\mu)= &\,  1 + C^{(n_l+1,1)}(Q,\mu) + C^{(n_l+1,2)}_{n_l+1}(Q,\mu) \nn \\
 & + \delta F^{(n_l+1,2)} (Q,m) \, ,
\end{align}
which is the one for the case $m<Q$ (scenarios II, III and IV) given in Eq.~(\ref{eq:hardcoeffII}). The massless result
at $\mathcal{O}(\alpha_s)$ reads
\begin{align}\label{eq:C1}
 & C^{(n_l+1,1)}(Q,\mu) \nn \\
 & =\frac{\alpha_s^{(n_l+1)}(\mu) C_{\!F}}{4\pi}  \left(-\,L_{-Q}^2 +3\, L_{-Q} - 8 + \frac{\pi^2}{6}\right)\, .
\end{align}

In the $n_l$-flavor scheme, we intend to implement the renormalization condition that the massive quark
corrections vanish for $Q \sim \mu\ll m$. Following the computation described in Sec.~\ref{sec:hardmatching},
we now do {\it not } include the scheme change contribution $\delta F^{{\rm OS} \rightarrow \MS}$, which
implies that we use $\alpha_s^{(n_l)}$, i.e.\ the $n_l$-flavor scheme for the strong coupling. The
resulting expressions for the counterterm and the renormalized current coefficient read
\begin{align}
 Z_C^{(n_l)}(Q,\mu) = & \, 1+ Z_C^{(n_l,1)}(Q,\mu) + Z_{C,n_l}^{(n_l,2)}(Q,\mu) \nn \\
 & + Z^{(n_l,2)}_{C,\rm OS}(Q,m,\mu) \,,  \label{eq:ZC_OS} 
 \end{align}
and 
\begin{align}
 C^{(n_l)}(Q,m,\mu) = & \, 1 + C^{(n_l,1)}(Q,\mu) + C^{(n_l,2)}_{n_l}(Q,\mu) \nn \\
 & + F^{(2,\rm OS)}_{\rm QCD}(Q,m) - F^{(2,\rm OS)}_{\rm SCET}(Q,m,\mu) \nn \\
 & - Z^{(n_l,2)}_{C,\rm OS}(Q,m,\mu)\, , \label{eq:C_OS}
\end{align}
where the one-loop terms $Z_C^{(n_l,1)}(Q,\mu)$ and $C^{(n_l,1)}(Q,\mu)$ are analogous to
Eq.~(\ref{eq:ZC1}) and~(\ref{eq:C1}),
respectively. The two-loop massless contributions $Z_{C,n_l}^{(n_l,2)}(Q,\mu)$, $C^{(n_l,2)}_{n_l}(Q,\mu)$ are given in
Eqs.~(\ref{eq:Zc})~and~(\ref{eq:C0}) with $n_{\!f} = n_l$, and the two-loop massive quark contributions
$F^{(2,\rm OS)}_{\rm QCD}(Q,m)$
and $F^{(2,\rm OS)}_{\rm SCET}(Q,m,\mu)$ are given in Eqs.~(\ref{eq:F_QCD}) and~(\ref{eq:F2_SCET}), respectively, with the
corresponding counterterm contribution denoted by $Z^{(n_l,2)}_{C,\rm OS}(Q,m,\mu)$. The condition of decoupling requires that
the massive quark contributions in Eq.~(\ref{eq:C_OS}) vanish for $m\rightarrow \infty$, so we obtain
\begin{align}\label{eq:ZC2_OS}
 Z^{(n_l,2)}_{C,\rm OS}(Q,m,\mu) = -\, F^{(2,\rm OS)}_{\rm SCET}(Q,m,\mu) \, .
\end{align}
Note that the QCD term $F^{(2,\rm OS)}_{\rm QCD}(Q,m)$ automatically decouples for $m \gg Q$, so that it does not
lead to any contributions in the counterterm. The renormalized current coefficient in this scheme is thus identical
to the result for $\mu_m>\mu_H$ given in Eq.~(\ref{eq:hardcoeffI}), where the effective theory scenario I was discussed.

It is now straightforward to determine the matching relation between the renormalized hard current coefficients
in the two schemes at the scale $\mu_m$. The matching accounts for the difference between the two schemes, thus
it is obtained by the relation 
\begin{align}\label{eq:MC_cont}
 \mathcal{M}_C(Q,m,\mu_m) & = \frac{C^{(n_l)}(Q,m,\mu_m)}{C^{(n_l+1)}(Q,m,\mu_m)} \nn\\
 & = \frac{Z_C^{(n_l+1)}(Q,\mu_m)}{Z_C^{(n_l)}(Q,m,\mu_m)} \, .
\end{align}
Since the difference in the factorization theorems for scenarios I and II in Eqs.~(\ref{eq:diffsigmaI}) 
and~(\ref{eq:diffsigmaII}) concerns just the current matching conditions and evolution, Eq.~(\ref{eq:MC_cont})
makes evident that the condition for the current mass mode matching coefficient automatically implements a
continuous transition between these two scenarios at $m \sim \mu_m \sim \mu_H \sim Q$. Comparing the
factorization theorems of scenarios I and II in this region we see that the same mass-shell contributions
are just swapped between the Wilson coefficient and the mass mode matching coefficient.

Since the expressions in Eq.~(\ref{eq:MC_cont}) are written in different schemes for $\alpha_s$ one has to
relate them by the decoupling relation for $\alpha_s$\,\footnote{Using the ratio of the counterterms instead
of the ratio of the renormalized matching coefficients in Eq.~(\ref{eq:MC_cont}) we need in Eq.~(\ref{eq:alphas_dec})
terms up to $\mathcal{O}(\epsilon^2)$. These can be easily determined from the result for $\Pi(m^2,0)$ in
Eq.~(\ref{eq:vacpolzero}) in $d$ dimensions. Otherwise the calculation is straightforward and completely
equivalent to the one based on the renormalized expressions.} ($L_m=\ln(m^2/\mu_m^2)$) 
\begin{align}\label{eq:alphas_dec}
 &\alpha_s^{(n_l)}(\mu_m) = \alpha_s^{(n_l+1)}(\mu_m) \nn \\
 &\times \left[ 1 + \frac{\alpha_s^{(n_l+1)}(\mu_m) T_F}{3\pi} L_m +\mathcal{O}(\alpha_s^2) \right] \, .
\end{align}
Using the structure of the Wilson coefficients in Eqs.~(\ref{eq:hardcoeffI}) and~(\ref{eq:hardcoeffII}),
we obtain at $\mathcal{O}(\alpha_s^2)$ in the fixed-order counting
\begin{align}\label{eq:MC_structure}
 & \mathcal{M}_C(Q,m,\mu_m)=  1  + F^{(n_l,2)}_{\rm QCD}(Q,m) \nn \\
 & - C^{(n_l+1,2)}_{n_{\!f}=1}(Q,\mu_m)- \delta F^{(n_l+1,2)} (Q,m,\mu_m) \nn \\
 &   + \frac{\alpha_s^{(n_l+1)}(\mu_m) T_F}{3\pi} L_m C^{(n_l+1,1)}(Q,\mu_m) +\mathcal{O}(\alpha_s^3) \, .
\end{align}
Inserting all explicit expressions gives at $\mathcal{O}(\alpha_s^2)$ in the fixed-order counting ($Q^2=Q^2+i\,0$)
\begin{align}
 &\mathcal{M}^{(2)}_{C}(Q,m,\mu_m)= \frac{\alpha_s^2 C_{\!F} T_F}{16\pi^2}
 \left\{\left[\frac{4}{3}\,L_m^2+\frac{40}{9}\,L_m \right. \right.\nn \\
 &+\left. \frac{112}{27}\right] \ln\bigg(\!\!-\frac{m^2}{Q^2}\bigg) -\frac{8}{9}\,L_m^3-\frac{2}{9}\,L_m^2  \nn \\
 &\left.+\bigg(\frac{130}{27}+
  \frac{2\pi^2}{3}\bigg)L_m +\frac{875}{54}+\frac{5\pi^2}{9}-\frac{52}{9}\,\zeta_3\right\} \, .
\label{eq:matchingIIb}
\end{align}
Since there are no $\mathcal{O}(\alpha_s)$ one-loop corrections the schemes of $\alpha_s$ and the mass appearing
in Eq.~(\ref{eq:matchingIIb}) do not need to be specified at this point. In Eq.~(\ref{eq:matchingIIb}) we see
explicitly the large rapidity logarithm $\ln\left(-\,m^2/Q^2\right)$ which enforces the counting
$\alpha_s\, \ln(m^2/Q^2) \sim 1$. One can set up a RG evolution in rapidity space as described 
in~\cite{Chiu:2011qc,Chiu:2012ir} to resum the associated higher order logarithms, which we postpone to a
later publication~\cite{Hoang:2015vua}.\footnote{The result will then depend on two rapidity scales which
should be varied independently of the invariant mass scales. This dependence can be easily restored in the result
of Eq.~(\ref{eq:matchingH3}) by replacements of the scales in the exponentiated logarithm $L_{-Q}$. The analogous
statement holds also for the jet and soft mass mode matching coefficients in Eqs.~(\ref{eq:matchingJ3}) 
and~(\ref{eq:matchingS3}).} For our purposes the outcome, namely that this logarithm exponentiates,
is sufficient. This allows us to determine the term of $\mathcal{O}(\alpha_s^4\, \ln^2(m^2/Q^2))\sim \mathcal{O}(\alpha_s^2)$.

For a complete analysis at N$^3$LL we would also need the term at
$\mathcal{O}(\alpha_s^3\, \ln(m^2/Q^2))\sim \mathcal{O}(\alpha_s^2)$. We can determine its $\mu_m$-dependent
contribution from the identity 
\begin{align}\label{eq:U_H}
 \mathcal{M}_C(Q,m,\mu_m)= & \, U_C^{(n_l+1)}(Q,\mu_m,m) \,\mathcal{M}_C(Q,m,m) \nn \\
 & \times U_C^{(n_l)}(Q,m,\mu_m) \, ,
\end{align}
or equivalently,
\begin{align}
 & \mu\,\frac{\df}{\df\mu}\, \mathcal{M}_C(Q,m,\mu) \nn \\
 & =  \left(\gamma_C^{(n_l)}(Q,\mu) - \gamma_C^{(n_l+1)}(Q,\mu)\right) \mathcal{M}_C(Q,m,\mu) \, .
\end{align}
Expanding consistently in $\alpha_s$ gives the perturbative result for the $\mu_m$-dependent terms. Including 
the relevant term at $\mathcal{O}(\alpha_s^3\, \ln(m^2/Q^2))$ in the exponent the structure of the mass mode matching 
coefficient reads ($\alpha_s^{(n_l+1)}=\alpha_s^{(n_l+1)}(\mu_m), m={\overline m}(\mu_m)={\overline m}^{(n_l+1)}(\mu_m)$) 
\begin{widetext}
\begin{align}\label{eq:matchingH3}
 & \mathcal{M}_C(Q,m,\mu_m)= \left\{ 1 + \frac{\big(\alpha_s^{(n_l+1)}\big)^2}{(4 \pi)^2}\!
 \left[\frac{1}{12}\, L_m^3\, \Gamma^C_0 \,\Delta \beta_0 + \frac{1}{4}\,L_m^2\! \left(\Delta \Gamma^C_1 +
 \gamma^C_0 \,\Delta \beta_0\right)+\frac{1}{2}\, L_m\!\left(\Delta \gamma^C_1+
 2\,\mathcal{M}^{C,+}_{2} \right)+\mathcal{M}^C_{2}\right]\right\}\nn\\
 & \times {\rm exp}\,\Bigg\{ \frac{\big(\alpha_s^{(n_l+1)}\big)^2}{(4 \pi)^2} L_{-Q}\!
 \left[-\,\frac{1}{4}\, L_m^2\, \Gamma^C_0 \,\Delta \beta_0 -
 \frac{1}{2}\, L_m \,\Delta \Gamma^C_1 - \mathcal{M}^{C,+}_{2}\right] +
 \frac{\big(\alpha_s^{(n_l+1)}\big)^3}{(4 \pi)^3} L_{-Q}\! \left[\,\frac{1}{6}\, L_m^3 \Gamma^C_0\,
 (\beta_0+\Delta \beta_0)\,\Delta \beta_0 \right. \nn \\
  & +\frac{1}{4}\, L_m^2\left(-\, \Gamma^C_0 \,\Delta \beta_1 -2\, \Gamma^C_1 \,\Delta \beta_0 +
  2\, (\beta_0 +\Delta \beta_0)\, \Delta \Gamma^C_1 +4 \,\Delta \beta_0\, \Gamma^C_0\, \gamma^m_0 \right) +
  \frac{1}{2}\,L_m\,\Big(\!- \Delta \Gamma^C_2 + 4\,\beta_0\,\mathcal{M}^{C,+}_{2} +
  c_{\rm dec}\, \Gamma^C_0  \nn \\
 &+2\, \Delta \Gamma^C_1 \gamma^m_0 \Big)-\mathcal{M}_3^{C,+} \bigg ]\Bigg\}\,.
\end{align}
\end{widetext}
Here $\Delta \eta \equiv \eta^{(n_l+1)} -\eta^{(n_l)}$ is the difference between an evolution constant
$\eta$ in the $(n_l+1)$- and $n_l$-schemes. The terms $\Gamma^C_i$, $\gamma^C_i$, $\gamma^m_i$ and $\beta_i$
denote the coefficients of the cusp and noncusp current anomalous dimensions, the mass anomalous dimension
and the beta function with $n_l+1$ light quarks, respectively, 
\begin{align}
 \frac{\mu}{C}\frac{\df C}{\df\mu} = &  \sum_{i\geq 0} 
 \bigg(\frac{\alpha_s^{(n_l+1)}}{4\pi}\bigg)^{\!\!i+1} \!\left[\,-\,\Gamma^C_i L_{-Q} + \gamma^C_i \,\right] \, , \\
 \frac{\mu}{\overline{m}}\frac{\df \overline{m}}{\df \mu}= & \, - 2 \sum_{i\geq 0}
 \bigg(\frac{\alpha_s^{(n_l+1)}}{4\pi}\bigg)^{i+1} \gamma^m_i  \, , \label{eq:gamma_m} \\
  \frac{\mu}{\alpha_s}\frac{\df \alpha_s}{\df \mu} = & \,- 2 \sum_{i\geq 0} 
  \bigg(\frac{\alpha_s^{(n_l+1)}}{4\pi}\bigg)^{i+1} \beta_i \, . \label{eq:beta}
\end{align}
In this notation we have e.g.\ for the one-loop terms $\Gamma^C_0= -\,4 \, C_F$,
$\gamma^C_0= -\,6  \,C_F$, $\gamma^m_0=3 \, C_F$ and $\beta_0=\frac{11}{3} \, C_A - \frac{4}{3} \, T_F (n_l+1)$
with $\Delta \beta_0= - \frac{4}{3} \, T_F$.
The terms $\mathcal{M}^{C,+}_{i}$ ($\mathcal{M}^{C}_{i}$) indicate the renormalization scale independent constants,
which multiply (do not multiply) the rapidity logarithm $\ln(-\,m^2/Q^2)$in the matching coefficient
$\mathcal{M}_C(Q,m,m)$ and $c_{\rm dec}$
is the coefficient of the two-loop correction in the decoupling relation between the strong couplings in the $n_l$-
and $(n_l+1)$-flavor schemes at the scale of the mass, that is being employed, i.e.\
\begin{align}\label{eq:alpha_dec}
\!\!\alpha_s^{(n_l)}(m)=\alpha_s^{(n_l+1)}(m)
 \bigg[1+\bigg(\frac{\alpha_s^{(n_l+1)}(m)}{4\pi}\bigg)^2 c_{\rm dec}\bigg] .
\end{align}
For the $\MS$ mass $m=\overline{m}(\mu_m)$ we have $c_{\rm dec}=22/9$. The inclusion of the $\mu_m$-dependent terms
at $\mathcal{O}(\alpha_s^3\, \ln(m^2/Q^2))$ can play an important role for obtaining the correct remaining
$\mu_m$-dependence in numerical predictions at N$^3$LL order. Inserting the values for
all of the constants and expanding Eq.~(\ref{eq:matchingH3}) using the logarithmic counting
$\alpha_s \ln(m^2/Q^2) \sim 1$ gives our final result in Eq.~(\ref{eq:matchingII}).

\subsection{Jet mass mode matching coefficient}\label{sec:jetmassmodematching}
The mass mode threshold factor $\mathcal{M}_J(s,m,\mu_m)$ arises when the RG evolution of the jet function
crosses the massive quark threshold. The derivation goes along the lines of the current mass mode threshold
factor, and we will again omit all $\mathcal{O}(\alpha_s^2 C_{\!F}^2)$ and $\mathcal{O}(\alpha_s^2 C_{\!F} C_{\!A})$ terms.
The bare and the renormalized jet functions $J^{(0)}(s,m,\mu)$ and $J^{(n_{\!f})}(s,m,\mu)$ are related to each
other via
\begin{align}
 J^{(0)}(s,m,\mu) =  \int\! \df s' \,Z_J^{(n_{\!f})}(s-s',m,\mu)\, J^{(n_{\!f})}(s',m,\mu) \, ,
\end{align}
where $Z_J^{(n_{\!f})}(s,m,\mu)$ is the counterterm in a $n_{\!f}$-flavor scheme. For scales $\mu> m$ we use the
$(n_l+1)$-flavor scheme, so we employ the $\MS$-subtractions for the UV divergent contributions to the strong
coupling and the jet function. The counterterm is mass independent and reads with the notation of
section~\ref{sec:massless} 
\begin{align}\label{eq:ZJ_MS}
 Z_J^{(n_l+1)}(s,\mu)= \delta(s)+  Z_J^{(n_l+1,1)}(s,\mu) +  Z_{J,n_l+1}^{(n_l+1,2)}(s,\mu) \, .
\end{align}
The $\mathcal{O}(\alpha_s^2 C_{\!F} T_F)$ contribution $Z_{J,n_l+1}^{(n_l+1,2)}(s,\mu)$ is given in
Eq.~(\ref{eq:ZJ}) with $n_{\!f} = n_l+1$, whereas the $\mathcal{O}(\alpha_s)$ term reads  ($\bar{s}=s/\mu^2$)
\begin{align}\label{eq:ZJ1}
 & \mu^2 Z_J^{(n_l+1,1)}(s,\mu)= \frac{\alpha_s^{(n_l+1)}(\mu) C_{\!F}}{4\pi}
 \left\{\left(\frac{8}{\epsilon^2}+\frac{6}{\epsilon}\right)\!\delta(\bar{s})\right.\nn \\
 & \left.-\frac{8}{\epsilon}\!\left[\frac{\theta(\bar{s})}{\bar{s}}\right]_+\right\}\,.
\end{align}
The renormalized jet function reads
\begin{align}
 J^{(n_l+1)}(s,m,\mu)& = \, \delta(s) + J^{(n_l+1,1)}(s,\mu) + J^{(n_l+1,2)}(s,\mu)\nn \\
 & \, \, \, + \delta J_m^{\rm dist}(s,m,\mu)+ \delta J_m^{\rm real}(s,m) \, ,
\end{align}
which is the one for the case $m^2<s\sim\mu_J$ (scenarios III and IV) given in Eq.~(\ref{eq:jetmassive}).
The massless result at $\mathcal{O}(\alpha_s)$ reads
\begin{align}\label{eq:J1}
 & \mu^2 J^{(n_l+1,1)}(s,\mu) = \frac{\alpha_s^{(n_l+1)}(\mu) C_{\!F}}{4\pi} \bigg\{\!\!
 \left(14-2\pi^2\right)\!\delta(\bar{s})\nn \\
 & - 6\!\left[\frac{\theta(\bar{s})}{\bar{s}}\right]_{+} +8\!\left[\frac{\theta(\bar{s})
 \ln \,{\bar{s}}}{\bar{s}}\right]_{+} \bigg\}\, .
\end{align}

In the $n_l$-flavor scheme, we intend to implement the renormalization condition that the massive quark
corrections vanish for $s \sim \mu^2\ll m^2$. Following the computation described in
Sec.~\ref{sec:jetfunction}, we now do {\it not } include the scheme change contribution
$\delta J^{{\rm OS} \rightarrow \MS}_m$, which implies that we use $\alpha_s^{(n_l)}$.
The resulting expressions for the counterterm and the renormalized jet function read
\begin{align}
 Z_J^{(n_l)}(s,m,\mu)=& \, \delta(s)+  Z_J^{(n_l,1)}(s,\mu) +  Z_{J,n_l+1}^{(n_l+1,2)}(s,\mu) \nn \\
 & + Z^{(2,\rm OS)}_{J}(s,m,\mu) \,,  \label{eq:ZJ_OS}
\end{align}
and
 \begin{align}
 J^{(n_l)}(s,m,\mu)= & \, \delta(s) + J^{(n_l,1)}(s,\mu) + J^{(n_l,2)}(s,\mu) \nn \\
 & + \delta J^{(\rm OS,\rm virt)}_{m}(s,m,\mu) + \delta J_m^{\rm real}(s,m) \nn \\
 & - Z^{(2,\rm OS)}_{J}(s,m,\mu)\, , \label{eq:J_OS}
\end{align}
where the one-loop terms $Z_J^{(n_l,1)}(s,\mu)$ and $J^{(n_l,1)}(s,\mu)$ are analogous to
Eqs.~(\ref{eq:ZJ1}) and~(\ref{eq:J1}), respectively. The two-loop massless contributions
$Z_{J,n_l}^{(n_l,2)}(s,\mu)$, $J^{(n_l,2)}_{n_l}(s,\mu)$ are given in
Eqs.~(\ref{eq:ZJ}),~(\ref{eq:J0}) with $n_{\!f} = n_l$, and the two-loop massive quark contributions
$\delta J^{(\rm OS,\rm virt)}_{m}(s,m,\mu)$ and $\delta J_m^{\rm real}(s,m)$ are given in
Eqs.~(\ref{eq:jet_OS}) and~(\ref{eq:J_real}), respectively, with the corresponding counterterm
contribution denoted by $Z^{(2,\rm OS)}_{J}(s,m,\mu)$. The condition of decoupling requires that
the RHS of Eq.~(\ref{eq:J_OS}) vanishes for $m\rightarrow \infty$, and we obtain
\begin{align}\label{eq:ZJ2_OS}
 Z^{(2,\rm OS)}_{J}(s,m,\mu) = \delta J^{(\rm OS,\rm virt)}_{m}(s,m,\mu) \, .
\end{align}
Note that the real radiation term $\delta J_m^{\rm real}(s,m)$ automatically decouples for
$4m^2 > s$, so that it does not lead to any contributions in the counterterm. The renormalized
jet function in this scheme is the one to be used for $m^2\gtrsim s$ (in scenarios I and II).
We note that $\delta J_m^{\rm real}(s,m)$ is part of the result and can contribute when kinematically allowed.

The matching procedure is accounting for the difference between the two schemes, thus the mass mode
matching coefficient is obtained by the relation
\begin{align}\label{eq:MJ_cont}
 & \mathcal{M}_J(s,m,\mu_m) \nn \\
 & = \int \!\df s' J^{(n_l)}(s-s',m,\mu_m)\! \left(J^{(n_l+1)}(s',m,\mu_m)\right)^{-1} \nn \\
 & = \int \!\df s' Z_J^{(n_l+1)}(s-s',\mu_m)\!  \left(Z_J^{(n_l)}(s',m,\mu_m)\right)^{-1} \, .
\end{align}
Since the difference in the factorization theorems for the scenarios II and III in
Eqs.~(\ref{eq:diffsigmaII}) and~(\ref{eq:diffsigmaIII}) concerns just the jet function and its evolution,
Eq.~(\ref{eq:MJ_cont}) shows that the matching condition for the jet function automatically implements a
continuous transition between these two scenarios at $m^2 \sim \mu_m^2 \sim \mu_J^2 \sim s$, since the real
radiation term $\delta J_m^{\rm real}(s,m)$ is fully included in both scenarios.

Relating the schemes of $\alpha_s$ via Eq.~(\ref{eq:alphas_dec}), we obtain at
$\mathcal{O}(\alpha_s^2)$ in fixed-order counting $\alpha_s^{(n_l+1)} = \alpha_s^{(n_l+1)}(\mu_m)$
\begin{align}\label{eq:MJ_structure}
 & \mathcal{M}_J(s,m,\mu_m)=  \delta(s) + \frac{\alpha_s^{(n_l+1)}T_F}{3\pi}\, L_m\, J^{(n_l+1,1)}(s,\mu_m) \nn \\
  & - J^{(n_l+1,2)}_{n_{\!f}=1}(s,\mu_m) - \delta J_m^{\rm dist}(s,m,\mu_m) +\mathcal{O}(\alpha_s^3) \, .
\end{align}
Note that using the renormalized jet functions for the matching calculation the real radiation terms cancel in
Eq.~(\ref{eq:MJ_cont}) and do not contribute to the threshold correction factor. Inserting all explicit
expressions gives at $\mathcal{O}(\alpha_s^2)$ in the fixed-order counting ($\bar{s}=s/\mu_m^2$)
\begin{align}
& \mu_m^2 \mathcal{M}^{(2)}_{J}(s,m,\mu_m)=\frac{\alpha_s^2 C_{\!F} T_F}{16\pi^2} 
\left\{\left[-\frac{16}{9}\,L_m^3-\frac{116}{9}\,L_m^2 \right.\right. \nn \\
& -\left.\left(\frac{932}{27}+\frac{8\pi^2}{9}\right)\!L_m -\frac{1531}{27} 
-\frac{20\pi^2}{27}+\frac{160}{9}\,\zeta_3\right]\!\delta(\bar{s})  \nn \\
&+\left.\left(\frac{16}{3}\,L_m^2+\frac{160}{9\,}L_m+\frac{448}{27}\right)\!\!
\left[\frac{\theta(\bar{s})}{\bar{s}}\right]_+\right\} \, .
\label{eq:matchingIIIb}
\end{align}
Since there are no $\mathcal{O}(\alpha_s)$ one-loop corrections the schemes of $\alpha_s$ and the mass
appearing in Eq.~(\ref{eq:matchingIIIb}) do not need to be specified at this point. Eq.~(\ref{eq:matchingIIIb})
contains a large logarithm $\ln(m^2/s)$, which can be better seen by using the invariant mass variable 
$\tilde{s}=s/\mu_J^2\sim \mathcal{O}(1)$ rather than $\bar{s}=s/\mu_m^2$. As for the current mass mode matching 
coefficient this is a rapidity logarithm which enforces the counting $\alpha_s\, \ln(m^2/s) \sim \mathcal{O}(1)$.
The logarithm is known to exponentiate which allows us to determine the terms of
$\mathcal{O}(\alpha_s^4\, \ln^2(m^2/s))\sim\mathcal{O}(\alpha_s^2)$. For a complete analysis at N$^3$LL
we also need the term at $\mathcal{O}(\alpha_s^3\, \ln(m^2/s))\sim\mathcal{O}(\alpha_s^2)$. We can determine its 
\mbox{$\mu_m$-dependent} contribution from the identity
\begin{align}\label{eq:U_J}
 & \mathcal{M}_J(s,m,\mu_m)= \int \!\df s'\! \int \!\df s'' \, U_J^{(n_l+1)}(s-s',m,\mu_m) \nn \\
 & \times \mathcal{M}_J(s'-s'',m,m)\, U_J^{(n_l)}(s'',\mu_m,m) \, .
\end{align}
or equivalently,
\begin{align}
 & \mu\frac{\df}{\df\mu} \mathcal{M}_J(s,m,\mu) = \int \!\df s' \, \mathcal{M}_J(s',m,\mu)\nn \\
 &  \times \left(\gamma_J^{(n_l)}(s-s',\mu) - \gamma_J^{(n_l+1)}(s-s',\mu)\right) \, .
\end{align}
Expanding consistently in $\alpha_s$ gives the perturbative result for the $\mu_m$-dependent terms.
Including the relevant term at $\mathcal{O}(\alpha_s^3\, \ln(s/m^2))$ in the exponent the structure of the mass
mode matching coefficient reads
($\alpha_s^{(n_l+1)}=\alpha_s^{(n_l+1)}(\mu_m), m = {\overline m}(\mu_m), \tilde{s}=s/\mu_J^2$) 
\begin{widetext}
\begin{align}\label{eq:matchingJ3}
 & \mu_J^2\,\mathcal{M}_J(s,m,\mu_m,\mu_J) =\!  \left\{\!\delta(\tilde{s}) +
 \frac{\big(\alpha_s^{(n_l+1)}\big)^2}{(4 \pi)^2}\,\delta(\tilde{s})\!\left[\frac{1}{12}\, L_m^3\, \Gamma^J_0
\, \Delta \beta_0 + \frac{1}{4}\,L_m^2\! \left(\Delta \Gamma^J_1 + \gamma^J_0\, \Delta \beta_0\right)\!+
 \frac{1}{2}\,L_m\!\left(\Delta \gamma^J_1 - 2\,\mathcal{M}^{J,+}_{2}\right)\!+\mathcal{M}^{J}_{2}\right]\right.\nn\\
& \, + \left. \frac{\big(\alpha_s^{(n_l+1)}\big)^2}{(4 \pi)^2}
\left[-\,\frac{1}{4}\,L_m^2\, \Gamma^J_0\, \Delta \beta_0 -
\frac{1}{2}\,L_m\, \Delta\Gamma^J_1 + \mathcal{M}^{J,+}_{2}\right]\!\!\left[\frac{\theta(\tilde{s})}{\tilde{s}}\right]_{+} \right\} \nn \\
 & \times {\rm exp}\left\{ \frac{\big(\alpha_s^{(n_l+1)}\big)^2}{(4 \pi)^2}\, \ln\bigg(\frac{\mu_J^2}{\mu_m^2}\bigg)\! \left[ 
 -\,\frac{1}{4}\, L_m^2\, \Gamma^J_0\, \Delta \beta_0 - \frac{1}{2}\,L_m \,\Delta \Gamma^J_1 +\mathcal{M}^{J,+}_{2}\right] 
 +\,\frac{\big(\alpha_s^{(n_l+1)}\big)^3}{(4 \pi)^3}\,\ln\bigg(\frac{\mu_J^2}{\mu_m^2}\bigg) \right. \nn \\
 & \hspace{1cm}\, \times \left[\frac{1}{6}\, L_m^3\, \Gamma^J_0 (\beta_0 + \Delta \beta_0)\,\Delta 
 \beta_0+\frac{1}{4}\,L_m^2\left(-\, \Gamma^J_0\, \Delta \beta_1 - 2\,\Gamma^J_1\, \Delta \beta_0 + 2\,(\beta_0 +\Delta \beta_0) 
 \,\Delta \Gamma^J_1 + 4\, \Delta \beta_0\, \Gamma^J_0\, \gamma^m_0 \right) \nn \right. \\
  & \hspace{1.5 cm} + \left.\left. \frac{1}{2}\, L_m\!\left(\!-\,\Delta \Gamma^J_2 
  - 4 \,\beta_0\,\mathcal{M}^{J,+}_{2} +  c_{\rm dec}\, \Gamma^J_0 +2\, \Delta \Gamma^J_1 \,
  \gamma^m_0\right)+\mathcal{M}^{J,+}_{3}\right]\right\} \, ,
\end{align}
The terms $\Gamma^J_i$ and $\gamma^J_i$ denote the coefficients
of the cusp and non-cusp jet function anomalous dimensions with $n_l+1$ flavors defined by
%
\begin{align}\label{eq:jet-an-dim}
&\mu\,\frac{\df}{\df\mu}\, J(s) = \sum_{i\geq 0} 
\bigg(\frac{\alpha_s^{(n_l+1)}}{4\pi}\bigg)^{\!\!i+1} \!\!\!\int\! \df s' 
\left[\,-\,\frac{\Gamma^J_i}{\mu^2} \left[\frac{\mu^2 \theta(s-s')}{s-s'}\right]_+  +
\gamma^J_i \delta(s-s') \right] J(s')  \, ,
\end{align}
\end{widetext}
i.e.\ with the one-loop terms $\Gamma^J_0= 16 \, C_F $ and $\gamma^J_0= 12 \, C_F$.
The terms $\gamma^m_i$ and $\beta_i$ denote the mass anomalous dimension and the beta function, respectively,
as defined in Eqs.~(\ref{eq:gamma_m}) and~(\ref{eq:beta}). Note that we have defined $J(s,\mu)$ as the thrust jet
function, so the terms in the anomalous dimension on the RHS of Eq.~(\ref{eq:jet-an-dim}) are twice the ones known
for the jet function of a single jet.
The terms $\mathcal{M}^{J,+}_{i}$ and $\mathcal{M}^{J}_{i}$
indicate the $\mu_m$-independent coefficients of the plus-distribution $1/m^2\,[\,m^2\theta(s)/s\,]_+$ and delta-distribution $\delta(s)$ in the matching coefficient $\mathcal{M}_J(s,m,m)$ (i.e. for $\mu_m=m$), respectively, and $c_{\rm dec}$ is
the mass scheme dependent two-loop decoupling constant for $\alpha_s$, see Eq.~(\ref{eq:alpha_dec}). Inserting the
values for all of the constants and expanding Eq.~(\ref{eq:matchingJ3}) using the logarithmic counting
$\alpha_s \ln(m^2/s) \sim \mathcal{O}(1)$ gives our final result in Eq.~(\ref{eq:matchingIII}).

\subsection{Soft mass mode matching}\label{sec:soft_matching}
The mass mode threshold factor $\mathcal{M}_S(\ell,m,\mu_m)$ arises when the RG evolution of the soft function
crosses the massive quark threshold. This does not happen in the RG setup we discussed in 
section~\ref{sec:massmode_setup}, since there the final renormalization scale has always been set to the soft
scale. However, if we choose a different final renormalization scale e.g.\ the jet scale $\mu_J$, we can get a 
factorization theorem depending on $\mathcal{M}_S(\ell,m,\mu_m)$. This happens e.g.\ in scenario III
($\mu_J>\mu_m>\mu_S$):
\begin{align} \label{eq:diffsigmaIIIB}
 & \frac{1}{\sigma_0}\frac{\df\sigma}{\df\tau}= Q\, \big|C^{(n_l+1)}(Q,m,\mu_H)\big|^2\,
 \big|U^{(n_l+1)}_{C}(Q,\mu_H,\mu_J)\big|^2 \nn \\
 & \times \int \!\df s  \int \!\df\ell\! \int\! \df\ell'\! \int \!\df\ell''\, J^{(n_l+1)}(s,m,\mu_J) \nn\\
 & \times U^{(n_l+1)}_S\Big(\ell-\frac{s}{Q},\mu_J,\mu_m\Big) \mathcal{M}_S(\ell'-\ell,m,\mu_m) \nn \\
 & \times U^{(n_l)}_S(\ell''-\ell',\mu_m,\mu_S) \,S^{(n_l)}(Q\,\tau-\ell'',\mu_S)\,.
\end{align}
The derivation of $\mathcal{M}_S(\ell,m,\mu_m)$ proceeds along the lines of the current and jet mass mode
threshold factor and we will again omit all terms at $\mathcal{O}(\alpha_s^2 C_{\!F}^2)$ and
$\mathcal{O}(\alpha_s^2 C_{\!F} C_{\!A})$. The bare and the renormalized soft functions $S^{(0)}(\ell,m,\mu)$ and 
$S^{(n_{\!f})}(\ell,m,\mu)$ are related to each other via
\begin{align}
 S^{(0)}(\ell,m,\mu) \,=\!  \int\! \df\ell'\, Z_S^{(n_{\!f})}(\ell-\ell',m,\mu)\, S^{(n_{\!f})}(\ell',m,\mu) \, ,
\end{align}
where $Z_S^{(n_{\!f})}(\ell,m,\mu)$ is the counterterm in a $n_{\!f}$-flavor scheme. For scales $\mu> m$ we use the 
$(n_l+1)$-flavor scheme, so we employ the $\MS$ subtractions for the UV divergent contributions to the strong
coupling and the soft function. The counterterm is mass independent and reads with the notation of 
section~\ref{sec:massless} 
\begin{align}\label{eq:ZS_MS}
 Z_S^{(n_l+1)}(\ell,\mu)= \delta(\ell)+  Z_S^{(n_l+1,1)}(\ell,\mu) +  Z_{S,n_l+1}^{(n_l+1,2)}(\ell,\mu) \, .
\end{align}
The $\mathcal{O}(\alpha_s^2 C_{\!F} T_F)$ contribution $Z_{S,n_l+1}^{(n_l+1,2)}(\ell,\mu)$ is given in
Eq.~(\ref{eq:ZS}) with $n_{\!f} = n_l+1$, whereas the $\mathcal{O}(\alpha_s)$ term reads~($\bar{\ell}=\ell/\mu$)
\begin{align}\label{eq:ZS1}
 &\mu \, Z_S^{(n_l+1,1)}(\ell,\mu) \nn \\
 &=\, \frac{\alpha_s^{(n_l+1)}(\mu) C_{\!F}}{4\pi} \left\{\!-\,\frac{4}{\epsilon^2}\,\delta(\bar{\ell})+
 \frac{8}{\epsilon}\left[\frac{\theta(\bar{\ell})}{\bar{\ell}}\right]_+\right\} \, .
\end{align}
The renormalized soft function reads
\begin{align}
 \hat{S}^{(n_l+1)}(\ell,m,\mu)= & \, \delta(\ell) + \hat{S}^{(n_l+1,1)}(\ell,\mu) + \hat{S}^{(n_l+1,2)}(\ell,\mu) \nn \\
 &+ \delta S_m^{\rm dist}(\ell,m,\mu)+ \delta S_m^{\rm real,\theta}(\ell,m) \nn \\
 & + \delta S_m^{\rm real,\Delta}(\ell,m) \, ,
\end{align}
and is the one for the case $m<\ell\sim \mu_S$ (scenario IV) given in Eq.~(\ref{eq:softmassive}).
The massless result at $\mathcal{O}(\alpha_s)$ reads
\begin{align}\label{eq:S1}
 &\mu \, \hat{S}^{(n_l+1,1)}(\ell,\mu) \nn \\
 &= \frac{\alpha_s^{(n_l+1)}(\mu) C_{\!F}}{4\pi} 
 \left\{\frac{\pi^2}{3}\,\delta(\bar{\ell})-16\!\left[\frac{\theta(\bar{\ell})
 \ln \, \bar{\ell}}{\bar{\ell}}\right]_{+}\right\}\, .
\end{align}

In the $n_l$-flavor scheme we intend to implement the renormalization condition that the massive quark
corrections vanish for $\ell \sim \mu\ll m$. Analogously to the computation of the hard current coefficient
and the jet function, we now do {\it not } include a corresponding scheme change contribution
$\delta S^{{\rm OS} \rightarrow \MS}_m$ (see Eq.~(14) in Ref.~\cite{Gritschacher:2013tza} for
the explicit expression), which implies that we use $\alpha_s^{(n_l)}$. The resulting expressions
for the counterterm and the renormalized soft function read
\begin{align}
 Z_S^{(n_l)}(\ell,m,\mu)=& \, \delta(\ell)+  Z_S^{(n_l,1)}(\ell,\mu) +  Z_{S,n_l}^{(n_l,2)}(\ell,\mu) \nn \\
 & + Z^{(2,\rm OS)}_{S}(\ell,m,\mu) \,,  \label{eq:ZS_OS}
 \end{align}
and
\begin{align}
\!\!\!\hat{S}^{(n_l)}(\ell,m,\mu)\,= & \, \delta(\ell) + \hat{S}^{(n_l,1)}(\ell,\mu) + \hat{S}^{(n_l,2)}(\ell,\mu) \nn \\
&\!\!\!\! + \delta S^{(\rm OS,\rm virt)}_{m}(\ell,m,\mu) + \delta S_m^{\rm real,\theta}(\ell,m) \nn \\
&\!\!\!\! + \delta S_m^{\rm real,\Delta}(\ell,m) - Z^{(2,\rm OS)}_{S}(\ell,m,\mu)\, , \label{eq:S_OS}
\end{align}
where the one-loop terms $Z_S^{(n_l,1)}(\ell,\mu)$ and $\hat{S}^{(n_l,1)}(\ell,\mu)$ are analogous to
Eqs.~(\ref{eq:ZS1}) and~(\ref{eq:S1}), respectively. The two-loop massless contributions
$Z_{S,n_l}^{(n_l,2)}(\ell,\mu)$, $\hat{S}^{(n_l,2)}_{n_l}(\ell,\mu)$ are given in
Eqs.~(\ref{eq:ZS}) and (\ref{eq:S0}) with $n_{\!f} = n_l$, and the two-loop massive quark contributions
$\delta S_m^{\rm real,\theta}(\ell,m)$, $\delta S_m^{\rm real,\Delta}(\ell,m)$ and
$\delta S^{(\rm OS,\rm virt)}_{m}(\ell,m,\mu)$ are given in
Eqs.~(\ref{eq:S_real}) and (\ref{eq:DeltaS_parametrization}), and by \cite{Gritschacher:2013tza}
\begin{align}\label{eq:soft_OS}
 & \mu \, \delta S^{(\rm OS, virt)}_{m}(\ell,m,\mu)= \frac{\alpha_s^2 C_{\!F} T_F}{16\pi^2} 
 \left\{\!\delta(\bar{\ell})\!\left[-\,\frac{4}{\epsilon^3} \right. \right. \nn \\
 & +\frac{1}{\epsilon^2}\left(\frac{16}{3}\,L_m+\frac{20}{9}\right)+\frac{1}{\epsilon}
 \left(\!-\,\frac{8}{3}\,L_m^2 +\frac{112}{27}-\frac{2\pi^2}{3}\right)  \nn \\
 & -\left. \frac{40}{9}\,L_m^2+\left(-\frac{448}{27}+\frac{8\pi^2}{9}\right)\!L_m-
 \frac{656}{27}+\frac{10\pi^2}{27}+8\,\zeta_3 \right] \nn \\
 & +\left[\frac{16}{3\epsilon^2}+\frac{1}{\epsilon}\left(-\frac{32}{3}L_m-\frac{80}{9}\right)+
 \frac{32}{3}L_m^2  \right.\nn \\ 
 &  \left.+\left. \frac{160}{9}\,L_m + \frac{448}{27}+
 \frac{8\pi^2}{9}\right]\!\!\left[\frac{\theta(\ell)}{\ell}\right]_{+}\!\right\} , 
\end{align}
respectively. The corresponding counterterm contribution is denoted by $Z^{(2,\rm OS)}_{S}(\ell,m,\mu)$.
The condition of decoupling requires that the RHS of Eq.~(\ref{eq:S_OS}) vanishes for $m\rightarrow \infty$
and we obtain
\begin{align}\label{eq:ZS2_OS}
 Z^{(2,\rm OS)}_{S}(\ell,m,\mu) = \delta S^{(\rm OS,\rm virt)}_{m}(\ell,m,\mu) \, .
\end{align}
Note that the real radiation term $\delta S_m^{\rm real,\theta}(\ell,m)$ automatically decouples for
$2m > \ell$ and $\delta S_m^{\rm real,\Delta}(s,m)$ vanishes for $m/\ell \rightarrow \infty$, so that these
terms do not lead to any contributions in the counterterm. The renormalized soft function in this scheme is
thus the one to be used for $m\gtrsim\ell$ (in scenarios I, II and III). We note that the real radiation terms
$\delta S_m^{\rm real,\theta}(\ell,m)$ and $\delta S_m^{\rm real,\Delta}(\ell,m)$ are part of the renormalized
soft function in both schemes and contribute when kinematically allowed.

The matching procedure has to take care of the difference between the two schemes, thus the mass mode matching
coefficient at the mass scale is obtained by the relation 
\begin{align}\label{eq:MS_cont}
 &\mathcal{M}_S(\ell,m,\mu_m)  \nn \\
 &= \int \!\df\ell'\, \hat{S}^{(n_l+1)}(\ell-\ell',m,\mu_m)\! \left(\hat{S}^{(n_l)}(\ell',m,\mu_m)\right)^{-1} \nn \\
 & = \int \!\df\ell'\, Z_S^{(n_l)}(\ell-\ell',m,\mu_m)\!  \left(Z_S^{(n_l+1)}(\ell',\mu_m)\right)^{-1} \, .
\end{align}
In the RG setup, where the final renormalization scale is the jet scale $\mu_J$, the difference in the
factorization theorems for scenario III, given in Eq.~(\ref{eq:diffsigmaIIIB}), and for scenario IV, given by

\begin{align}
 & \frac{1}{\sigma_0}\frac{\df\sigma}{\df\tau}= \, Q\, \big|C^{(n_l+1)}(Q,m,\mu_H)\big|^2 \,
 \big|U^{(n_l+1)}_{C}(Q,\mu_H,\mu_J)\big|^2 \nn \\
 & \times\! \int \!\df s\!  \int\! \df\ell \, J^{(n_l+1)}(s,m,\mu_J)\, U^{(n_l+1)}_S\Big(\ell-\frac{s}{Q},\mu_J,\mu_S\Big) \nn\\
 &\times S^{(n_l+1)}\left(Q\,\tau-\ell,m,\mu_S\right)\,,
\label{eq:diffsigmaIVB}
\end{align}
concerns just the soft function and its evolution. Thus Eq.~(\ref{eq:MS_cont}) shows that the condition for the
soft function automatically implements a continuous transition between these two scenarios in the region
$m \sim \mu_m \sim \mu_S \sim \ell$, when the real radiation terms $\delta S_m^{\rm real,\theta}(\ell,m)$
and $\delta S_m^{\rm real,\Delta}(\ell,m)$ are included in both scenarios. From the viewpoint of the factorization 
theorems in the top-down evolution to the soft scale given in Eqs.~(\ref{eq:diffsigmaIII}) and~(\ref{eq:diffsigmaIV}), 
respectively, the continuity seems less obvious. However, since the final renormalization scale is unphysical, these two 
RG setups are related to each other via consistency relations, which we discuss in Sec.~\ref{sec:consistency}.

Relating the schemes of $\alpha_s$ via Eq.~(\ref{eq:alphas_dec}), we obtain at $\mathcal{O}(\alpha_s^2)$ in the 
fixed-order counting 
\begin{align}\label{eq:MS_structure}
 & \mathcal{M}_S(\ell,m,\mu_m)=  \delta(\ell) - \frac{\alpha_s^{(n_l+1)} T_F}{3\pi}\,
 L_m\, \hat{S}^{(n_l+1,1)}(\ell,\mu_m) \, \nn \\
 & + \hat{S}^{(n_l+1,2)}_{n_{\!f}=1}(\ell,\mu_m) + \delta S_m^{\rm dist}(\ell,m,\mu_m) +\mathcal{O}(\alpha_s^3) \, .
\end{align}
Note that the real radiation terms cancel in the ratio in Eq.~(\ref{eq:MS_cont}) and do not contribute to the
threshold correction factor. Inserting all explicit expressions, this gives at $\mathcal{O}(\alpha_s^2)$ in the
fixed-order counting ($\bar{\ell}=\ell/\mu_m$)
\begin{align}\label{eq:matchingIVb}
& \mu_m\, \mathcal{M}^{(2)}_{S}(\ell,m,\mu_m)=\frac{\alpha_s^2 C_{\!F} T_F}{16\pi^2} 
\left\{\left[-\,\frac{8}{9}\,L_m^3-\frac{40}{9}\,L_m^2 \right. \right. \nn \\
& +\left. \left(-\frac{448}{27}+\frac{4\pi^2}{9}\right)\!L_m-\frac{656}{27}+\frac{10\pi^2}{27}+
\frac{56}{9}\,\zeta_3 \right]\!\delta(\bar{\ell})\nn\\
 & + \left.\left[\frac{16}{3}L_m^2+\frac{160}{9}\,L_m+\frac{448}{27}\right]\!\!
 \left[\frac{\theta(\bar{\ell})}{\bar{\ell}}\right]_{+}\right\} .
\end{align}
Since there are no $\mathcal{O}(\alpha_s)$ corrections the schemes of $\alpha_s$ and the mass appearing in
Eq.~(\ref{eq:matchingIVb}) do not need to be specified at this point. Eq.~(\ref{eq:matchingIVb}) contains a
large logarithm, which can be better seen using the rescaled soft energy variable
$\tilde{\ell}=\ell/\mu_S \sim \mathcal{O}(1)$ rather than $\bar{\ell}=\ell/\mu_m$. As for the current mass
mode matching coefficient this is a rapidity logarithm which enforces the counting
$\alpha_s\, \ln(m/\ell) \sim \mathcal{O}(1)$. This logarithm is known to exponentiate, which allows us to determine
the terms of $\mathcal{O}(\alpha_s^4\, \ln^2(m/\ell))\sim\mathcal{O}(\alpha_s^2)$. For a complete analysis at
N$^3$LL we also need the term at $\mathcal{O}(\alpha_s^3\, \ln(m/\ell))\sim\mathcal{O}(\alpha_s^2)$.
We can determine its $\mu_m$-dependent contribution from the identity 
\begin{align}\label{eq:U_S}
 & \mathcal{M}_S(\ell,m,\mu_m)= \!\int \!\df\ell' \!\int \!\df\ell'' \, U_S^{(n_l)}(\ell-\ell',m,\mu_m) \nn \\
 &\times \mathcal{M}_S(\ell'-\ell'',m,m)\, U_S^{(n_l+1)}(\ell'',\mu_m,m) \, ,
\end{align}
or equivalently,
\begin{align}
 & \mu\,\frac{\df}{\df\mu}\, \mathcal{M}_S(\ell,m,\mu) =  \int \!\df\ell'\, \mathcal{M}_S(\ell',m,\mu) \nn \\
 &\times \left(\gamma_S^{(n_l+1)}(\ell-\ell',\mu) - \gamma_S^{(n_l)}(\ell-\ell',\mu)\right) .
\end{align}
Expanding consistently in $\alpha_s$ gives the perturbative result for the $\mu_m$-dependent terms.
Including the relevant term at $\mathcal{O}(\alpha_s^3\, \ln(s/m^2))$ in the exponent the structure of the mass
mode matching coefficient reads ($\alpha_s^{(n_l+1)}=\alpha_s^{(n_l+1)}(\mu_m), m=\bar{m}(\mu_m), \tilde{\ell}=\ell/\mu_S$)
\vspace*{-1.2cm}
\begin{widetext}
\begin{align}\label{eq:matchingS3}
 & \mu_S\,\mathcal{M}_S(\ell,m,\mu_m,\mu_S)= \left\{\delta(\tilde{\ell}) +
 \frac{\big(\alpha_s^{(n_l+1)}\big)^2}{(4\pi)^2}\,\delta(\tilde{\ell})\!\left[-\,\frac{1}{12}\, L_m^3\,
 \Gamma^S_0\, \Delta \beta_0 - \frac{1}{4}\,L_m^2 \!\left(\Delta \Gamma^S_1 +
 \gamma^S_0\, \Delta \beta_0\right) \right.\right.\nn\\
& \, \left. -\left.\frac{1}{2}\,L_m\!\left(\Delta \gamma^S_1 + \mathcal{M}^{S,+}_{2}\right)\!+\mathcal{M}^{S}_{2}\right] + 
\frac{\big(\alpha_s^{(n_l+1)}\big)^2}{(4 \pi)^2}\left[\frac{\theta(\tilde{\ell})}{\tilde{\ell}}\right]_{+}
\left[\frac{1}{2}\,L_m^2\, \Gamma^S_0\, \Delta \beta_0+ L_m\, \Delta\Gamma^S_1+\mathcal{M}^{S,+}_{2}\!\right] \right\} \nn \\
 & \times {\rm exp}\left\{ \frac{\big(\alpha_s^{(n_l+1)}\big)^2}{(4 \pi)^2}\, \ln\bigg(\frac{\mu_S^2}{\mu_m^2}\bigg)\!
 \left[\frac{1}{4}\, L_m^2\, \Gamma^S_0\, \Delta \beta_0+\frac{1}{2}\,L_m\, \Delta \Gamma^S_1 
 +\frac{1}{2}\,\mathcal{M}^{S,+}_{2}\!\right] +\frac{\big(\alpha_s^{(n_l+1)}\big)^3}{(4 \pi)^3}\,
 \ln\bigg(\frac{\mu_S^2}{\mu_m^2}\bigg) \right. \nn \\
 & \hspace{1cm}\, \times \!\left[-\,\frac{1}{6}\, L_m^3\, \Gamma^S_0 (\beta_0 + \Delta \beta_0)\,\Delta \beta_0+
 \frac{1}{4}\,L_m^2\!\left(\Gamma^S_0\, \Delta \beta_1 + 2\, \Gamma^S_1 \,\Delta \beta_0 -
 2\,(\beta_0 +\Delta \beta_0)\, \Delta \Gamma^S_1 -4 \,\Delta \beta_0 \,\Gamma^S_0 \,\gamma^m_0 \right) \nn \right. \\
  & \hspace{1.5 cm} + \left.\left. \frac{1}{2}\, L_m\!\left(\Delta \Gamma^S_2 -
  2\,\beta_0\, \mathcal{M}^{S,+}_{2} -  c_{\rm dec}\, \Gamma^S_0 -
  2\, \Delta \Gamma^S_1 \,\gamma^m_0\right)\!+\frac{1}{2}\mathcal{M}^{S,+}_{3}\right]\right\} .
\end{align}
The terms $\Gamma^S_i$ and $\gamma^S_i$ denote the coefficients of the cusp
and noncusp soft function anomalous dimensions with $n_l+1$ flavors given by

\begin{align}
 &\mu\frac{\df}{\df\mu} S(\ell)=  \sum_{i\geq 0} 
 \bigg(\frac{\alpha_s^{(n_l+1)}}{4\pi}\bigg)^{\!\!i+1} \!\!\!\int \!\df\ell' \left[\,-\,\frac{2 \,\Gamma^S_i}{\mu} 
 \left[\frac{\mu \, \theta(\ell-\ell')}{\ell-\ell'}\right]_+  + \gamma^S_i\, \delta(\ell-\ell') \right]\!
 S(\ell') \, ,
\end{align}
i.e.\ with the one-loop terms $\Gamma^S_0= -\,8 \, C_F $ and $\gamma^S_0= 0$.
The terms $\gamma^m_i$ and $\beta_i$ denote the mass anomalous dimension and the beta function, respectively,
as defined in Eqs.~(\ref{eq:gamma_m}) and~(\ref{eq:beta}). The terms $\mathcal{M}^{S,+}_{i}$ and $\mathcal{M}^{S}_{i}$
indicate the $\mu_m$-independent coefficients of the plus-distribution $1/m\,[\,m \,\theta(\ell)/\ell\,]_+$ and
delta-distribution $\delta(\ell)$ in the matching coefficient $\mathcal{M}_S(\ell,m,m)$ (i.e. for $\mu_m=m$), respectively, and $c_{\rm dec}$
is the mass scheme dependent two-loop decoupling constant for $\alpha_s$, see Eq.~(\ref{eq:alpha_dec}). Inserting
the values for all of the constants and expanding Eq.~(\ref{eq:matchingS3}) using the logarithmic counting
$\alpha_s\, \ln(m^2/\ell^2) \sim \mathcal{O}(1)$ gives our final result,
\begin{align}
& \mu_S\, \mathcal{M}^{(2)}_{S}(\ell,m,\mu_S,\mu_m)= \delta(\tilde{\ell})+\!
\left[\frac{\big(\alpha_s^{(n_l+1)}\big)^2 C_{\!F} T_F}{(4 \pi)^2}\, \delta(\tilde{\ell}) 
\,\ln\bigg(\frac{\mu_S^2}{\mu_m^2}\bigg)\!\left\{\frac{8}{3}\,L_m^2+
\frac{80}{9}\,L_m+\frac{224}{27}\right\}\right]_{\mathcal{O}(\alpha_s)} \nn \\
& + \left[ \frac{\big(\alpha_s^{(n_l+1)}\big)^2 C_{\!F} T_F}{(4 \pi)^2}\left\{\delta(\tilde{\ell})
\left[-\,\frac{8}{9}\,L_m^3-\frac{40}{9}\,L_m^2+\left(\!-\,\frac{448}{27}+\frac{4\pi^2}{9}\right)\!L_m 
-\frac{656}{27}+\frac{10\pi^2}{27}+\frac{56}{9}\,\zeta_3\right] \right.\right. \nn \\
& + \left. \left[\frac{16}{3}L_m^2+
\frac{160}{9}L_m+\frac{448}{27}\right]\!\bigg[\frac{\theta(\tilde{\ell})}{\tilde{\ell}}\bigg]_{+}\right\}
  + \frac{\big(\alpha_s^{(n_l+1)}\big)^3 C_{\!F} T_F}{(4 \pi)^3} \,
\delta(\tilde{s}) \,\ln\bigg(\frac{\mu_S^2}{\mu_m^2}\bigg)\! 
\left\{L_m^3\left[-\,\frac{176}{27}\,C_{\!A}+\frac{128}{27}\,T_F+\frac{64}{27}\,T_F\, n_l\right] \right. \nn \\
& +L_m^2\!\left[\left(\frac{184}{9}-\frac{16}{9} \pi ^2\right)\! C_{\!A} -24\, C_{\!F} +\frac{320}{27}\,T_F \right]+ 
L_m\!\left[\left(\frac{1240}{81}-\frac{160\pi^2}{27}+\frac{224}{3}\, \zeta_3\right)\!C_{\!A} 
+\left(\frac{8}{3}-64\,\zeta_3\right)\!C_{\!F} \right.\nn \\
 & +\left.\left.\frac{2176}{81}\,T_F\, n_l +\frac{1984}{81}\,T_F\right] +
 \frac{\mathcal{M}^{S,+}_{3}}{2 C_{\!F} T_F} \right\}+\frac{\big(\alpha_s^{(n_l+1)}\big)^4 C_{\!F}^2 T_F^2}{(4 \pi)^4}\, \delta(\tilde{s})
\,\ln^2\bigg(\frac{\mu_S^2}{\mu_m^2}\bigg)\left\{\frac{32}{9}\,L_m^4+\frac{640}{27}\,L_m^3+\frac{1664}{27}\,L_m^2 \right.\nn \\
 &+\left.\left.\frac{17920}{243}\,L_m+\frac{25088}{729}\right\}\right]_{\mathcal{O}(\alpha_s^2)} \, .
\label{eq:matchingIV}
\end{align}
\end{widetext}

\subsection{Consistency relations}\label{sec:consistency}
As already discussed in the introduction of Sec.~\ref{sec:RGconsistency}, the mass mode threshold 
factors $\mathcal{M}_C$ for the hard current mass mode matching, $\mathcal{M}_J$ for the jet mass mode matching
and $\mathcal{M}_S$ for the soft mass mode matching are related by consistency of RG running in analogy to the
well-known relation between the evolution factors and anomalous dimensions shown in Eqs.~(\ref{eq:consistency_ML})
and~(\ref{eq:consistency_ML2}), respectively. This consistency relation can be easily read off 
Eqs.~(\ref{eq:diffsigmaIII}) and~(\ref{eq:diffsigmaIIIB}), which show the factorization theorems for
$Q > Q\lambda > m > Q \lambda^2$ (scenario III) with the final
renormalization scale $\mu$ set equal to the soft and the jet scale, respectively. It reads
\begin{align}\label{eq:consistency_M}
\mathcal{M}_S(\ell,m,\mu_S,\mu)= & \, Q \left|\mathcal{M}_C(Q,m,\mu_H,\mu)\right|^2  \nn \\
& \times \mathcal{M}_J(Q \ell,m,\mu_J,\mu) \, .
\end{align}
The relation implies in particular that the rapidity logarithms (and singularities) that arise in the hard,
collinear and soft sectors are intrinsically related to each other.
Relation~(\ref{eq:consistency_M}) holds identically at each finite order for $\mu_J = \sqrt{\mu_S \mu_H}$
using the counting explained in Eqs.~(\ref{eq:matchingH3}), (\ref{eq:matchingJ3}), and (\ref{eq:matchingS3}).
For $\mu_J \sim \sqrt{\mu_S \mu_H}$ the relation
holds up to terms at higher order. The consistency condition further
implies that the currently unknown constants $\mathcal{M}^{J,+}_3$, $\mathcal{M}^{S,+}_3$, and
$\mathcal{M}^{C,+}_3$, which are enhanced by a rapidity logarithm but not constrained by the $\mu$
anomalous dimension are related by
\begin{align}
 \mathcal{M}^{J,+}_{3} =\mathcal{M}^{S,+}_{3}= 4\, \mathcal{M}^{C,+}_{3}\, .
\end{align}
Using the structure of the matching coefficients given in Eqs.~(\ref{eq:MC_structure}),~(\ref{eq:MJ_structure}) 
and~(\ref{eq:MS_structure}), the consistency relation at $\mathcal{O}(\alpha_s^2)$ in the fixed-order
expansion has the explicit form
\begin{widetext}
\begin{align}
&2 \,\textnormal{Re}\!\left[F^{(n_l,2)}_{\rm QCD}(Q,m)-C^{(n_l+1,2)}_{n_{\!f}=1}(Q,\mu)-
\delta F^{(n_l+1,2)}(Q,m,\mu)\right]\!\delta(\tau) \nn \\
&-Q^2\! \left[J^{(n_l+1,2)}_{n_{\!f}=1}(Q^2\tau,\mu)+ \delta J^{\rm dist}_m(Q^2 \tau ,m,\mu)\right]-
Q\! \left[\hat{S}^{(n_l+1,2)}_{n_{\!f}=1}(Q\tau,\mu) + \delta S^{\rm dist}_m(Q \tau,m,\mu)\right]\nn \\
&+\frac{\alpha_s T_F}{3 \pi}\,L_m\left\{2\, \textnormal{Re}\!\left[C^{(n_l+1,1)}(Q,\mu)\right]\!\delta(\tau)+
Q^2 J^{(n_l+1,1)}(Q^2 \tau ,\mu)+Q\, \hat{S}^{(n_l+1,1)}(Q \tau, \mu)\right\}= 0\, . \label{eq:consistency_virt}
\end{align}
\end{widetext}
Here $C^{(n_l+1,2)}_{n_{\!f}=1}(Q,\mu)$, $J^{(n_l+1,2)}_{n_{\!f}=1}(s,\mu)$ and $\hat{S}^{(n_l+1,2)}_{n_{\!f}=1}(\ell,\mu)$
are the $\mathcal{O}(\alpha_s^2 C_{\!F} T_F)$ massless contributions to the hard current coefficient, jet and soft
function for one single flavor corresponding to the expressions in Eqs.~(\ref{eq:C0}),~(\ref{eq:J0})
and~(\ref{eq:S0}), respectively. The corrections due to the quark mass given by
$F^{(n_l,2)}_{\rm QCD}(Q,m)$, $\delta F^{(n_l+1,2)}(Q,m,\mu)$, $\delta J^{\rm dist}_m(s ,m,\mu)$ and
$\delta S^{\rm dist}_m(\ell ,m,\mu)$ can be found in Eqs.~(\ref{eq:F_QCD}),~(\ref{eq:F_II}),~(\ref{eq:J_virt})
and~(\ref{eq:S_virt}). These do not involve terms related to real radiation of heavy quarks and yield together
with the massless terms the virtual quark contributions to the SCET current, the jet function and the soft
function (in the $\MS$ scheme). Finally also the one-loop terms $C^{(n_l+1,1)}(Q,\mu)$, $J^{(n_l+1,1)}(s,\mu)$
and $\hat{S}^{(n_l+1,1)}(\ell, \mu)$ given in Eqs.~(\ref{eq:C1}),~(\ref{eq:J1})
and~(\ref{eq:S1}) appear due to the virtual heavy quark contributions to the strong coupling encoded in the
decoupling relation of $\alpha_s$ between the $n_l$- and ($n_l+1$)-flavor scheme.
Eq.~(\ref{eq:consistency_virt}) relates the virtual quark contributions to the hard current coefficient, the jet
function, the soft function and $\alpha_s$ to one another and is a consequence of the consistency of the mass mode
setup. Moreover, Eq.~(\ref{eq:consistency_virt}) is also the analytic relation behind the fact that the transition
between the factorization theorems in Eqs.~(\ref{eq:diffsigmaIII}) and~(\ref{eq:diffsigmaIV}) for scenarios III
and IV, respectively, is continuous.\,\footnote{Note that the gap parameter in the soft model function and the 
renormalon subtractions to the partonic soft function also change. However, they compensate each other for
$\mu_m\sim \mu_S$ due to the matching relation in Eq.~(\ref{eq:gap_decoupling}).}

We emphasize again that the form of the threshold factors and the validity of the consistency relation in
Eq.~(\ref{eq:consistency_M}) are not restricted to thrust, but arise in analogous form for other observables,
which exhibit factorization theorems with a similar structure, i.e.\ involving a hard current coefficient,
a jet function and a soft function as building blocks.

\subsection{Fixed-order QCD result}
The factorization theorems discussed in the previous sections each contain all information about the singular 
$\mathcal{O}(\alpha_s^2 C_{\!F} T_F)$ secondary massive quark corrections to the thrust distribution in the fixed-order
expansion in full QCD. As these fixed-order corrections have not been made available in literature in an explicit
form we give them in the following. Besides the virtual contributions the singular fixed-order corrections consist
of the singular collinear and soft real radiation contributions which arise for $\tau\sim m^2/Q^2\ll 1$ and
$\tau\sim m/Q\ll 1$, respectively, in the dijet regime.\footnote{Note that at the corresponding thresholds
$\tau = 4m^2/Q^2$ and $\tau = 2m/Q$, respectively, in the fixed-order expansion the nonsingular contributions
can be numerically larger than the singular terms $\delta J^{\rm{real}}_{m}$ and $\delta S^{{\rm real},\theta}_{m}$
since the latter vanish at the respective thresholds. This feature was already discussed in 
Ref.~\cite{Gritschacher:2013pha} for the case of massive gauge bosons at one loop.}
Setting $\mu=\mu_H=\mu_J=\mu_S$, using the $n_l$-flavor scheme for $\alpha_s$ and ignoring the gap
subtraction we obtain
\begin{align}
& \left.\frac{1}{\sigma_0} \frac{\df\sigma}{\df\tau}\right|_{\mathcal{O}(\alpha_s^2 C_{\!F} T_F)}= 2\,{\rm Re}\!
\left[F^{(n_l,2)}_{\rm QCD}(Q,m)\right]\!\delta(\tau) \nn \\
& +Q^2\, \delta J^{\rm{real}}_{m}(Q^2 \tau,m) + Q\,  \delta S^{{\rm real},\theta}_{m}(Q\tau,m) \nn \\
& +Q\, \delta S^{{\rm real},\Delta}_{m}(Q\tau,m) \, ,
\label{eq:diffsigmafo}
\end{align}
where $F^{(n_l,2)}_{\rm QCD}(Q,m)$, $\delta J^{\rm{real}}_{m}(s,m)$, $ \delta S^{{\rm real},\theta}_{m}(\ell,m)$
and $\delta S^{{\rm real},\Delta}_{m}(\ell,m)$ have been given in Eqs.~(\ref{eq:F_QCD}), (\ref{eq:J_real}),
(\ref{eq:S_real}) and (\ref{eq:DeltaS_parametrization}), respectively. Writing out Eq.~(\ref{eq:diffsigmafo})
explicitly for the convenience of the reader we get
\begin{widetext}
\begin{align}
& \left.\frac{1}{\sigma_0} \frac{\df\sigma}{\df\tau}\right|_{\mathcal{O}(\alpha_s^2 C_{\!F} T_F)} = 
\bigg(\!\frac{\alpha_s^{(n_l)}(\mu) }{4\pi}\!\bigg)^{\!\!2}\!C_{\!F} T_F \nn \\
& \times \left\{\delta(\tau)\!\left[\!\left(\!-\,r^4+2 r^2+\frac{5}{3}\right)\!\!
\left(\!4\,\Li_3\bigg(\frac{r-1}{r+1}\bigg)+\frac{1}{3}\,\ln^3\bigg(\frac{r-1}{r+1}\bigg)-
\frac{2\pi^2}{3}\,\ln\bigg(\frac{r-1}{r+1}\bigg) - 4\,\zeta_3\right)\right.\right.\nn\\
& \hspace{1.5cm} + \left.\left(\frac{46}{9}\,r^3+\frac{10}{3}\,r\right)\!\!
\left(\!4\,\Li_2\bigg(\frac{r-1}{r+1}\bigg)+
\ln^2\bigg(\frac{r-1}{r+1}\bigg)-\frac{2\pi^2}{3}\right)+\left(\frac{220}{9}+
\frac{400}{27}\, r^2\!\right)\!\ln\bigg(\frac{1-r^2}{4}\bigg)+ \frac{476}{9}\,r^2+\frac{2426}{81}\right] \nn \\
& + \frac{1}{\tau} \,\theta\!\left(\!\tau-\frac{4m^2}{Q^2}\right) \!\!\left[- \frac{64}{3}\,
\Li_2\bigg(\frac{b-1}{b+1}\bigg)+\frac{32}{3}\ln\bigg(\frac{1-b^2}{4}\bigg)\ln\bigg(\frac{1-b}{1+b}\bigg)
-\frac{16}{3}\,\ln^2\!\left(\frac{1-b}{1+b}\right) \right.\nn\\
& \hspace{3cm} +\left.\left(b^4-2b^2+\frac{241}{9}\right)\!\ln\bigg(\frac{1-b}{1+b}\bigg)-
\frac{10}{27}\,b^3+\frac{482}{9}\,b-\frac{16\pi^2}{9} \right] \nn \\
& + \frac{1}{\tau}\, \theta\!\left(\!\tau-\frac{2m}{Q}\right)\!\! \left[\frac{64}{3}\,\Li_2\bigg(\frac{w-1}{w+1}\bigg)-
\frac{32}{3}\,\ln\bigg(\frac{1-w^2}{4}\bigg)\ln\bigg(\frac{1-w}{1+w}\bigg) +\frac{16}{3}\,\ln^2
\bigg(\frac{1-w}{1+w}\bigg)-\frac{160}{9}\, \ln \bigg(\frac{1-w}{1+w}\bigg) \right. \nn \\
 & \hspace{3cm}+ \left.\left. \frac{64}{27}\,w^3-\frac{320}{9}\,w+\frac{16\pi^2}{9} \right]\right\} 
  + Q \,\delta S^{{\rm real},\Delta}_{m}(Q\tau,m) \, ,
\end{align}
\end{widetext}
with
\begin{align}
 r \equiv \sqrt{1+\frac{4m^2}{Q^2}}\, , \, b\equiv \sqrt{1-\frac{4m^2}{Q^2\tau}}\, , \,
 w\equiv \sqrt{1-\frac{4m^2}{Q^2\tau^2}} \, .
\end{align}

An important difference in the SCET setup to this fixed-order QCD expansion is that in the factorization theorems
the various components are calculated in different flavor number schemes to allow for the summation of logarithms 
involving ratios of the scales $Q$, $Q\lambda$, $Q\lambda^2$ and $m$. A maybe even more notable difference is that
the consistent and IR-safe definitions of the jet and the soft functions entail that virtual corrections have 
non-vanishing support for finite values of $\tau$, so that they do not only arise in coefficients of $\delta(\tau)$,
but also in coefficients involving plus-distributions $(\ln^n\tau/\tau)_+$. In contrast, the fixed-order expansion 
contains only real radiation corrections for finite values of $\tau$ and virtual corrections proportional to 
$\delta(\tau)$ each of which is individually IR-regular for $m\to 0$. The rearrangement of virtual and mass-singular 
corrections, which is intrinsically connected to the consistency relation of Eqs.~(\ref{eq:consistency_M}) 
and~(\ref{eq:consistency_virt}), is the basis of rendering the hard coefficient and the jet and the soft functions in the 
factorization theorems infrared-safe in the limit $m\rightarrow 0$. This may provide a guideline to understanding the 
factorization from the point of view of the fixed-order expansion.

\section{Numerical analysis}\label{sec:analysis}
In the following we investigate the numerical effects of secondary bottom and
top quarks in the thrust distribution related to the mass-dependent
factorization theorems we have described and presented in the
previous sections. The emphasis is on a comparison to the predictions where the mass of
the secondary quark is neglected. In Ref.~\cite{Gritschacher:2013pha} a
similar examination was carried out which, however, did not account for nonperturbative effects (described by 
the soft model function $F$), for the renormalon subtractions and the
associated gap formalism.\,\footnote{The soft model function and the gap
subtraction lead to significant changes in the thrust distribution and affect
the secondary quark mass effects as well as the contributions from the massless
quarks. Our partonic results are in complete agreement with the results shown in
Ref.~\cite{Gritschacher:2013pha}.} We note that our analysis does not include the
nonsingular contributions which might be sizeable in the tail and the far-tail
region, so some of the conclusions concerning the tail region, e.g.\ concerning
the scale variations, are preliminary and final conclusions are postponed to 
a complete phenomenological analysis which also includes the effects of primary
massive quark production.

The results in our analysis are calculated at N$^3$LL order in the usual SCET
counting, so we use the beta function and the cusp 
anomalous dimension up to four loops and the non-cusp anomalous dimensions
including R-evolution up to
three loops.\,\footnote{For $\Gamma^{\rm cusp}_3$ we use the Pad\'e approximation of
Ref.~\cite{Abbate:2010xh}. The remaining missing ingredients for a
complete N$^3$LL analysis are logarithmic enhanced coefficients ${\cal M}^{C,+}_3$ and
${\cal M}^{J,+}_3$ in the mass mode threshold factors at $\mathcal{O}(\alpha_s^3)$ (which we set to zero) and the
massive  R-anomalous dimension at $\mathcal{O}(\alpha_s^3)$ (for which we use
the massless approximation, see discussion in Sec.~\ref{sec:gap}).}
The perturbative corrections to the matrix elements and
matching conditions are included up to $\mathcal{O}(\alpha_s^2)$ and expanded
out within the factorized expressions to avoid higher order cross terms. 
There are many ingredients concerning the RG-evolution and matrix elements
involving massless quarks which we have not discussed in detail in this paper,
but which are used in our analysis. For these contributions we employed the
results given in Ref.~\cite{Abbate:2010xh} with top-down evolution.
We have further checked that the massless limit of the factorization theorem for
scenario~IV agrees with the N$^3$LL thrust distribution given in
Ref.~\cite{Abbate:2010xh} for massless quarks up to implementation-dependent
higher order corrections.

For the renormalization scales of the individual structures and the renormalon
subtraction scale we use the $\tau$-dependent profile functions for the hard,
jet, soft and $R$ scale given in Ref.~\cite{Abbate:2010xh}, which contain an
appropriate generic scaling and smoothly interpolate
between peak, tail and far-tail regimes. Adding an additional profile for the
$\tau$-independent mass mode matching scale the profile functions have the form
\begin{align}
\mu_H &=\,e_H\,Q \, , \label{eq:mu_H} \\[0.1cm]
\mu_S(\tau) &=\left\{\begin{array}{ll}\mu_0+\frac{b}{2 t_1} \tau^2 \, ,\hskip1.6cm
    & 0\leq \tau\leq t_1 \, ,\\[0.2cm]
  b \, \tau+d \, ,   & t_1\leq \tau\leq  t_2 \, , \\[0.1cm]
  \mu_H-\frac{b}{1-2 t_2}(\frac{1}{2}-\tau)^2 \, , &
    t_2\leq \tau\leq \frac{1}{2} \, ,\end{array} \right. \label{eq:mu_S} \\[0.1cm]
\mu_J(\tau) &=\bigg(1+
e_J\Big(\frac{1}{2}-\tau\Big)^2\bigg)\,\sqrt{\mu_H\,\mu_S(\tau)} \, , \label{eq:mu_J} \\[0.1cm]
R(\tau)&= \left\{\begin{array}{ll} R_0 + \mu_1 \tau + \mu_2 \tau^2 \, ,\hskip 0.8 cm
    & 0\leq \tau\leq t_1 \, ,\\[0.2cm]
 \mu_S(\tau) &  t_1 \leq \tau \leq \frac{1}{2} \, , \end{array} \right.  \label{eq:mu_R} \\[0.1cm]
 \mu_m & =\, e_m \,m_b \, .
\end{align}
As default values we use
\begin{align}
 \!\!\!& e_H=e_m=1 \, , \, e_J=0 \, , \, \mu_0=2 \, {\rm GeV} \, , \, R_0 = 0.85\, \mu_0 \, ,  \nn \\[0.1cm]
 \!\!\!& t_2= 0.25 \, , \, n_1\equiv\frac{Q\, t_1}{1 \, {\rm GeV}} = \left\{\begin{array}{ll} 5 \, ,\hskip 0.8cm
    \!\!\!& Q \geq \frac{5\, {\rm GeV}}{t_2} ,\\[0.2cm]
    \frac{Q\, t_2}{1 \, {\rm GeV}} , & Q \leq \frac{5\, {\rm GeV}}{t_2} \, ,\end{array} \right. 
\end{align}
and $b$, $d$, $\mu_1$ and $\mu_2$ are fixed by demanding smoothness of the
profiles as in Ref.~\cite{Abbate:2010xh}. Compared to Ref.~\cite{Abbate:2010xh} we
have modified the default value of the parameter $n_1$ for the case $Q \leq 5 \,
{\rm GeV}/t_2$ such that one always has $t_1 \leq t_2$, and the profiles can remain
smooth for small values of $Q$. 

We perform the convolution with the soft model function directly in momentum space since the mass-dependent
corrections to the jet and the soft functions cannot be easily treated in Fourier space. This requires a thorough
treatment of fractional plus-distributions of the form $[\ln^n(x)/x^{1+\omega}]_+$, where $\omega$ can be larger
than 0. For convenience the corresponding rules are given in the Appendix. As a model function we use
\begin{align}
F(\ell)=\frac{128\,\ell^3}{3\lambda^4}\, e^{-4 \ell/\lambda} \, ,
\end{align}
which is properly normalized to unity. The parameter $\lambda$ is a measure for the width of the
function and therefore contributes as in Eq.~(\ref{eq:omega1}) together with the
gap parameter to the first nonperturbative moment, $\Omega_1 = \bar{\Delta}+\lambda/2$. 
As a default we use the following parameters,
\begin{align}
 & \Omega_1^{(5)}(13 \, {\rm GeV},13 \, {\rm GeV})=0.5 \, {\rm GeV} \,, \,\, \lambda=0.65 \, {\rm GeV} \, ,\nn \\
 & \alpha_s^{(5)}(m_Z)=0.114 \, , \,\,\, \overline{m}_b(\overline{m}_b) = 4.2 \, {\rm GeV} \,, \nn \\
 & \overline{m}_t(\overline{m}_t) = 163 \, {\rm GeV} \, .
\end{align}
We have checked that the basic characteristics of the results for the mass effects
are rather weakly depending on these parameters within their known accuracy and
on details of the shape of the soft model function, so our observations
represent generic properties of the mass effects from secondary massive quarks.

\begin{figure}
 \centering
 \includegraphics[width= \linewidth]{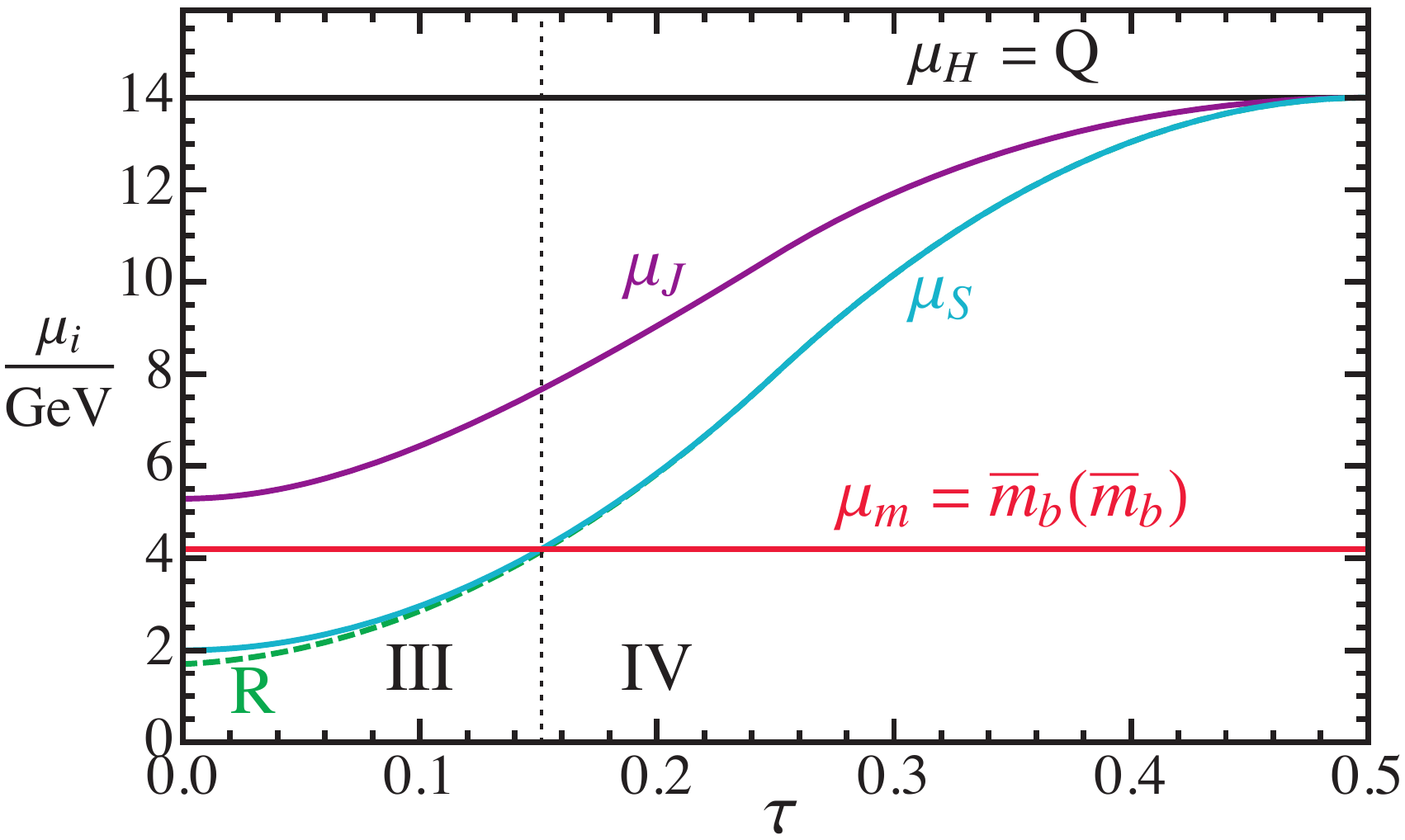}
 \caption{Default profiles for $Q=14$ GeV. The transition value for $\tau$ between the scenarios III and
 IV is indicated by the dotted line. \label{fig:profiles14}}
\end{figure}
\begin{figure}
 \centering
 \includegraphics[width= \linewidth]{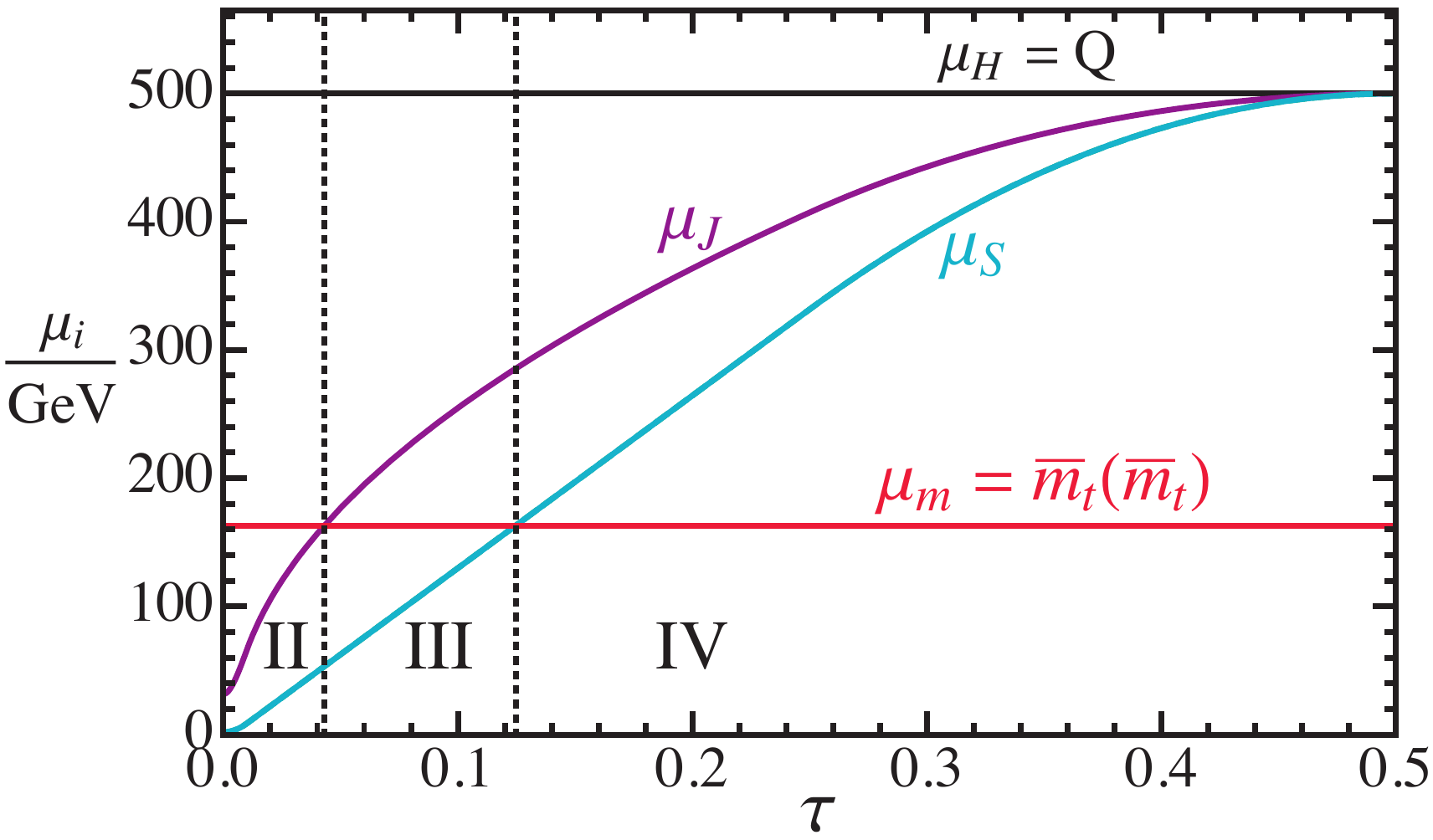}
 \caption{Default profiles for $Q=500$ GeV. The transition values for $\tau$ between the scenarios II and
 III and between the scenarios III and IV are indicated by dotted lines. \label{fig:profiles500}}  
\end{figure}
An important aspect of the practical implementation of the VFNS concerns the
prescription how the predictions within the various scenarios discussed in the
previous sections are patched together to obtain the complete spectrum of the
thrust distribution. As described in Sec.~\ref{sec:massmode_setup}, one switches
between neighboring scenarios when the mass scale is close to one of the
kinematic scales related to the hard coefficient and the jet and the soft
functions. It is therefore natural to tie the prescription for the 
transition to the $\tau$-dependent profile functions for $\mu_m$, $\mu_H$, $\mu_J$
and $\mu_S$. In Figs.~\ref{fig:profiles14} and~\ref{fig:profiles500} the
default profile functions (including the subtraction scale $R$) are shown for
$Q=14$ GeV with $\mu_m=\overline{m}_b(\overline{m}_b)$ and $Q=500$~GeV with
$\mu_m=\overline{m}_t(\overline{m}_t)$, respectively. The prescription we adopt
is that the transition concerning the scenarios II, III and IV is carried out when $\mu_m$
is equal to $\mu_J$ or $\mu_S$,  
respectively. For each choice of the profile functions and the mass mode
matching scale  this leads to a unique value of $\tau$ for the transition. The
resulting $\tau$ regions for the scenarios II, III and IV are indicated in
Figs.~\ref{fig:profiles14} and~\ref{fig:profiles500} by the black dotted
vertical lines. We note that the general freedom to choose the transition value
for $\tau$ in some range causes 
variations in the predictions that are related to higher order corrections in
the same way as changes of the renormalization and matching scales $\mu_m$,
$\mu_H$, $\mu_J$ and $\mu_S$ in their respective ranges. Our prescription ties
the choices made for their profile functions to the range in $\tau$ of the
scenarios.

The practical implementation of the factorization theorems from the different
scenarios at N$^3$LL involves a treatment of perturbative terms at higher orders
that arise from cross terms of the perturbative series for the hard, jet and
soft functions and the mass mode threshold factors. As mentioned
above, we use 
the common approach to expand out the perturbative terms in the matrix elements
and matching factors to $\mathcal{O}(\alpha_s^2)$, but to keep the RG evolution
factors multiplying all expanded terms to the highest order. This approach has
been proven advantageous 
to avoid spurious higher order corrections in the fixed-order expansion and to
obtain reliable information on the remaining renormalization scale dependence at
the corresponding order. This approach is also crucial to achieving a good
numerical agreement of the factorization theorems in overlap regions where two
different scenarios can be employed. In the same spirit, to avoid gaps at the
transition points between neighboring scenarios related to spurious higher
order terms and to obtain a continuous distribution, we also adopt the approach
to expand in the series (at the scale $\mu_m$) for the decoupling relations of
the strong coupling and the gap parameter $\bar{\Delta}$ up to
$\mathcal{O}(\alpha_s^2)$, see Eqs.~(\ref{eq:alphas_dec})
and (\ref{eq:gap_decoupling}). 

\begin{figure}
 \centering
 \includegraphics[width=\linewidth]{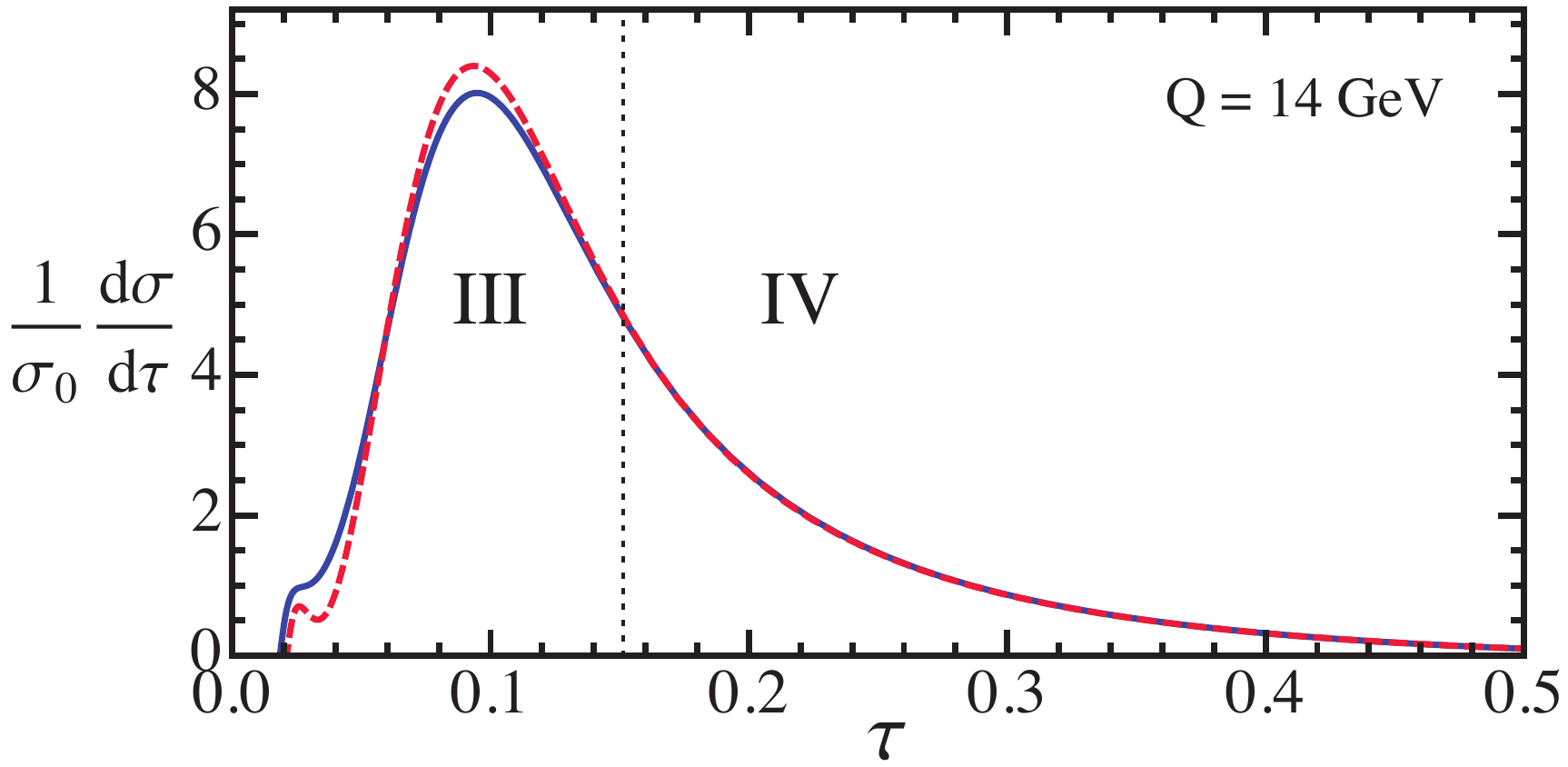}
 \caption{The thrust distribution at $Q=14$ GeV including secondary massive bottom effects (blue, solid)
 compared to keeping the bottom quark massless (red, dashed). \label{fig:Q14}}
\end{figure}
\begin{figure}
 \centering
 \includegraphics[width=\linewidth]{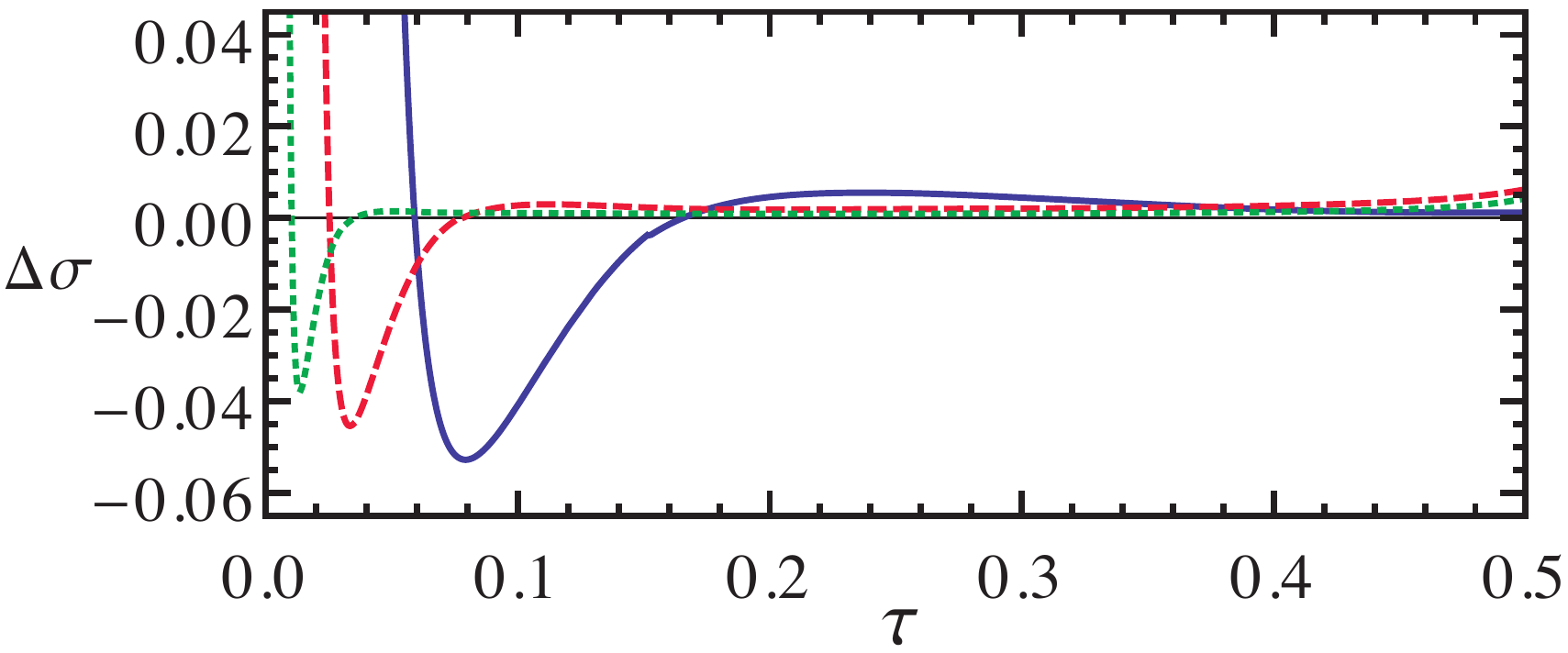}
 \caption{Relative secondary massive bottom effects for \mbox{$Q=14$ GeV} (blue, solid), $Q=35$ GeV
 (red, dashed) and $Q=m_Z$ (green, dotted). \label{fig:DifferentQ}}  
\end{figure}
In Fig.~\ref{fig:Q14} the thrust distribution (for primary production of the
four light quark flavors and secondary production of the light flavors and the
bottom quark) normalized to the Born cross
section $\sigma_0$ is shown for $Q=14$~GeV at N$^3$LL order, based on the
factorization theorems for secondary massive bottom quarks with
$\mu_m=\overline{m}_b(\overline{m}_b)$ (blue, solid line) and in the massless
approximation (red, dashed line). We see that the finite bottom mass effects are
significant at and below the peak, but small in the tail region. Overall the
secondary quark mass effects lead to a significant decrease in the peak cross
section. Interestingly at the peak the deviations are only weakly depending on
the value of $Q$. This is illustrated in Fig.~\ref{fig:DifferentQ}, where we
display for $Q=14$ GeV (blue, solid curve), $Q=35$ GeV (red, dashed curve) and
$Q=m_Z$ (green, dotted curve) the relative change due to the finite mass of
the secondary massive bottom quarks $\Delta \sigma(m_b)$, with  
\begin{align}
\Delta \sigma(m)\, \equiv\, \frac{\frac{\df\sigma}{\df\tau}(m)- \frac{\df\sigma}{\df
    \tau}(m=0)}{\frac{\df\sigma}{\df\tau}(m=0)} \, . 
\end{align}
Here $\frac{\df \sigma}{\df \tau}(m)$ is the complete thrust distribution at N$^3$LL
for primary massless quarks and for secondary quark production which includes the proper number of light quarks and
an additional flavor with mass $m$. We see that in the peak region $\Delta \sigma$ is up to
$\sim -5 \%$ and depends only weakly on the $Q$ value. In the tail region the mass
effects from secondary bottom quarks amount to relative corrections below $1 \%$
for $Q=14$ GeV, and they quickly decrease for larger values of $Q$. 

\begin{figure}
\begin{center}
 \includegraphics[width=\linewidth]{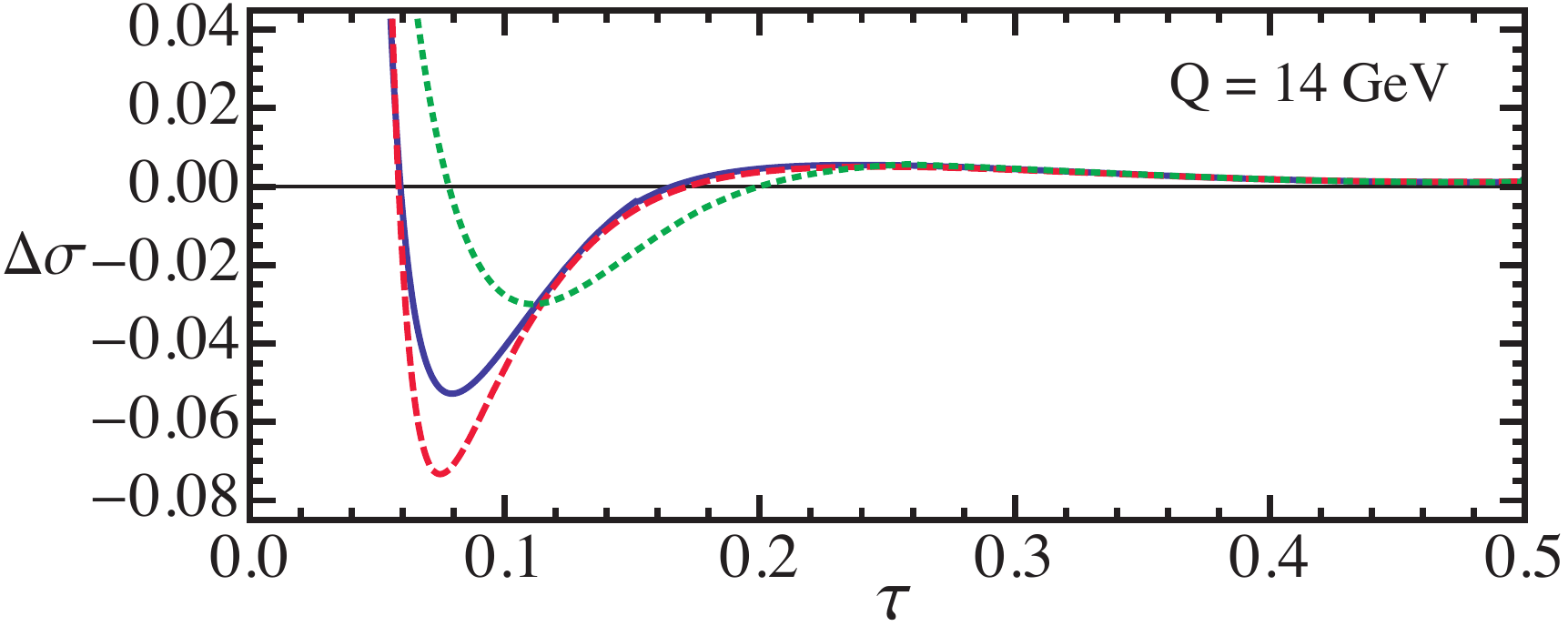}
   \end{center}
 \caption{Dependence of $\Delta\sigma$ on the mass mode matching scale
   $\mu_m$ for secondary bottom quarks for $Q=14$~GeV: $\mu_m=m_b$ (blue, solid), $\mu_m=m_b/2$ (red, dashed),
   $\mu_m=2m_b$ (green, dotted). \label{fig:mum}}
\end{figure}
It is also important to discuss the scale variations of the mass correction
$\Delta \sigma$. In Fig.~\ref{fig:mum} the impact of the variations of $\mu_m$
is illustrated for the bottom quark case for $Q=14$ GeV. The curves are for
$\mu_m$ equal to the $\MS$ bottom mass and for variations by factors of two and
one half. The $\mu_m$ dependence is quite small in the tail and far-tail
regions. In the peak region, on the other hand, the variation of  $\Delta \sigma$
increases to $4\%$ and grows even further below the
peak, where $\Delta \sigma$ changes sign. This behavior is generic for the
bottom quark case and very similar for other values of $Q$.
This behavior might appear formidable for $\Delta\sigma$, but it should be judged taking into account
that in the peak region the dependence on $\mu_m$ is related to missing higher
order corrections of the complete thrust distribution and not only to $\Delta
\sigma$. In other words, the finite mass of the secondary heavy
quark leads to a modification of the peak behavior which represents a property
of the complete thrust distribution and not just of $\Delta \sigma$
itself. The result shows that variations of $\mu_m$ need to
be accounted for when estimating perturbative uncertainties in the peak region.

\begin{figure*}
  \begin{center}
  \subfigure{\epsfig{file=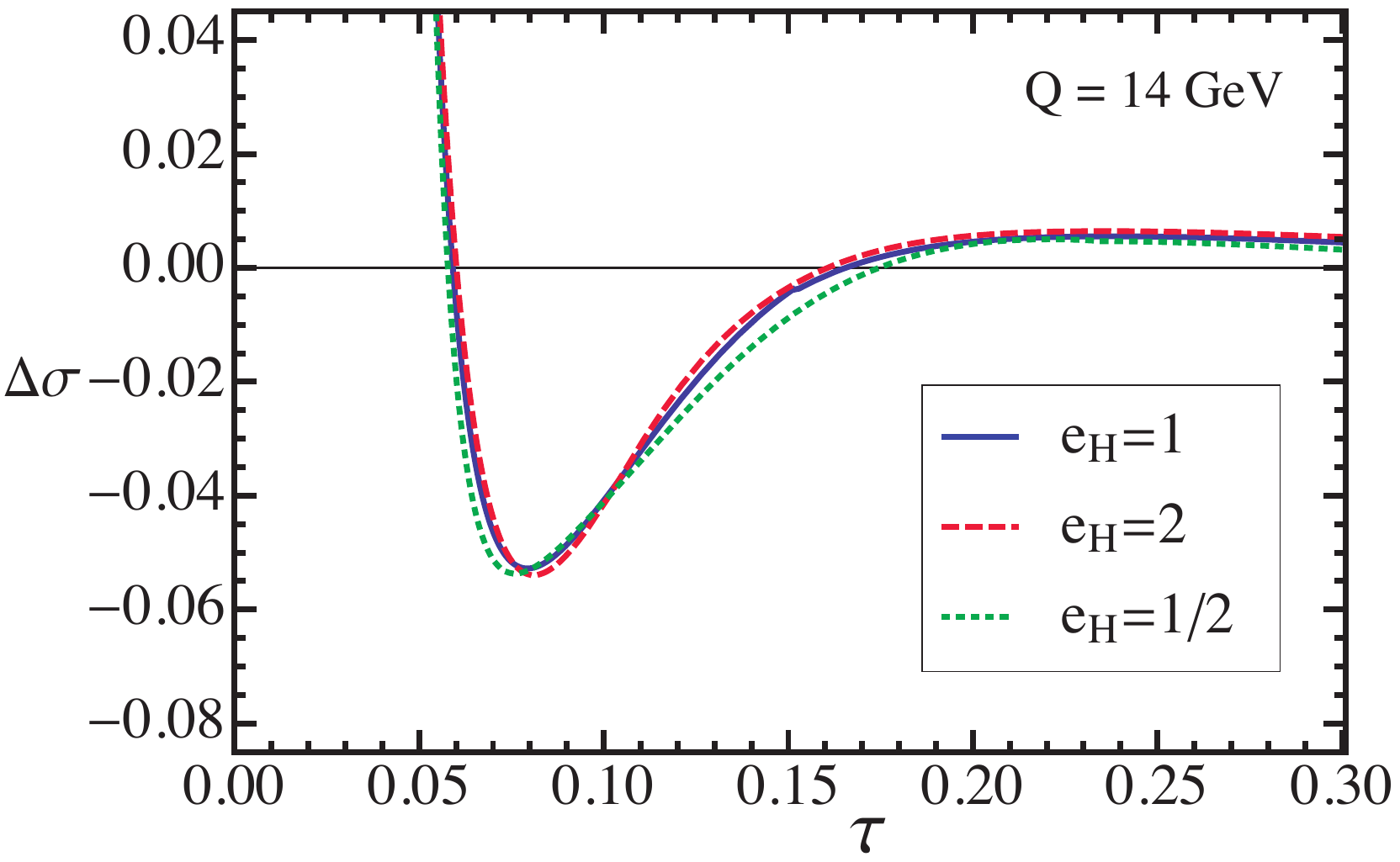,width=0.325\linewidth,clip=}} \hfill
  \subfigure{\epsfig{file=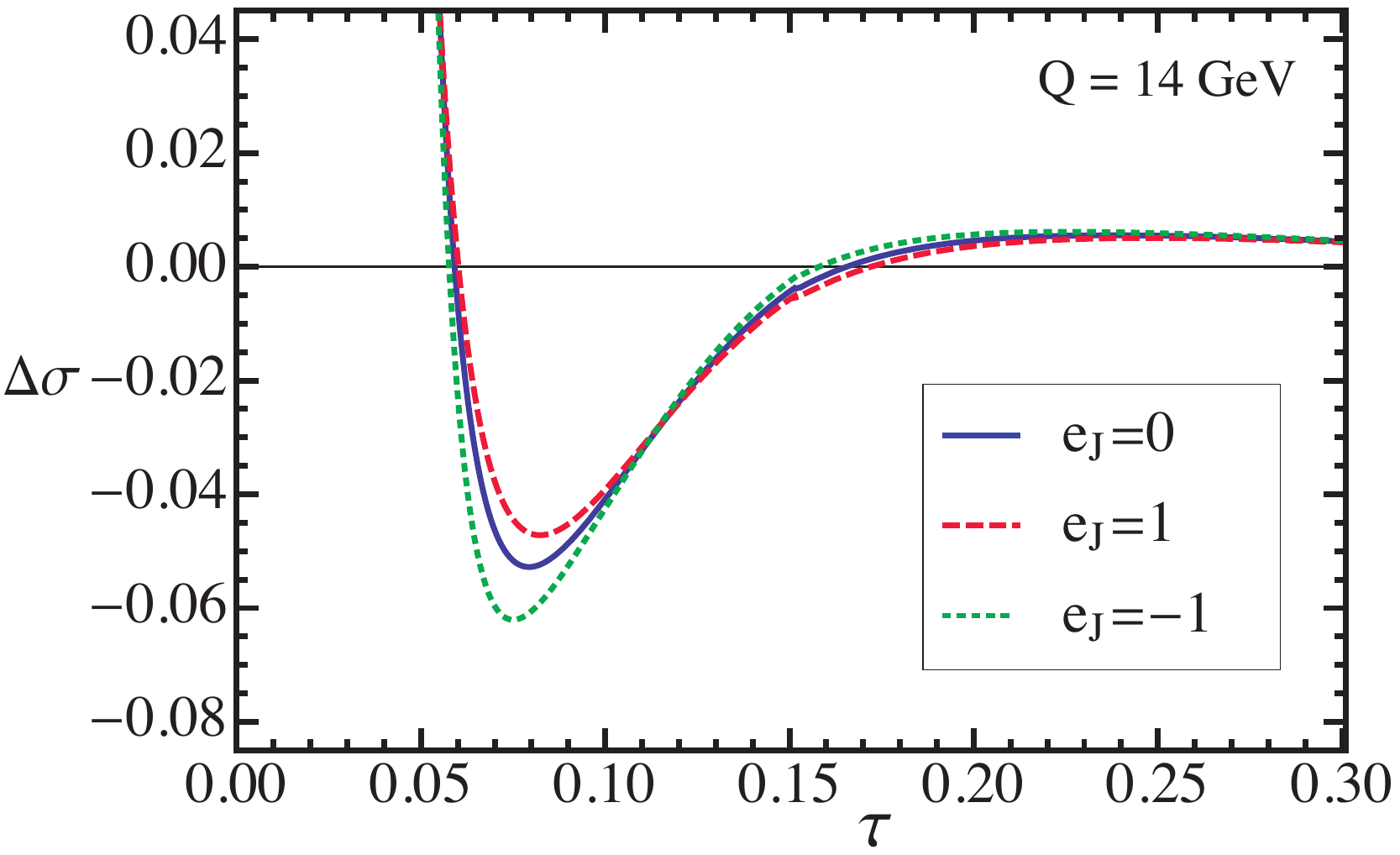,width=0.325\linewidth,clip=}} \hfill
  \subfigure{\epsfig{file=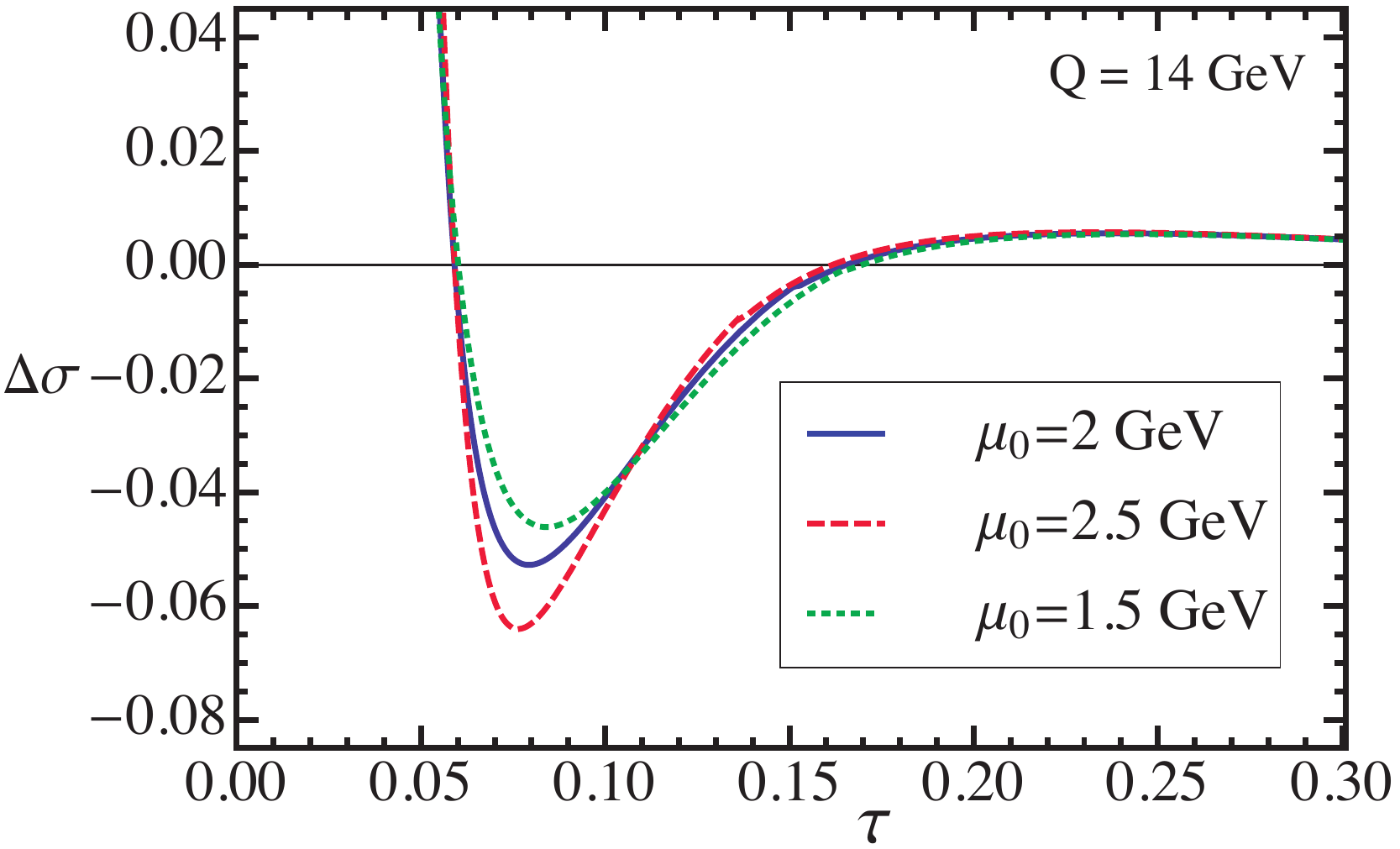,width=0.325\linewidth,clip=}} \hfill
   \subfigure{\epsfig{file=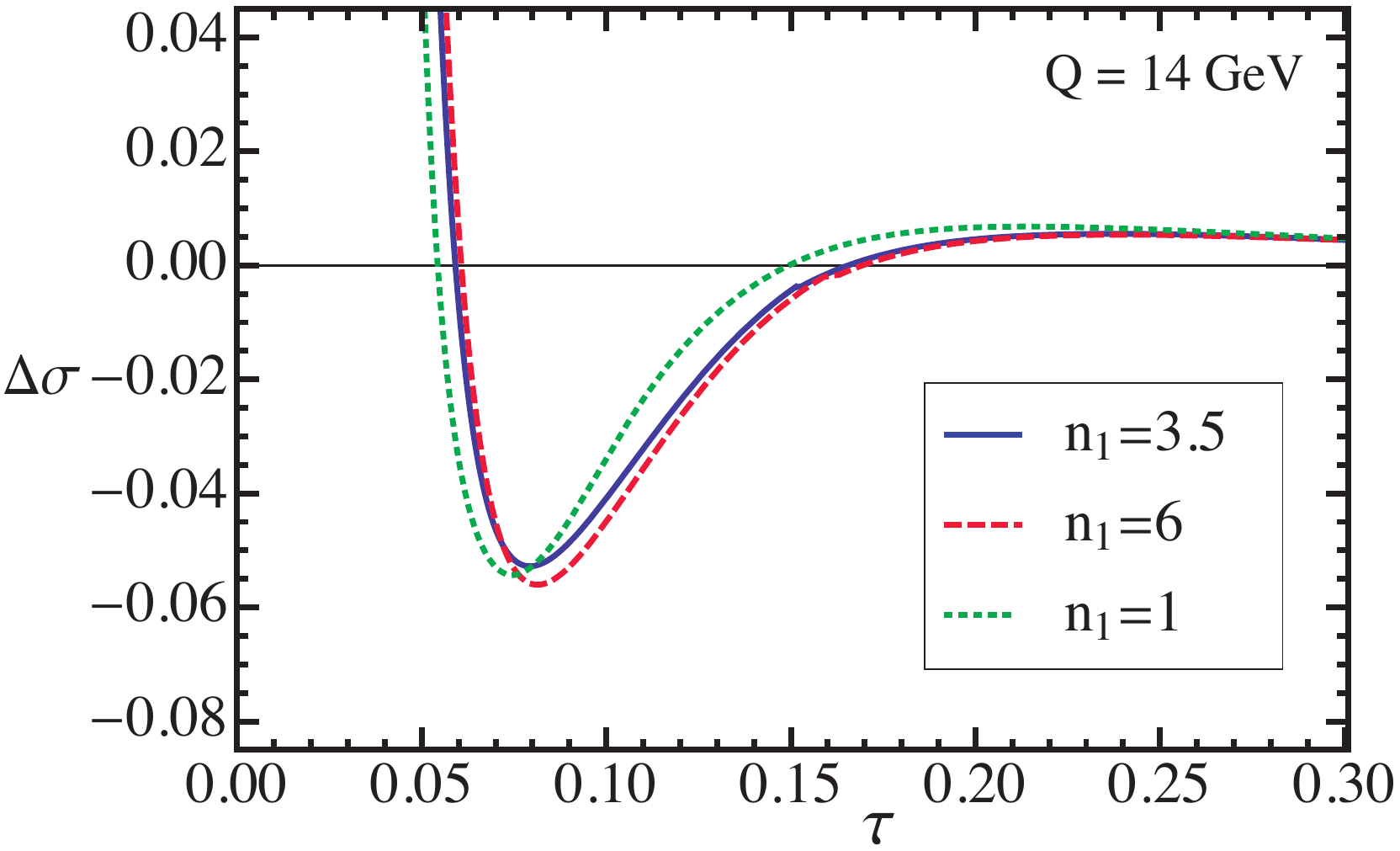,width=0.33\linewidth,clip=}}
    \subfigure{\epsfig{file=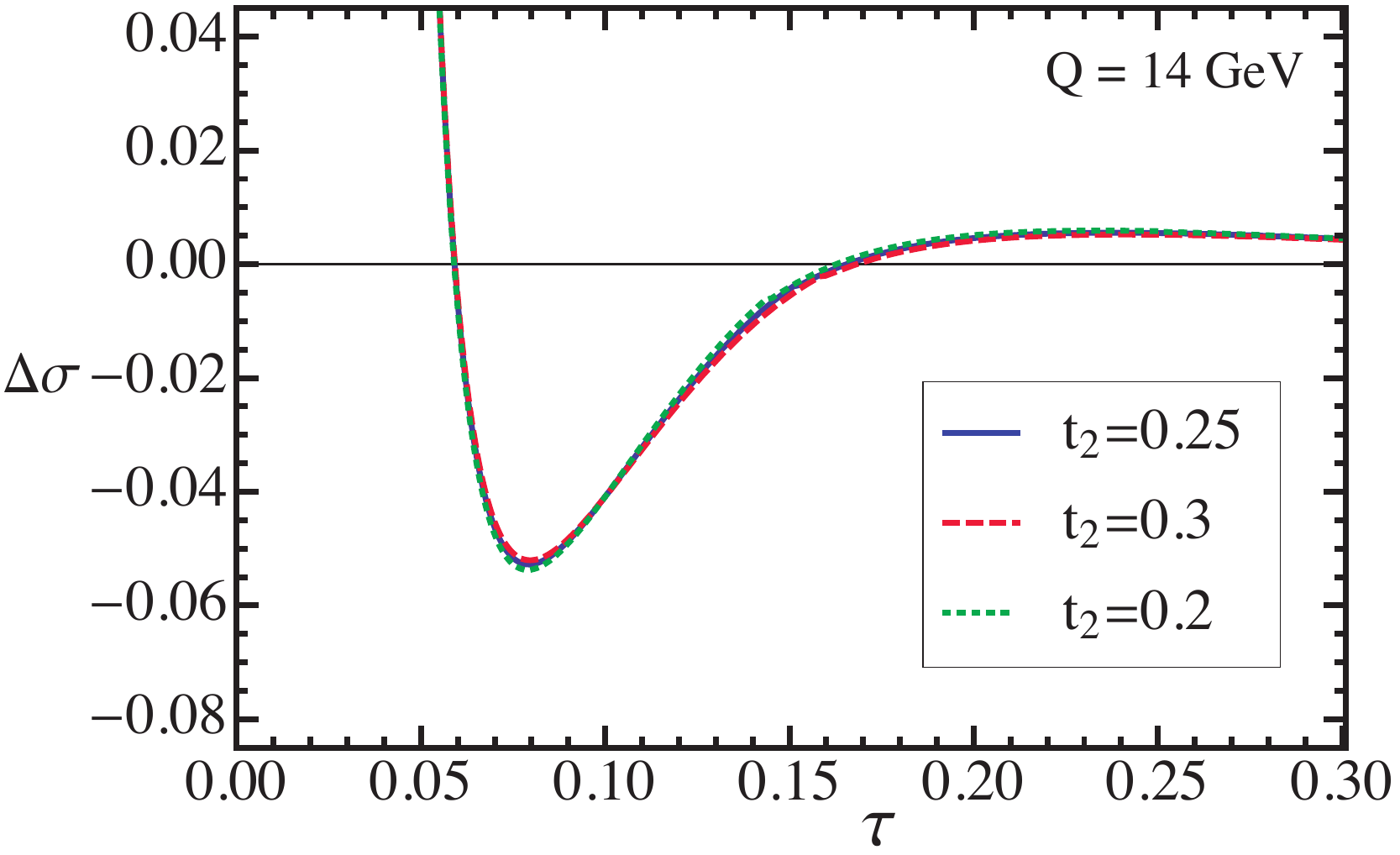,width=0.33\linewidth,clip=}}
   \caption{Relative secondary bottom mass effects for $Q=14$ GeV under variation of the profile
   parameters for the hard, jet and soft scales.}
  \label{fig:profile_vary}
  \end{center}
\end{figure*}

In this context it is also important to examine the variation of $\Delta \sigma$
due to changes of the profile functions for $\mu_H$, $\mu_J$, $\mu_S$ as well as
for $R$. In Fig.~\ref{fig:profile_vary} $\Delta \sigma$ is shown for the bottom
quark case  at $Q=14$~GeV for variations of the parameters $e_H$, $e_J$, $\mu_0$, $n_1$ and
$t_2$, see Eqs.~(\ref{eq:mu_H})-(\ref{eq:mu_R}), which parametrize changes of
the profile functions. The ranges of variation are described in the figure
caption and are identical to the ones used
for the thrust analysis of Ref.~\cite{Abbate:2010xh} (except for $n_1$ which
requires a lower range for low $Q$ values). These variations induce visible
changes in $\Delta \sigma$, but they are in general much smaller 
than the dependence on $\mu_m$ discussed just above. This
outcome is again generic for other values of $Q$ and also for the top quark case
and shows that independent variations of the
profile functions {\it and} of $\mu_m$ are essential for a thorough assessment of the
scale variations of the complete thrust distribution.

\begin{figure}
 \centering
 \includegraphics[width=\linewidth]{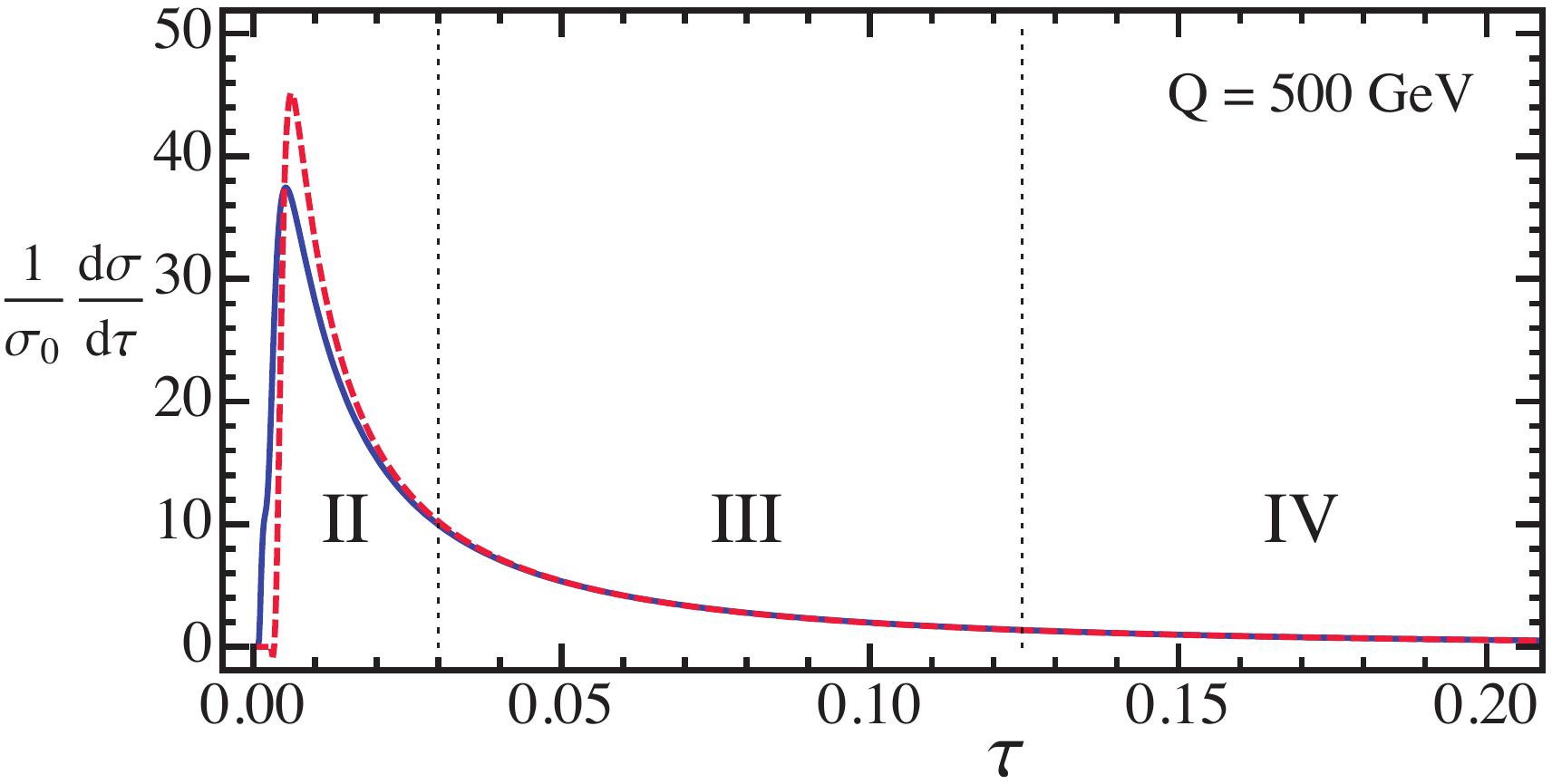}
 \caption{The thrust distribution at $Q=500$ GeV including secondary massive top effects (blue, solid)
 compared to keeping the top quark massless (red, dashed). \label{fig:Q500}}  
\end{figure}

\begin{figure}
 \centering
 \includegraphics[width=\linewidth]{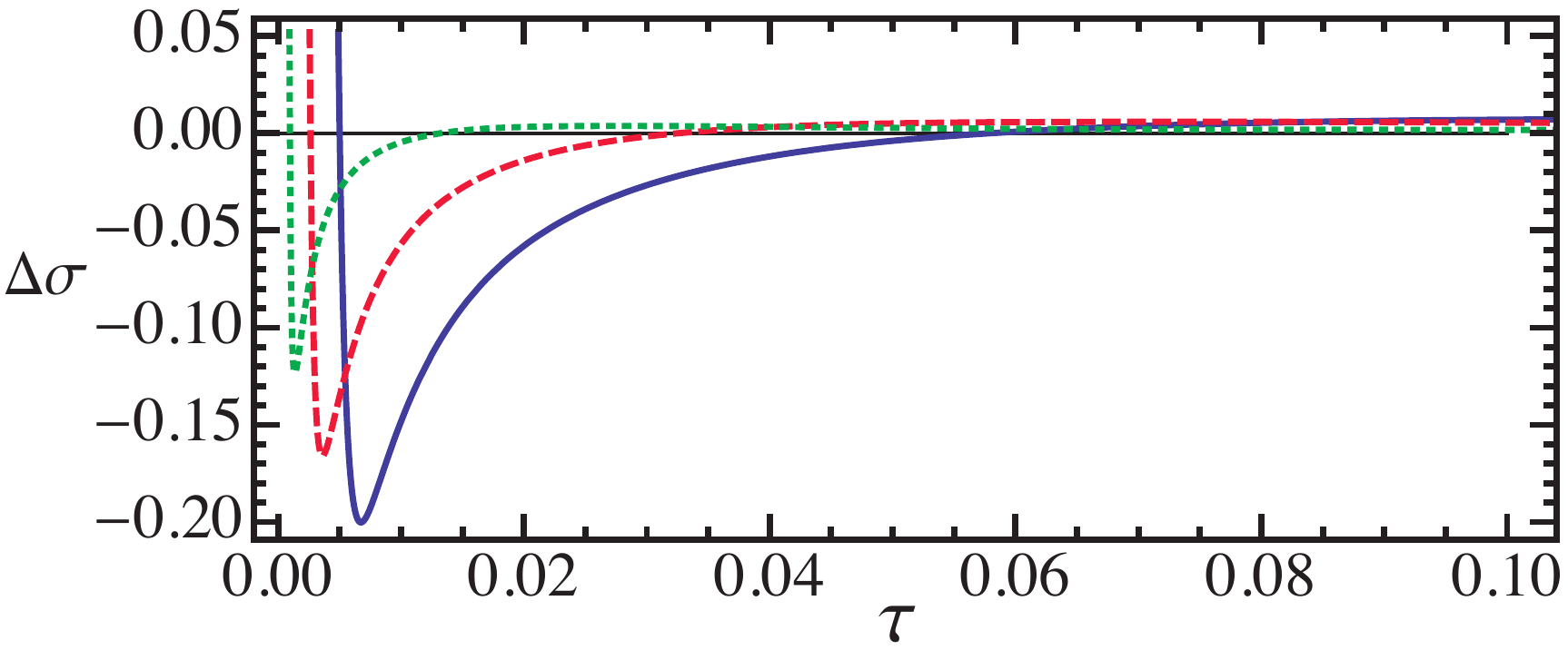}
 \caption{Relative secondary massive top effects for \mbox{$Q=500$ GeV} (blue, solid), $Q=1000$ GeV
 (red, dashed) and $Q=3000$ GeV (green, dotted). \label{fig:DifferentQ2}}
\end{figure}

We complete our analysis by showing in Fig.~\ref{fig:Q500} the thrust
distribution for primary light quark production at $Q=500$~GeV with $n_l=5$
massless flavors and a secondary massive top quark (blue, solid line). The
figure also shows the prediction for the case where the secondary top quark is
treated as massless (red, dashed line). For all parameters the default values
mentioned above are used. It is visible that the finite top quark mass
causes, besides a reduction of the distribution at the peak, as we have observed
in the bottom quark case, also a shift of the peak to lower $\tau$ values. This
effect is related to the top quark mass effects in the R-scale dependence of the
gap parameter, which is significantly modified below the massive threshold as
described in Sec.~\ref{sec:gap}.
In Fig.~\ref{fig:DifferentQ2} $\Delta \sigma$ is shown for the top quark case
for $Q=500$~GeV (blue, solid line), $Q=1000$ GeV (red, dashed line) and $Q=3000$
GeV (green, dotted line). We find again sizable mass effects at and below the
peak of the distribution. In the tail region, on the other hand, the top mass
effects are relatively small. At the peak the mass effects amount to $10-20 \%$ 
and remain significant even for large values of $Q$. In contrast to the bottom quark case, the size of the top mass
effects is larger than the variations due to changes 
of the mass mode matching scale $\mu_m$ which amount to $1$ to $2\%$ in the total cross section. We
note that in the top quark case and for these c.m.\ energies the decoupling
limit, i.e.\ the thrust distribution with just 5~massless flavors and a decoupled
top quark, is at the peak much closer to the VFNS prediction than the massless
top approximation (shown in the red dashed curve). This is because in the peak
region the top mass is significantly larger than the jet and the soft scales
such that the decoupling approximation is more appropriate than the massless one.

\section{Conclusions}\label{sec:conclusions}
In this work we have provided a variable flavor number scheme (VFNS) for inclusive final state jets taking the production
of secondary massive quarks (see Fig.~\ref{fig:QCDdiag2}) for the thrust distribution as the concrete application. In the 
dijet limit the singular terms factorize into a hard function given by a Wilson coefficient for the dijet production
current, a jet function describing the hard collinear radiation within the jet and a soft function describing soft
radiation between the jets. The factorization is based on the fact that the typical invariant masses of the fluctuations 
described by these factors are widely separated, where the size of these scales depends on the thrust variable $\tau$. 

Including the radiation of secondary massive quarks the situation becomes more complicated since the quark mass adds
another $\tau$-independent scale to the situation that can lead to different kinds of hierarchies or relations w.r.\ to
the hard jet and soft scales. Apart from a more complicated analytic structure one has to deal with potential
mass singularities and the summation of logarithms of the mass -- all with the requirement that the massive quark
decouples in the infinite mass limit and that one obtains the well-known massless quark description in the limit
that the quark mass vanishes.


The VFNS we provide is based on the hierarchy between the hard, jet and soft scales and accounts in addition for the
quark mass, for which, however, no assumption concerning a hierarchy w.r.\ to the other scales has to be imposed.
The effective theory description was explained in detail in Ref.~\cite{Gritschacher:2013pha} using the field theoretic 
analogy between secondary massive quarks and the radiation of ``massive gluons". The four emerging effective theories are related to the four hierarchical regions of the mass 
w.r.\ to the hard, jet and soft scales, and deal with collinear and soft massless quarks and gluons as well as 
corresponding ``mass modes". The treatment of these mass modes differs for
each of the four effective theories and leads to modifications of the known factorization for massless quarks.
These (i) add a quark mass dependence in the hard coefficient, the jet and soft functions, (ii) affect the RG evolution
which is carried out with different flavor numbers above and below the quark mass scale and (iii) lead to additional massive 
threshold correction factors when the RG evolution of one of the structures crosses the mass scale.
The factorization scale where this crossover is performed (``mass mode matching scale") can be varied in analogy to
the renormalization scales. An essential aspect of the VFNS is that in a transition region between two neighboring 
effective theories both of their descriptions can be used which ensures that the transition is continuous (up to
perturbative terms from beyond the order that is employed in the description). 

An important outcome of our mass mode treatment is that the way in which the massive quark contributes to the hard coefficient,
the jet function and the soft function as well as to their mass mode threshold factors can be determined for each of them
individually without having to deal with the factorization theorem for the thrust distribution as a whole. This is related
to the fact that the hard coefficient, the jet function and the soft function are by themselves well-defined
field theoretical quantities that can be renormalized consistently. If the description includes the small mass case
(including the massless limit) the $\MS$ renormalization prescription is employed for the secondary massive quark
corrections. On the other hand, if the description includes the large mass case (including the decoupling limit)
the on-shell (low momentum-subtraction) renormalization condition is employed for the secondary massive quark
corrections. For sufficiently large scales hierarchies it may be possible to use massless quark results with
the appropriate number of flavors for some of the structures as a good approximation. The transition regions are
located where the quark mass is of the order of the hard, jet or soft scales,
and the difference of the renormalized  quantities constitutes the mass mode threshold correction factors.

We have discussed the numerical impact of the secondary quark mass effects on the thrust distribution.
These turn out to be small corrections in the tail region, but sizable at the peak, so that a phenomenological
analysis of this region will have to take them into account. For the assessment of the renormalization scale
dependence it is crucial to account for changes of the mass mode matching scale.

In this work we have demonstrated the concept of a VFNS for final state jets for the secondary massive quark
effects in the thrust distribution. In subsequent publications our proposed VFNS shall be applied also to the primary 
production of massive quarks~\cite{Butenschoen:2014}, where new subtleties arise as well as to other processes including deep inelastic 
scattering~\cite{Hoang:2015iva}, where the relation to VFNS for initial state massive quarks is elucidated.

\begin{acknowledgments}
We thank the Erwin-Schrödinger Institute (ESI) for partial support in the framework
of the ESI program ``Jets and Quantum Fields for LHC and Future Colliders''. We thank Daniel Samitz for pointing out an error in Eq.~(\ref{eq:M_3+}) in the previous version of this paper. P.\,P. would like to thank Bahman Dehnadi for helpful discussions and cross checks
in parts of the numerical analysis.\newline
\end{acknowledgments}

{\bf Note added:} 
After initial submission we received Ref.~\cite{Ablinger:2014vwa} where the $\mathcal{O}(\alpha_s^3)$ non-singlet 
corrections to the heavy flavor matching of the parton distribution functions were calculated. Considerations in DIS for 
large $x$ in analogy to this work~\cite{Hoang:2014ira,Hoang:2015iva} lead to a consistency relation similar to Eq.~(\ref{eq:consistency_M}) involving this threshold correction, which implies
\begin{align}\label{eq:M_3+}
& \mathcal{M}^{C,+}_{3} = \frac{1}{4} \mathcal{M}^{J,+}_{3} = \frac{1}{4} \mathcal{M}^{S,+}_{3} \\
 & =  -C_{\!F} T_F \left\{C_{\!F}\! \left(\frac{1417}{27}- \frac{604}{9}\, \zeta_3 + \frac{8 \pi^4}{15} -
 \frac{32}{3} B_4 \!\right) \right. \nn \\
  & + C_{\!A} \!\left(\!-\,\frac{17726}{729}+ \frac{824 \pi^2}{243} + 30\, \zeta_3  - \frac{88 \pi^4}{135} +
  \frac{16}{3} B_4 \!\right) \nn \\
  & + \left.  T_F n_l \!\left(\!-\,\frac{12032}{729} + \frac{256}{27}\,\zeta_3\!\right) +
  T_F \!\left(\frac{6032}{729} - \frac{448}{27}\, \zeta_3\!\right)\! \right\} ,\nonumber
\end{align}
where
\begin{align}
\!\!\!B_4 = \frac{2}{3} \,\ln^4(2) -\frac{2\pi^2}{3} \,\ln^2(2) -
 \frac{13\pi^4}{180} + 16 \,\Li_4\Big(\frac{1}{2}\Big)\, .
\end{align}
For $n_l=4$ and $n_l=5$ light flavors one obtains the numerical values $\mathcal{M}^{C,+}_{3}=28.2337$ and
$\mathcal{M}^{C,+}_{3}=29.9362$, respectively.\footnote{Note that the published version of this paper contains an error in Eq.~(\ref{eq:M_3+}) and the associated numerical values.} We note that the numerical impact of the constants
$\mathcal{M}^{C,+}_{3}$ and $\mathcal{M}^{J,+}_{3}$ in the analysis of Sec.~\ref{sec:analysis} is negligibly small.

\appendix

\section{Plus-distributions}\label{sect:plusdist}
We give the definition and integral prescription for the plus-distributions appearing in the nonlocal evolution factors,
denoted by $[\ln^n(x)/x^{1+\omega}]_+$, for arbitrary (non-vanishing) $\omega$.
The prescription of these plus-distributions is based on
an analytic continuation to a domain with a well-behaved convergence following  Ref.~\cite{Fleming:2007xt}
\begin{align}
 \left[\frac{\theta(x) \ln^n(x)}{x^{1+\omega}}\right]_+ \stackrel{\omega< 0}{\longrightarrow} & \,\,\,
 \frac{\theta(x)\, \ln^n(x)}{x^{1+\omega}} \, .
\end{align}
\begin{widetext}
This leads to a definition based on the subtraction of strictly divergent terms at $x=0$ that can be generalized to arbitrary values of $\omega$ in a straightforward way: 
\begin{align}
 & \left[\frac{\theta(x) \ln^n(x)}{x^{1+\omega}}\right]_+ =  \lim_{\epsilon \rightarrow 0} 
 \left[\frac{\theta(x-\epsilon) \ln^n(x)}{x^{1+\omega}} -\sum_{k=0}^\infty \delta^{(k)}(x) \,\frac{(-1)^k}{k!}\,  \frac{\Gamma(n+1,(\omega-k)\ln(\epsilon))}{(\omega-k)^{n+1}}\right] \, .
\end{align}
This expression can be rewritten as an integral prescription
\begin{align}
& \int_0^X \df x\left[\frac{\theta(x) \ln^n x}{x^{1+\omega}}\right]_+ f(x) = \int_0^X \df x\,\frac{\ln^n(x)}{x^{1+\omega}}\left[f(x)-\sum_{k=0}^{\infty} f^{(k)}(0) \frac{x^k}{k!}\right] 
 -\sum_{k=0}^{\infty} f^{(k)}(0) \frac{1}{k!} \frac{\Gamma(n+1,(\omega-k)\ln(X))}{(\omega-k)^{n+1}} \, ,
\end{align}
where the sums can be truncated for $k=N$ if $\omega<N+1$.
\end{widetext}

\bibliography{VFNSthrust}{}
\bibliographystyle{my_bibstyle}
 
\end{document}